\documentclass[useAMS,usenatbib]{mn2e}

\def\arcs{\rlap{.}$^{\prime\prime}$}
\def\arc{$^{\prime\prime}$}

\title[Galaxy Secular Mass Flow Rate Determination]
{Galaxy Secular Mass Flow Rate Determination
Using the Potential-Density Phase Shift Approach}
\author[Xiaolei Zhang and Ronald J. Buta]{X. Zhang$^{1}$\thanks{E-mail: xzhang5@gmu.edu} and R. J.  Buta$^{2}$\\
$^{1}$Department of Physics and Astronomy, George Mason University, 
4400 University Drive, Fairfax, VA 22030, USA\\
$^{2}$Department of Physics and Astronomy, University of Alabama, 514 University Blvd E, Box 870324, Tuscaloosa, AL 35487, USA}

\begin{document}

\date{Accepted . Received ; in original form }

\maketitle

\label{firstpage}

\begin{abstract}
We have carried out an initial study of a small sample of nearby spiral
and barred galaxies with a broad distribution of Hubble types in order
to have a first estimate of the magnitude of their secular mass
accretion/excretion rates in the context of bulge building and
morphological evolution along the Hubble sequence. The stellar surface
density maps of the sample galaxies were derived from the archival data
of the {\it Spitzer} Infrared Nearby Galaxies Survey (SINGS), 
as well as the Sloan Digital Sky Survey (SDSS).  The
corresponding molecular and atomic surface densities
were derived from archival CO(1-0) and HI interferometric observations of
the BIMA SONG, THINGS, and VIVA surveys.  The method used for
determining the radial mass flow rates follows from our previous work using
the potential-density phase shift approach, which utilizes a
volume-type torque integral to calculate the angular momentum exchange
rate between the basic state disk matter and the density wave modes.
This volume-type torque integral is shown to contain the contributions
from both the gravitational torque couple and the advective torque
couple at the nonlinear, quasi-steady state of the wave modes, in sharp
contrast to its behavior in the linear regime, where it contains only
the contribution from the gravitational torque couple as was
shown by Lynden-Bell \& Kalnajs in 1972.  The phase-shift/volume-torque
method is shown to yield radial mass flow rates and angular momentum
transport rates several times higher than similar rates estimated using
the traditional method of gravitational torque couple of Lynden-Bell
and Kalnajs, and this difference reflects the dominant role played by
collisionless shocks in the secular evolution of galaxies containing
extremely non-linear, quasi-steady density wave modes.  It is shown
that these non-linear modes maintains their quasi-steady state at the
expense of a continuous radial mass flow and the resulting
morphological transformation of galaxies throughout their lifetime,
not only for late-type and intermediate-type disk galaxies but, under favorable
conditions, also for earlier types including S0s and disky ellipticals.
Enabling this transformation along the entire Hubble sequence, we show
here for the first time using observational data that {\em stellar}
mass accretion/excretion is just as important, and oftentimes much more
important, than the corresponding accretion/excretion processes in the
{\em gaseous} component, with the latter being what had been emphasized
in most of the past secular evolution studies.  We address the
implications of our result in the context of the cosmological evolution
of galaxies, as well as the challenges it poses to the standard 
Lambda Cold Dark Matter (LCDM) paradigm.  We also point out the
connection between the large-scale collective dissipation process 
enabled by the galactic density wave modes to similar processes happening
in fluid turbulence and in symmetry breaking processes
in fundamental physics, which places the galaxy evolution study in the context
of the general studies of complex, nonequilibrium systems.

\end{abstract}

\begin{keywords}
galaxies: evolution; galaxies: formation; galaxies: spiral \end{keywords}

\section{INTRODUCTION}

The possibility that galaxy morphologies can transform 
significantly over their lifetime, not only through violent 
episodes such as merger and satellite accretion, but also 
through slow and steady internal secular dynamical processes, 
is an idea that has been gaining acceptance in recent decades.  
In the past, work on secular evolution has been focused on 
gas accretion in barred galaxies and the growth of
pseudo (disk-like) bulges (Kormendy \& Kennicutt 2004, hereafter KK04,
and references therein).  This is partly due to the long-held
notion, naturally expected in the classical mechanics of few-body
systems developed since Newton's time and refined throughout the 18th and 
19th century, that gas is the only mass component capable of dissipation, while
the stellar component is adiabatic, and generally does not lose or gain
energy and angular momentum as stars orbit around the center of a
galaxy.  This view, however, is expected to be challenged in view
of parallel developments in other physical systems such as atmospheric
convection flow, chemical clocks, and turbulence in fluid dynamics, where 
the interaction of the multitudes of degree-of-freedom in these 
complex, nonlinear and nonequilibrium systems leads to the emergence of 
chaos and coorperative behavior (i.e. self-organized, large-scale 
non-equilibrium patterns that break the symmetry of the parent systems).
The emergence of self-organized global patterns are often 
accompanied by the emergence of new meta-laws governing the 
evolution of these complex systems, a characteristic that has been
revealed not only in the classical complex systems,
but more prevalently in the quantum domain which
dominates the studies of the foundations of physics.

The indication that there is the possibility of significant secular
{\it stellar mass redistribution} in galaxies in fact already appeared in the 
seminal work of Lynden-Bell and Kalnajs (1972, hereafter LBK), who 
showed that a trailing spiral density wave possesses a 
gravitational torque that over time can transport angular
momentum outward. LBK were interested in the angular momentum 
transport phenomenon because they were seeking a generating 
mechanism for the spiral density waves, thought at the time
to have a longevity of the order of $10^9$ years because of
the radial group velocity discovered by Toomre (1969), which
resulted in their progagation and eventual absorption at the
at the inner Lindblad resonance (thus the reason for the title 
of the LBK paper: ``On the Generating Mechanism of Spiral Structure'').
Since the density wave is considered to possess negative 
energy and angular momentum density inside corotation relative
to the basic state (i.e. the axisymmetric disk, see, e.g. Shu [1992], Chapters
11 and 12 for the fundamentals of density wave theory), an 
outward angular momentum transport from the inner disk would 
encourage the spontaneous growth of the wave trains.  LBK
at that time were not interested in the secular morphological evolution of
the basic state of the disks.  In fact, in the same paper, they showed that
for WKBJ (tightly wrapped) waves, the long-term energy and angular momentum
exchange between a {\em quasi-steady} wave train and the basic state is 
{\it identically zero} away from the wave-particle resonances, thus 
much of the basic state of the disk does not experience secular mass 
redistribution according to their theory.  
That this zero-exchange between the wave and the disk matter is possible, 
in the presence of the outward angular momentum transport by gravitational 
torque couple, is because there is a second type of torque couple, the so-called
advective torque couple (due to lorry transport, or Reynolds stress,
see \S 6 of LBK paper for more detailed explanation), 
that opposes the gravitational
torque couple, and the sum of the two types of torque couples is a constant
independent of galactic radius (LBK, see also Binney \& Tremaine 2008, 
hereafter BT08, pp. 492).  The total torque couple, which is equal to the 
rate of angular momentum flux, is thus a constant in their theory during 
the outward radial transport of angular momentum, 
and there is no interaction of the wave and the basic state
except at the wave-particle resonances (i.e., LBK thought the wave picks
up angular momentum from the basic state at the inner Lindblad resonance
and dumps it at the outer Lindblad resonance, and {\em en route} of this radial
transport the total angular momentum flux remains constant).  
Strickly speaking, LBK proved these results only for WKBJ 
(tightly-wrapped) waves, and they have also taken the zero-growth (steady
wave) limit.  

Zhang (1996,1998,1999, hereafter Z96, Z98, Z99) 
showed that the classical theory of LBK ignored
an important collective dissipation process\footnote{
The phrases ``collective dissipation'' or ``collective
effects'' we adopted in describing the galactic density 
wave dynamics attempt to describe a general tendency for non-equilibrium,
self-organized coherent structures to spontaneously generate
emergent global dynamics through the local interactions
of subsystems.   Other phrases used in this
context in the literature, ranging in applications including
condensed matter physics, fluid dynamics, economics and social sciences,
include ``cooperative effect'', ``coupling and interaction
among the degrees-of-freedom of a complex system''. A not entirely
fitting analogy of collective versus hierarchical control is
the difference between a democracy and a dictatorship.  
A democracy functions through the interactions of sub-entites but
ultimately is controlled by the innate dynamical forces of
society and history, whereas a dictatorship in its ultimate form
exerts a top-down control with no lateral influence among
the subunits. In the case of a galaxy possessing a density
wave pattern, a top-down control would be for orbits to respond
entirely {\em passively} to an {\em applied} potential, whereas 
the collective point of view allows us to take into account new
physics emergent from the mutual interactions of N-particles
in the galaxy, which on the other hand is implicitly
determined by the innate dynamics of such a system to
spontaneously form self-organized global instabilities when
the boundary condition is right.}
present in the gravitational
N-body disks possessing {\em self-organized, or spontaneously-formed, density
wave modes}. This process is mediated by collisionless shocks at the
density wave crest, which breaks the adiabaticity or the conservation
of the Jacobi integral condition
-- a condition which was shown to be valid only
for a {\em passive} orbit under an {\em applied} steady spiral
or bar potential (BT08, p. 179), and is now shown not to be 
obeyed by orbits undergoing collective dissipation as is applicable
for a self-sustained or self-organized spiral or bar modes.
The overall manifestation of the collective dissipation
process is an azimuthal phase shift between the potential and the density
distributions of the density wave pattern, and for a self-sustained mode,
the radial distribution of phase shift is such that it is
positive inside corotation and negative outside (Z96, Z98).
The presence of the phase shift means that for every annulus of the galaxy, there
is a secular torque applied by the density wave on the disk matter in the
annulus, and the associated energy and angular momentum exchange between the wave
mode and the basic state of the disk at the nonlinear, quasi-steady state of
the wave mode (in the linear regime, the same phase shift distribution
leads to the spontaneous emergence of the wave mode. See Z98).  
As a result, the disk matter inside corotation
(both stars and gas) loses energy and angular momentum to the wave, and
spirals inward, and the disk matter outside corotation gains energy
and angular momentum from the wave and spirals outward.  {\it This energy
and angular momentum exchange between the wave and the basic state
of the disk thus becomes the ultimate driving mechanism for the secular
evolution of the mass distribution of the basic state of galaxy disks.}
The energy and angular momentum received by the wave from the basic
state, incidentally, also serve as a natural damping mechanism 
for the spontaneously
growing unstable mode, allowing it to achieve quasi-steady state
at sufficiently nonlinear amplitude Z98). In other words, the density
wave pattern achieves a state of {\em dynamical equilibrium} at the expense of a 
continuous flux of matter, energy, and entropy through the system,
with the local entropy production balanced by entropy export by the
pattern (thus no local entropy increase, as evidenced by the steady
wave amplitude), but with the overall entropy increase of the combined system
of the pattern plus the environment: a condition satisfied by similar kind of
nonequilibrium dissipative structures in nature (Nicolis \& Prigogine
1977).

In Z98, a set of analytical expressions for the secular mass
accretion/excretion rate was derived, and was confirmed quantitatively
in the N-body simulations presented in the same paper.  However, due to
the two-dimensional nature of these simulations, where the bulge and halo were
assumed to be spherical and inert, the simulated wave mode has an average
arm-to-interarm density contrast of 20\% and potential contrast of 5\%,
both much lower than the commonly observed density wave contrast in
physical galaxies, so the simulated disk did not evolve much (Z99),
despite the fact that these low evolution rates conform exactly
to the analytical formula's prediction of the mass flow rate for the
corresponding (low) wave amplitude (Z98, section 5.2).
Incidentally, the evolution rate given by this theory is proportional
to the wave amplitude squared, so a change from 10\% average wave amplitude
to 30\% average wave amplitude is almost a factor of 10 difference in
effective secular mass flow rate.

Since state-of-the-art N-body simulations have yet to reproduce the
extremely nonlinear density wave amplitudes in observed physical
galaxies, an alternative to confirming the magnitude of the secular
evolution effect using numerical simulations,
is to use observed near- and mid-infrared images
directly as tracers of galaxy mass, and apply the analytical formalism
of Z96, Z98 to these observationally-derived surface density maps to derive
secular mass flow rates.  As a first
step, Zhang \& Buta (2007, hereafter ZB07) and Buta \& Zhang (2009,
hereafter BZ09) used near-infrared
images of more than 150 observed galaxies to derive the radial
distribution of the azimuthal potential-density phase shifts, and to
use the positive-to-negative (P/N) zero crossings of the phase shift curve to
determine the corotation radii (CRs) for galaxies possessing
spontaneously-formed density wave modes.  This approach works because
the alternating positive and negative humps of phase shift distribution
lead to the correct sense of energy and angular momentum exchange
between the wave mode and the disk matter to encourage the spontaneous
emergence of the mode (Z98).  In ZB07 and BZ09,
we have found good correspondence between the predicted CRs
using the potential-density phase shift approach, and the resonance
features present in galaxy images, as well as with results from other
reliable CR determination methods within the range of validity of these
methods.  Besides CR determination, an initial test for mass flow rate 
calculation, for galaxy NGC 1530, was also carried out in ZB07 using the
same volume-torque-integration/potential-density phase shift approach.
We found there that since NGC 1530 has exceptionally high
surface density, as well as extremely-large arm-to-interarm contrast,
mass flow rates of more than 100 solar masses per year were obtained across
much of the disk of this galaxy. This level of mass flow
rate, if sustainable, is more than sufficient to transform the Hubble
type of a late type galaxy to an early type (i.e. S0 or disky elliptical)
within a Hubble time.

Recently, we have applied the potential-density phase shift/volume-torque
method to an initial sample of disk galaxies of varying Hubble
types in order to study various issues relating to the secular
mass redistribution and Hubble-type transformation.
In what follows, we apply the potential-density phase shift/volume-torque 
method to our initial sample of disk galaxies with varying Hubble types
in order to estimate the mass flow rates for both stars and
gas, and to compare their relative importance.  

\section{THEORETICAL BASIS FOR APPLYING THE POTENTIAL-DENSITY PHASE SHIFT
APPROACH TO THE MASS FLOW RATE CALCULATION}

The detailed discussion on the collective dissipation mechanism responsible 
for the secular redistribution of both stellar and gaseous mass in galaxies 
can be found in Z96, Z98, Z99.  In ZB07 
and BZ09, procedures for calculating the radial distribution 
of potential-density phase shift of the density wave patterns using observed
infrared images are described.  A similar procedure, with slight modification, will
be used for the calculation of the mass flow rates in the current paper.
We now briefly summarize the theoretical derivations relevant
to these calculations.

For pedagogical purposes we will derive the equation for mass flow rate through
the orbit decay rate of an average star under the action of a density wave mode.
The (inward) radial mass accretion rate at a galactic radius R is related to the
mean orbital decay rate $-dR/dt$ of an average star through
\begin{equation}
{dM(R) \over dt} = - {dR \over dt} 2 \pi R \Sigma_0(R)
\end{equation}
where $\Sigma_0(R)$ is the mean surface density of the basic state
of the disk at radius R.

We also know that the mean orbital decay rate of a single star is related to its
angular momentum loss rate $dL^*/dt$ through
\begin{equation}
{dL^* \over dt} = - V_c M_* {dR \over dt}
\end{equation}
where $V_c$ is the mean circular velocity at radius R, and $M_*$ the mass of the relevant star.

Now we have also
\begin{equation}
{dL^* \over dt} = {\overline{ {{dL} \over {dt}}}} (R)
{M_*  \over \Sigma_0}
\end{equation}
where ${\overline{ {{dL} \over {dt}}} (R)}$
is the angular momentum loss rate of the basic
state disk matter per unit area at radius R.

Since
\begin{equation}
{\overline{ {{dL} \over {dt}}}} (R)
= { 1 \over {2 \pi}} \int_0^{2 \pi}
\Sigma_1 { {\partial {\cal V}_1} \over {\partial \phi}} d \phi
\label{eq:eq4}
\end{equation}
(Z96), we have finally

\begin{equation}
{dM (R) \over dt} = 
{R \over {V_c}} \int_0^{2 \pi} 
\Sigma_1  {{\partial {\cal V}_1} \over {\partial \phi}} d \phi
\label{eq:eq5}
\end{equation}
where the subscript 1 denotes the perturbation quantities.
This result, even though derived through the stellar orbital decay rate, 
is in fact general, and can be applied to the mass accretion rate
of both stars and gas, as long as the relevant perturbation surface
density is used.  And we note that the potential perturbation to be
used for the calculation of either the stellar or gaseous accretion
needs to be that of the total potential of all the mass components,
since the accretion mass cannot separate the forcing field component,
and responds only to the total forcing potential.

In the above derivation we have used a volume-type of torque $T_1(R)$,
\begin{equation}
T_1(R) \equiv R \int_{0}^{2 \pi}
\Sigma_1  {{\partial {\cal V}_1} \over {\partial \phi}} d \phi,
\end{equation}
which was first introduced in Z96 and Z98 in the context 
of the self-torquing of the disk matter
in a unit-width annulus at R by the potential of
the associated spontaneously-formed density wave mode.  
The volume torque is equal to
the time rate of angular momentum exchange between the density wave and the
disk matter in a unit-width annulus at R, for wave modes in approximate
quasi-steady state.  In the past, two other types of 
torque-coupling integrals have also been used (LBK; BT08).  These
are the gravitational torque couple $C_g(R)$
\begin{equation}
C_g(R) = {R \over {4 \pi G}} \int_{- \infty}^{\infty} \int_0^{2 \pi}
{{\partial {\cal V}} \over {\partial \phi}}
{{\partial {\cal V}} \over {\partial R}}  
d \phi dz ,
\end{equation} 
and the advective torque couple $C_a(R)$
\begin{equation}
C_a(R) = R^2 \int_0^{2 \pi} \Sigma_0 V_R V_{\phi} d \phi ,
\end{equation}
where $V_R$ and $V_{\phi}$ are the radial and azimuthal velocity perturbation
relative to the circular velocity, respectively. 

In the classical theory, the volume torque integral $T_1(R)$
can be shown to be equal to $dC_g/dR$ in the linear regime
(see Appendix A2 of Z98, original derivation due to 
S. Tremaine, private communication).
However, for spontaneously-formed density wave modes,
when the wave amplitude is significantly nonlinear and the
importance of collisionless shocks at the density wave crest
begins to dominate, it was shown in Z98 that one of the
crucial conditions in the proof of the $T_1(R) = dC_g/dR$ relation,
that of the validity of the differential form of the Poisson 
equation, is no longer valid.  At the quasi-steady state (QSS)
of the wave mode, it was shown in Z99 that 
in fact $T_1(R) = d(C_a+C_g)/dR \equiv dC/dR$!
An intuitive derivation of this equality can be given as
follows: $ d(C_a+C_g)/dR \equiv dC/dR$ is the gradient of total
wave angular momentum flux in the Eulerian picture, 
and $T_1(R)$ is the rate of angular momentum
loss by the disk matter to the wave potential
in a unit-width annulus located at R in the
Lagrangian picture.  At the quasi-steady
state of the wave mode, these two need to balance each other so the wave
amplitude does not continue to grow (i.e., all the negative angular momentum
deposited by the wave goes to the basic state of the disk matter
and none goes to the wave itself,
so that the wave amplitude does not continue to grow, as required by the
condition for the quasi-steady state of the wave mode).

The past calculations of the secular angular momentum redistribution
rate (i.e. Gnedin et al. [1995], Foyle et al. [2010]) considered only
the contribution from gravitational torque couple and ignored the contribution
of the advective torque couple (which cannot be directly estimated using the
observational data, except through our round-about way of estimating the
total torque using the volume-type of torque integral $T_1(R)$, and then
subtracting from it the contribution of the gravitational torque to
arrive at the contribution of the advective torque).
Later in this paper, we will show that for the kind of density
wave amplitudes usually encountered in observed galaxies, the advective contribution
to the total torque in fact is several times larger than the
contribution from the gravitational torque.  Furthermore, in the nonlinear
regime the advective torque couple is of the same sense
of angular momentum transport as the gravitational torque couple -- another
characteristic unique to the nonlinear mode.  
Past calculations of gas mass accretion near the central region
of galaxies (e.g. Haan et al. 2009) are likely to have
significantly underestimated the gas mass flow rate for the same
reason.

Later in this paper, we will also show that the density wave modes
in observed galaxies show predominantly
a kind of two-humped shape for the volume torque distribution (with zero crossing 
of the two humps located at CR).  This two-humped distribution of the volume
torque is equivalent to the two-humped distribution of the phase shift
because the phase shift is defined through the volume torque (Z98).
In ZB07 and BZ09, we found that galaxies often possess nested modes of
varying pattern speeds, and each modal region has its own well-defined
two-humped phase shift distribution.  The constant pattern speed regime
is delineated between the negative-to-positive (N/P) crossings of the
adjacent modes, with the P/N crossing, or CR, sandwiched in between.
The two-humped volume-torque or phase shift distribution
is consistent with (and a direct result of) the bell-shaped torque 
couplings previously found in both N-body simulations 
(Z98, for gravitational, advective, and total
torque couplings), and in observed galaxies (Gnedin et al. 1995\footnote{A little
clarification on the chronologies of these early works appear in order.  
In the late 80's Zhang carried out initial studies of the possible
dynamical mechanisms for the secular redistribution of galaxy matter
and summarized the results of these early work in her PhD thesis (Zhang 1992)
During the 1994 the Hague IAU general assembly she discussed 
this work with the then Princeton graduate student J. Rhodes, 
who requested preprints from her.  She sent Rhodes the three draft papers 
that eventually published as Z96, Z98, Z99) in ApJ.  Rhodes showed 
these papers to his professors Jeremy Goodman and David Spergel, 
who both showed great interests.  Goodman subsequently asked his 
then graduate student O. Gnedin to begin a study of spiral torques 
in galaxy M100 using the approach of LBK, which was later published 
as Gnedin et al. (1995).  The three papers by Zhang,
however, experienced significant delays (including resubmissions)
in the refereeing process, and thus came out after the Gnedin et al. 
paper, even though the latter work was carried out due to the direct 
influence of the former.} for gravitational torque coupling only).  These
characteristic distributions are another important piece of evidence
that the density waves present in disk galaxies are in fact spontaneously-formed
unstable modes in the underlying basic state of the disks (Z98 \S3.4).

\begin{figure}
\vspace{310pt}
\centerline{
\includegraphics{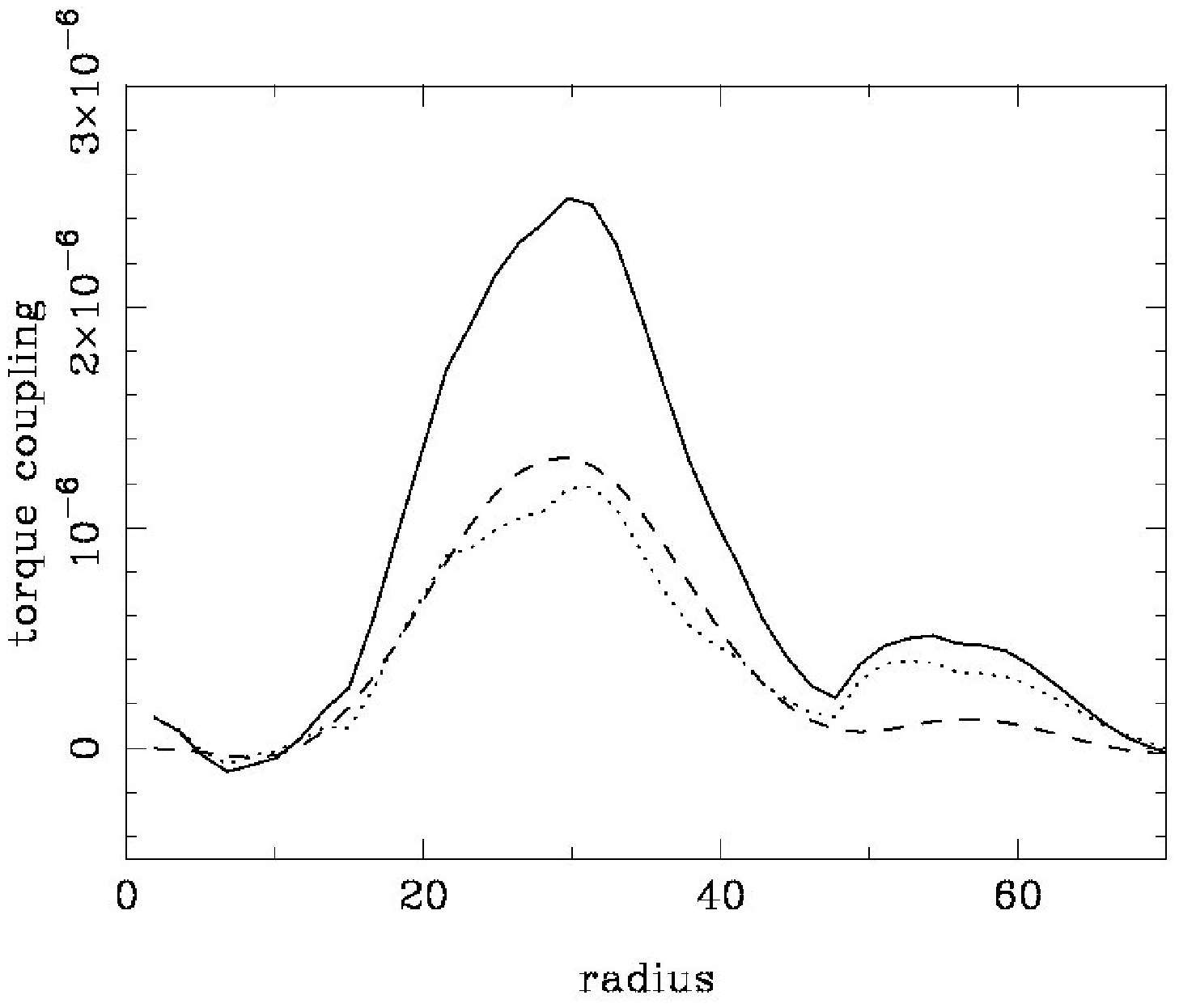}
\includegraphics{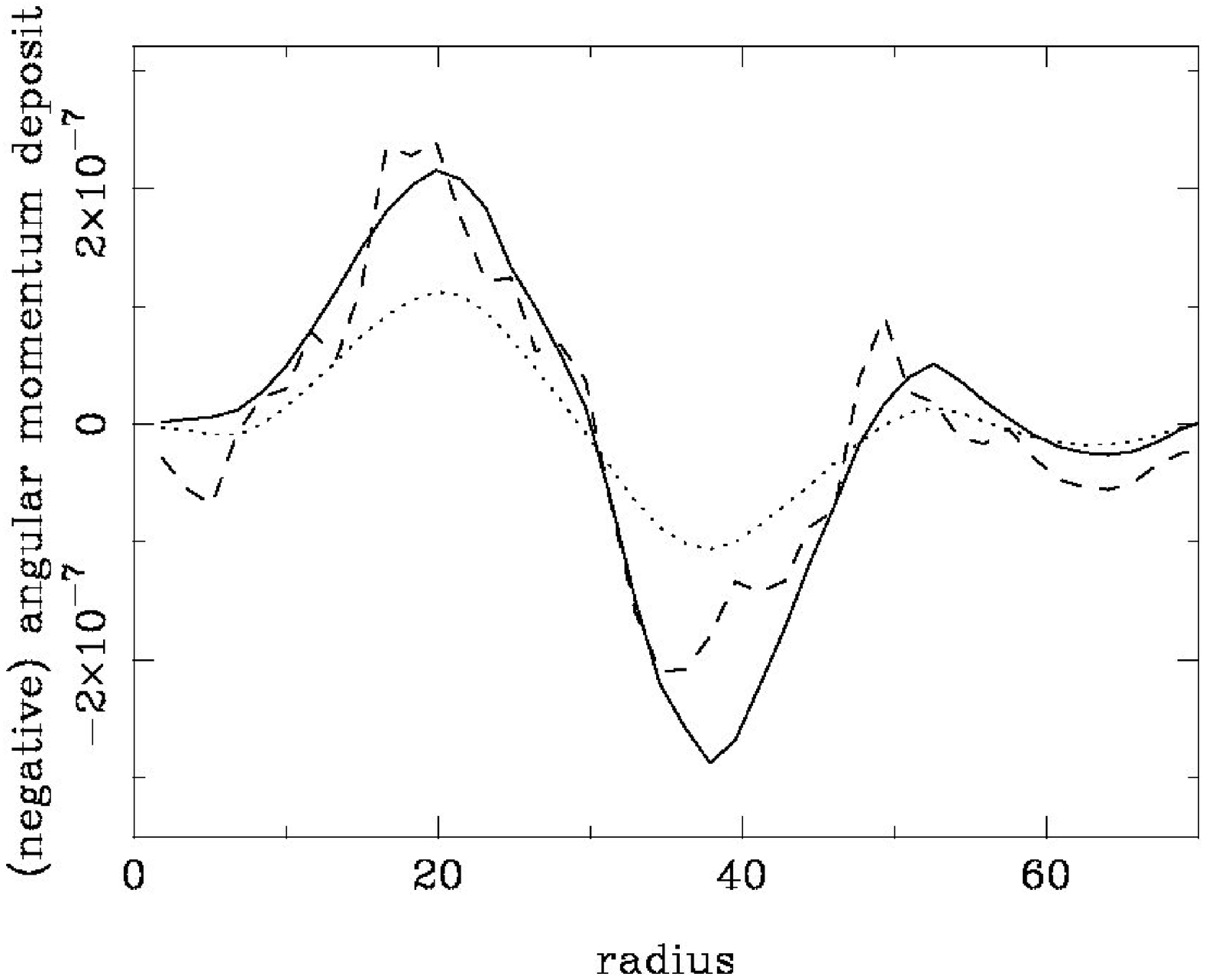}
}
\caption{{\it Top}: Gravitational (dotted), advective (dashed), and total (solid) torque couples
from the N-body simulations of Z96, Z98). 
{\it Bottom}: Gradient of gravitational (dotted) and total (dashed) torque couples,
and the volume torque  $T_1$ (solid), from the same N-body simulations (previously unpublished).}
\label{fg:Fig1}
\end{figure}

In Figure \ref{fg:Fig1}, we show the result of an N-body simulation from
Z96, Z98. The top frame, which displays the various torque
couplings, shows curves which have 
the characteristic bell-shapes that we have commented above.  
The peaks of the bell curves are near the CR of the dominant spiral mode 
in this simulation, at r=30.  In the bottom frame, we show the calculated 
{\em gradient} of the gravitational and total torque couples, 
and compare them with $T_1(R)$, using quantities from the same
set of N-body simulation.  It is clear that $T_1(R) \neq
dC_g/dR$, and rather is closer to $dC/dR$, though the equality is not yet
exact because this particular simulated N-body mode never achieved true steady
state. The shallower second bell-shape in the top frame (which corresponds 
to the second positive-followed-by-negative-hump feature in the bottom frame) 
is due to a spurious edge mode.
In physical galaxies, as our examples below will show, the difference between
$T_1(R)$ and $dC_g/dR$ becomes even more pronounced than in these N-body
simulations, due to the higher degree of nonlinearity of the wave modes
in physical galaxies.

\section{OBSERVATIONAL MOTIVATIONS FOR STUDYING
SECULAR EVOLUTION ALONG THE HUBBLE SEQUENCE USING
NEARBY SAMPLES OF DISK GALAXIES}

As is well known, the original Hubble sequence was motivated 
by Hubble's early work in classifying nearby galaxies in order to find
evolutionary connections among them according to the Jeans theory
of the morphological evolution of galaxies (Jeans 1928).
The original Jeans theory, however, got the evolution 
direction backwards, which apparently influenced Hubble's choice
of the names for ``early'' and ``late'' type galaxies. With the observational
evidence accumulated over the past few decades, we know now that
galaxy morphological transformation generally proceeds along the
Hubble sequence in the reverse direction (i.e. from late to early types),
and this evolution generally takes a good fraction or longer of the
entire age of the universe.  The Hubble Deep and Ultra Deep Fields
as well as many follow-on observations showed that high-z objects 
in general do not, at first sight, seem to conform to the classical 
Hubble classification.  One might thus wonder: what is the rationale
for studying the evolution along the Hubble sequence of nearby
galaxies, since it is not clear that it would be relevant to
the evolution of galaxies we observe across cosmic time?

We want to say at the outset that what we are trying 
to explore and establish is a broad trend that as the
universe evolves, the galaxy population as a whole progressively evolves
from disk-dominated systems to bulge-dominated systems.
There is plenty of observational evidence now that
the disk fraction was higher at $z=1$ and has since decreased
(Lilly et al. 1998).  This is most naturally explained by secular evolution 
since the number density of galaxies of all types has not evolved much over the
same time range (Cohen 2002), and the merger fraction since z=1 is low
(Lopez-Sanjuan et al. 2009).
  
The next question is, what would serve as a 
natural starting point of secular evolution along
the Hubble sequence?  An obvious choice would be
the very late type disks, i.e., Sd/Sc's, and low surface brightness
galaxies (LSBs), since these are at the tail end of the
Hubble sequence.  One subtle point here is that the present-day LSB and 
late-type galaxies could not have been the ancestors of the
present-day early-type galaxies, both due to their formation time
difference and their statistical size/mass difference.  We want,
however, to bring up an analogy of the galaxy evolution scenario
with the Darwinian evolution theory.  According to this theory,
humans evolved from an ape-like ancestor, but this is not the same as
saying modern-day apes will one-day evolve into modern-day humans. 
This difference is the well-known distinction between an evolution-tree
(which has historic connotation) and a ladder-of-life (which is
used for present-day classification of species).
The two phrases have close connections, but should not be confused with
each other.   For the problem of the secular evolution of galaxies,
the same principle applies: we can fully envision that the ancestors
of the present-day early-type galaxies were likely to be disk
systems much larger in size than present-day late-type disks,
and they appear much earlier in the evolution history of the universe.
Some of these early large disks may be the recently-discovered
rotationally-supported high-z disks (Genzel et al. 2006; 
Ceverino et al. 2011), as well as the large rotationally-supported
Damped Lyman Alpha (DLA) systems (Wolfe et al. 2005).  Many of the fainter
outer disks of these high-z systems may lie beyond detection
(Hopkins et al. 2009), and thus we need to also take into 
account possible selection effects in interpreting high-z results (this
aspect is related to the difficulty in detecting the high-z quasar
host galaxies).  The observed ``down-sizing'' trend of galaxy
mass assembly (Cimatti et al. 2006; Nelan et al. 2005) 
also gives support to the idea that the first
disk galaxies formed are likely to be larger ones. 

There are theoretical reasons why galaxy evolution 
tend to proceed through a disk-phase, i.e., in a dissipative self-gravitating
system the natural step of relaxation is first to settle onto a disk 
with the axis of rotation corresponding to the axis of 
maximum momentum of inertia (Zeldovich 1970).  Such a system subsequently 
continues its entropy-increasing evolution towards the direction
of increasing central concentration together with an extended
outer envelope, and it gets rid of the excess angular momentum
during this secular evolution process through the formation of 
large-scale collective instabilities (Z96). Besides considering
purely gravitational processes for galaxy formation,
it is essential to point out that we need also to take into account 
evidences that early galaxy formation may involve a significant
role played by turbulence and supersonic shocks (Ozernoy 1974; Bershadskii \& 
Screenivasan 2003; Zhang 2008 and the references therein). Incidentally, 
the primordial turbulence scenario for galaxy formation helps to eliminate
one of the key reasons for the involvement of CDM in early galaxy formation,
i.e. the smallness ($10^{-5}$) of initial seeds of perturbation on the cosmic
microwave background (CMB), since the turbulence
shock compression will form galaxies bypassing
the slow purely gravitational collapse, and this scenario may
underlie the observed clumpy high-z disks that appears to
be in a violent kinematic state (Genzel et al. 2006).
This more violent disk formation scenario is different from the likely
formation process of the quiescent LSB galaxies in the nearby universe, which 
appear to have formed mostly from gravitational collapse alone.
Despite the heterogeniety of the initial conditions, which allow some
galaxies to by-pass certain stages of the evolution along the Hubble
sequence, or else allow the galaxies to fast-evolve through the initial
stages to build up a substantial bulge relatively early in time, the
general trend of galaxy formation and evolution still appears to be
through a disk-dominated configuration (van Dokkum et al. 2011).
This is only reasonable because the large-scale gravitational
instabilities which form spontaneously in disk galaxies can accelerate
the speed of entropy evolution many orders of magnitudes higher compared
to a smooth distribution (Zhang 1992), and nature, as it appears,
has always chosen her evolution configuration that maximizes the
entropy production rate.  The support for this argument in a 
broader context is the theory on ``dissipative structures''
(Nicholis \& Prigogine 1977): 
the large-scale coherent pattens present in nonlinear,
far-from-equilibrium complex systems exist not merely as
pretty and impressive veneers, they serve an important
dynamical role of greatly accelerating the entropy evolution of the
underlying non-equilibrium system -- in the case of a 
self-gravitating system, the direction of entropy evolution is towards
configurations of ever-increasing central concentration, together
with an extended outer envelope, and in the case of galaxies,
this is the same direction as evolution along the Hubble sequence
from late to early Hubble types.

It is for these reasons that we study the evolution of nearby galaxies
along the Hubble sequence as a template for what could have
possibly happened in galaxies in the more distant universe, even
knowing that for individual galaxies the trend
may not be all smooth and universal, i.e., the initial conditions may not
all have been that of a thin disk, and the smooth disk-dominated 
evolution could be interrupted by a merger or satellite accretion.
We also point out that the same phase-shift induced collective
dissipation process is likely to operate in configurations that
have a vertical extent as well: i.e., halos and the thick inner
regions of galaxies may possess skewness which lead to the same kind
of potential-density phase shift as in thin disks, 
and thus will have accelerated evolution
according to the same basic dynamical process we describe for
disk galaxies.

\section{ANALYSES OF INDIVIDUAL GALAXIES}

In this section, we analyze the morphology and kinematics of six
bright, nearby galaxies, to set the stage for further evolutionary
studies of these galaxies in the next section.  The parameters of these
six galaxies are given in Table 1. The total-mass maps derived using
methods described below and in Appendix A are shown in Figure~\ref{fg:Fig2}.

\begin{table*}
\caption{Adopted parameters for sample galaxies.$^a$}
\halign{%
\rm#\hfil&
\qquad\rm\hfil#&
\qquad\rm\hfil#&
\qquad\rm\hfil#&
\qquad\rm\hfil#&
\qquad\rm\hfil#&
\rm#\hfil\cr
Galaxy & $i$ & $\phi_n$ & $h_R$ & $h_z$ & Distance & ~~~~~References for \cr
& (degrees) & (degrees) & (arcsec) & (arcsec) & (Mpc) & ~~~~~($i$, $\phi_n$) \cr
\noalign{\vskip 10pt}

NGC 628  & 6 & 25 & 64 (3.6) & 7.1 & 8.2 & ~~~~~Shostak et al. (1983)\cr
NGC 3351 & 40.6 & 13.3 & 44 (3.6) & 8.9 & 10.1 & ~~~~~3.6$\mu$m isophotes \cr
NGC 3627 & 60 & 173 & 66 (3.6) & 13.3 & 10.1 & ~~~~~RC2, Zhang et al. (1993) \cr
NGC 4321 & 31.7 & 153 & 63 (3.6) & 12.6 & 16.1 & ~~~~~Hernandez et al. (2005) \cr
NGC 4736 & 30 & 116 & 135 ($i$) & 27.1 & 5.0 & ~~~~~3.6$\mu$m,$g$,$i$ isophotes; Buta (1988) \cr
NGC 5194 & 20$\pm$5 & 170$\pm$3 & 50 ($K_s$) & 10.0 & 7.7 & ~~~~~Tully (1974)\cr
}
$^a$Col. 1: galaxy name; col. 2: adopted inclination; col. 3: adopted
line of nodes position angle; col. 4: adopted radial scale length, not based on
decomposition but from slope of azimuthally-averaged surface brightness
profile (filter used in parentheses); col. 5: adopted vertical scale
height derived as $h_z = h_R/5$ except for NGC 628 where $h_z = h_R/9$; col. 6: mean redshift-independent distance from NED; typical uncertainty
$\pm$1-2Mpc; col. 7: source of adopted orientation parameters.
\end{table*}

\begin{figure}
\vspace{310pt}
\includegraphics{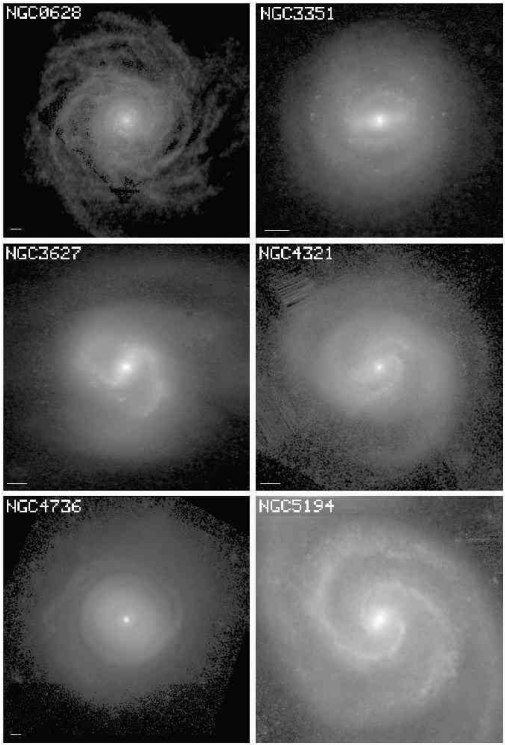}
\caption{Total mass maps of the six sample galaxies.}
\label{fg:Fig2}
\end{figure}

\subsection{NGC 4321 (M100)}

\begin{figure}
\vspace{160pt}
\includegraphics{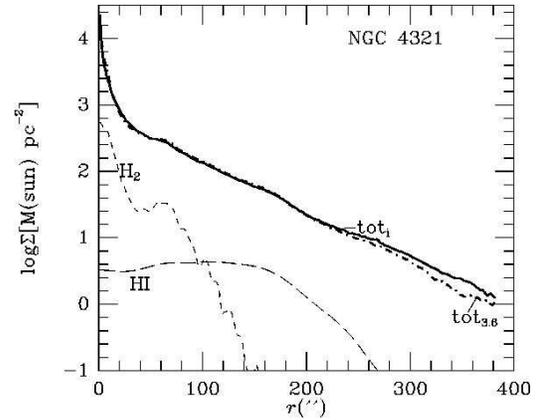}
\caption{Azimuthally-averaged surface mass density profiles of NGC
4321, based on the atomic (HI), molecular (H$_2$), and total (3.6$\mu$m
+ HI + H$_2$ and $i$ + HI + H$_2$) mass maps. The profile based on the
3.6$\mu$m image is called tot$_{3.6}$ while that based on the $i$-band
is called tot$_i$.}
\label{fg:Fig3}
\end{figure}

NGC 4321 (M100) is the bright Virgo spiral, of mid-infrared (MIR) RC3
type SAB(rs,nr)bc (Buta et al. 2010).  It lies at a distance of
approximately 16 Mpc. The procedure for calculating the surface density
distributions for the stellar and gaseous components using the near-
and mid-IR observations, as well as atomic and molecular
interferometric observations of galaxies, is described in detail in
Appendix A. Two stellar mass maps were derived, one based on a SINGS
(Kennicutt et al. 2003) 3.6$\mu$m IRAC image, and the other based on
a SDSS $i$-band image.  The same HI and H$_2$ gas maps were added
onto these stellar surface density maps to obtain the total disk
surface density maps.  The surface densities of the gaseous mass
components alone, HI and H$_2$, derived from VIVA (Chung et al.
2009) and BIMA SONG (Helfer et al. 2003),
respectively, and the total mass profiles using the IRAC 3.6$\mu$m
image and the SDSS $i$-band image (``tot$_{3.6}$" and ``tot$_i$",
respectively), are illustrated in Figure \ref{fg:Fig3}.  This shows
that the 3.6$\mu$m and $i$-band total mass maps give very similar
average surface density distributions for M100.

\begin{figure}
\vspace{330pt}
 \includegraphics{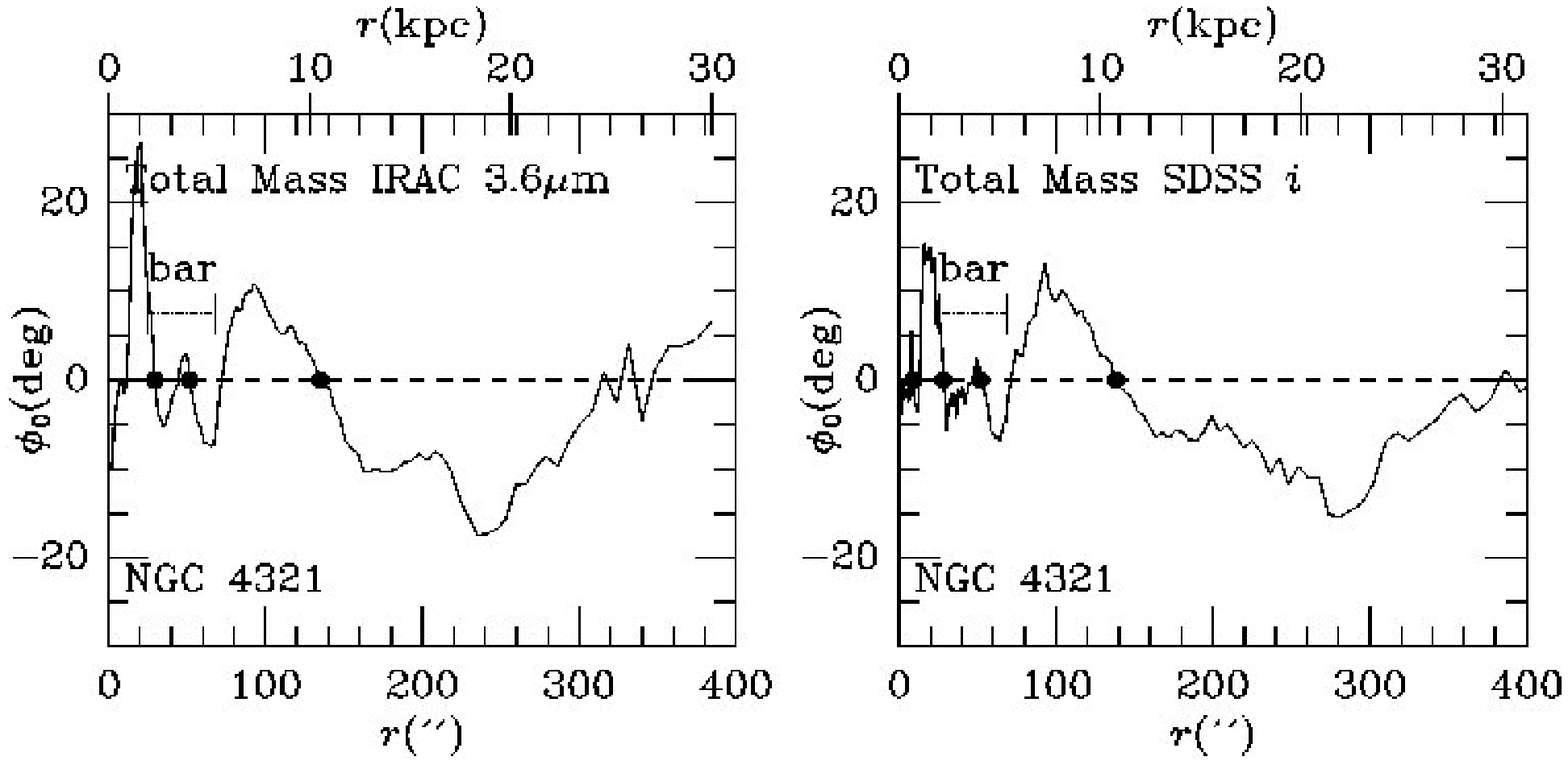}
 \includegraphics{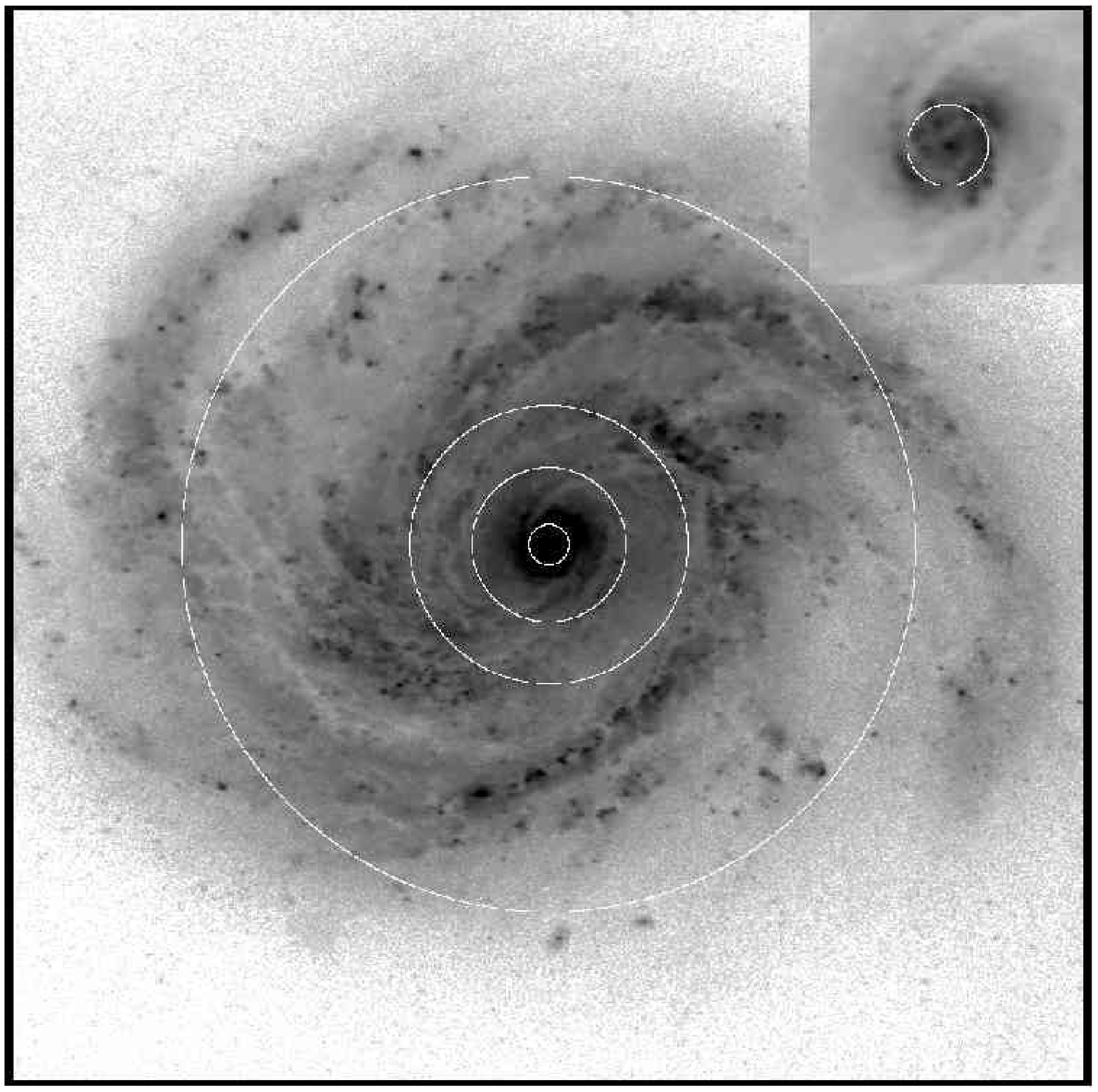}
\caption{{\it Top:} Potential-density phase shift versus galactic radius
for NGC 4321 derived using the total mass map with IRAC and SDSS data.
CR radii are indicated by the filled circles. The radius range of the
bar from ZB07 is indicated.
{\it Bottom:} Corotation circles overlaid on a deprojected
SDSS $g$-band image of NGC 4321.
The main frame covers an area 6\rlap{.}$^{\prime}$4 square, while the inset is
0\rlap{.}$^{\prime}$8 square. The units are mag arcsec$^{-2}$.}
\label{fg:Fig4}
\end{figure}

In ZB07 we carried out a phase shift analysis for this galaxy using the
3.6$\mu$m SINGS image, and assumed a constant mass to light ratio
($M/L$) without gas. Four sets of corotation (CR) resonances were
found, and they were shown to correspond to well-defined morphological
features.  Figure~\ref{fg:Fig4}, top, shows the radial-dependence of
potential-density phase shift derived using the total stellar plus
gaseous maps for the IRAC 3.6 $\mu$m band (left frame) and the SDSS
i-band (right frame).  We can see that these two maps give similar
CR predictions (in terms of the positive-to-negative (P/N) zero
crossings of the phase shift curves), except near the very center of
the galaxy where the factor of 2 better spatial resolution of the SDSS
map allows the resolving of a possible 4th pattern in the nuclear
region.  We have also tried to calculate phase shift distributions
using images from the different optical and NIR wave bands that we have
access to assuming constant M/L, and found that the phase shifts
derived from the above total mass maps (calibrated with radial
dependent M/L) show the best coherence, most likely due to the best
kinematical and dynamical mutual consistency between the potential and
density pair used for the total mass analysis, as is the case for
physical galaxies. Table 2 lists the (unweighted) average CR radii from the
3.6$\mu$m and $i$-band maps.

\begin{table*}
\caption{Corotation radii from potential-density phase shifts}
\halign{%
\rm#\hfil&
\qquad\rm\hfil#&
\qquad\rm\hfil#&
\qquad\rm\hfil#&
\qquad\rm\hfil#&
\qquad\rm\hfil#&
\rm#\hfil\cr
Galaxy & CR$_1$ & CR$_2$ & CR$_3$ & CR$_4$ & CR$_5$ & ~~~~~Filters \cr
& (arcsec) & (arcsec) & (arcsec) & (arcsec) & (arcsec) & ~~~~~$$ \cr
\noalign{\vskip 10pt} 

NGC 628  &  7.4 & 39.2 & 81.5 & 253.0\rlap{:} & .... & ~~~~~3.6 \cr
NGC 3351 & 26.2$\pm$0.4 & 86.5$\pm$5.1 & .... & .... & .... & ~~~~~3.6,$i$\cr
NGC 3627 & 78.6$\pm$3.5 & ... & .... & .... & .... & ~~~~~3.6,$i$\cr
NGC 4321 & 8.3$\pm$0.7 & 28.8$\pm$0.6 & 51.9$\pm$0.0 & 136.9$\pm$1.6 & .... & ~~~~~3.6,$i$\cr
NGC 4736 & 27 & 39 & 120 & 275 & 385 & ~~~~~ $i$  \cr
NGC 5194 & 21.3$\pm$3.1 & 110.1$\pm$0.1 & .... & .... & .... & ~~~~~3.6,$i$\cr
}
\end{table*}

Figure~\ref{fg:Fig4}, bottom shows the four CR circles for M100
superposed on the $g$-band SDSS image. The inset shows only the inner
region, which in the $g$-band includes a nuclear pseudoring and in the
longer wavelength bands includes a nuclear bar. The CR$_1$ radius
determined from the new maps, (8\arcs 3$\pm$0\arcs 7) is smaller than
what we obtained previously (13\arc) in ZB07, but turns out to better
correspond to the features of the resolved inner bar. The nuclear bar
seems to extend either a little beyond or just up to its CR and
terminates in a broad, swept-up spiral section (a nuclear pseudoring) at
the location indicated by the next N/P crossing. The next CR, CR$_2$,
appears to be related to an inner spiral that breaks from near the
nuclear pseudoring. The curved dust lanes in this spiral are on the
leading sides of the weak bar. CR$_3$ appears to lie completely within
the main spiral arms and could be the CR of the bar itself. The radius
of CR$_3$, 52\arc, is slightly less than that determined by ZB07,
59\arc, and may lie a little inside the ends of the primary bar.
CR$_4$, at 137\arc, lies within the outer arms. These values should be
compared with r(CR) = 97\arc $\pm$15\arc\ obtained by Hernandez et al.
(2005) using the Tremaine-Weinberg method applied to an H$\alpha$
velocity field.  Their value is close to the average of our CR$_3$ and
CR$_4$ radii. We believe our two-outer-CR result is more reasonable
than the single outer CR result from the TW method for this galaxy,
because on the $g$-band overlay image one can clearly discern the dust
lanes moving from the inner/leading edge of the spiral to
outer/trailing edge across CR$_4$, indicating the location where the
angular speed of disk matter and density wave switch their relative
magnitude. CR$_3$ on the other hand is tied to the main bar rather
than the spiral.

It appears that the patterns surrounding CR$_1$ and CR$_3$ are best
described as super-fast bars (ZB07, BZ09) of a dumb-bell shape with
their thick ends created by the different pattern speeds interacting in
the region of their encounter (see also the classification in Appendix
B). Super-fast bars are a peculiar implication of the phase shift
method, because they contradict the results of passive orbit analyses
that imply that a bar cannot extend beyond its own CR due to a lack of
orbital support (Contopoulos 1980).

\subsection{NGC 3351 (M95)}

\begin{figure}
\vspace{160pt}
\includegraphics{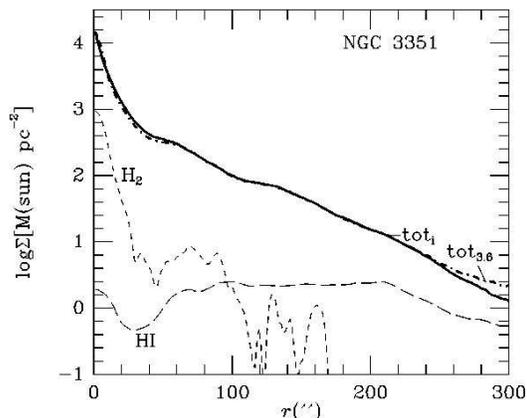}
\caption{Azimuthally-averaged surface mass density profiles of NGC
3351. The layout is the same as in Figure 3.}
\label{fg:Fig5}
\end{figure}

\begin{figure}
\vspace{330pt}
 \includegraphics{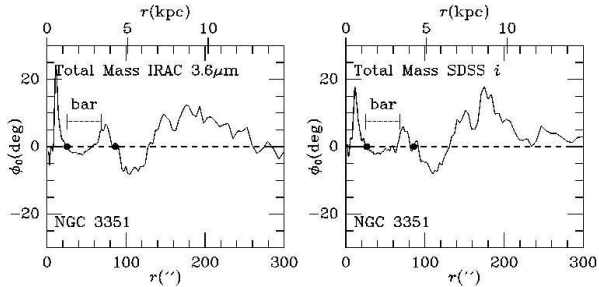}
 \includegraphics{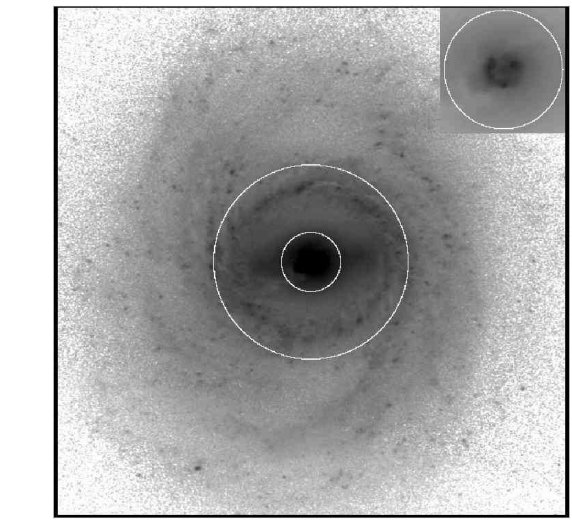}
\caption{{\it Top:} Potential-density phase shift versus galactic
radius for galaxy NGC 3351 derived using the total mass map with IRAC
and SDSS data. {\it Bottom:} Deprojected $g$-band image of NGC 3351 in
units of mag arcsec$^{-2}$, with major phase shift crossings superposed
as solid circles. The main frame covers an area 7\rlap{.}$^{\prime}$5
square, while the inset is 0\rlap{.}$^{\prime}$93 square.}
\label{fg:Fig6}
\end{figure}

NGC 3351 (M95) is a barred spiral galaxy of MIR RC3 type (R')SB(r,nr)a
(Buta et al. 2010).  It lies at a distance of approximately 10 Mpc.
The surface densities of the different mass components for this galaxy
derived using IRAC, SDSS, BIMA SONG and THINGS data are shown in Figure
\ref{fg:Fig5}.
Phase shift analyses for NGC 3351 were carried out using
the IRAC and SDSS mass surface density images based
on the 3.6$\mu$m SINGS and $i$-band SDSS images, 
plus the gas maps (Figure~\ref{fg:Fig6}, top).  
The phase shift distributions for both the IRAC and SDSS total mass
maps show a major CR near $r$ = 86\arc, and also another CR at
$r$=26\arc.  

Figure~\ref{fg:Fig6}, bottom shows the two CR radii as solid circles
superposed on the $g$-band image.  Two prominent N/P crossings follow
these CR radii at 65\arc\ and 128\arc. For comparison, the inner ring
of NGC 3351 has dimensions of 71\arcs\ 0 $\times$ 67\arcs\ 8 and lies
close to the first N/P crossing, implying that the ring/spiral is a
separate mode whose CR is CR$_2$ at 86\arc. In Figure~\ref{fg:Fig6},
top also, the extent of the main part of the bar is indicated, and both
mass maps show the same thing: the phase shifts are negative across the
main part of the bar, implying (as discussed in ZB07 and BZ09) that the
corotation radius of the bar is CR$_1$, not CR$_2$. Thus, NGC 3351 is
also a case of a super-fast bar.

The inner pseudoring at the location of the first major N/P
crossing would, in passive orbit analysis in a galaxy with a bar and
non-self-gravitating clouds, be considered an inner 4:1 resonance ring
as identified by Schwarz (1984) and Buta \& Combes (1996). This ring
would be directly related to the bar and have the same pattern speed as
the bar. Our phase shift analysis, however, suggests that the apparent
ring/pseudoring could be due to a ``snow plough'' effect of two sets of
inner/outer patterns having different pattern speeds, and which thus
accumulate mass at that radius.  The strong and symmetric arms which
end a little within the second CR circle might be related to what
Patsis/Contopoulos/Grosbol had advocated: that symmetric patterns
sometimes end at the inner 4:1 resonance (counter-examples are
discussed by ZB07 and BZ09). The spiral patterns outside the second CR
circle in this case are more fragmented.  It is not yet clear what
different dynamics would make some spirals extend all the way to OLR
and others end mostly within CR or inner 4:1.  Perhaps those that have
CRs intersecting the arms (such as NGC 5247 analyzed in ZB07) tend to
be Sb, Sc or later types, such that the arms are going through the
transitional phase from a skewed long inner bar to either a bar-driven
spiral or an inner organized spiral plus outer diffused arms.
Therefore, the ones where the spiral ends at the inner 4:1 resonance or
CR (i.e. the NGC 3351 type) should tend to be more mature and more
steady types, i.e., earlier-type disks in general than those with arms
being crossed by the CR circle.

\subsection{NGC 5194 (M51)}

\begin{figure}
\vspace{160pt}
\includegraphics{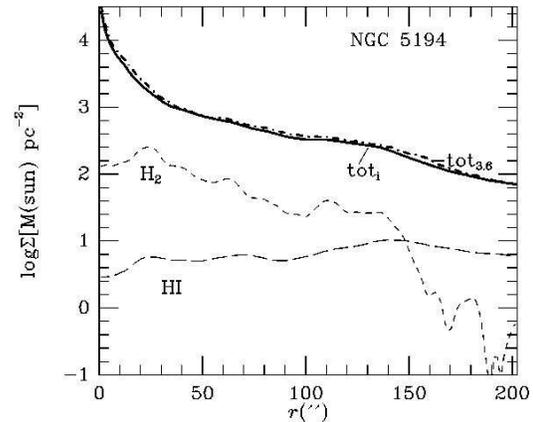}
\caption{Azimuthally-averaged surface mass density profiles of NGC
5194. The layout is the same as in Figure 3.}
\label{fg:Fig7}
\end{figure}

NGC 5194 (M51) is the proto-typical interacting grand design spiral, of
MIR RC3 type SAB(rs,nr)bc (Buta et al. 2010).  It lies at a distance of
approximately 8.4 Mpc.  The surface densities of the different mass
components for this galaxy derived using IRAC, SDSS, BIMA SONG and
THINGS data are shown in Figure \ref{fg:Fig7}.  It can be seen that the
hot dust-corrected total 3.6$\mu$m surface density profile and that for
the $i$-band data are very similar.

\begin{figure}
\vspace{330pt}
 \includegraphics{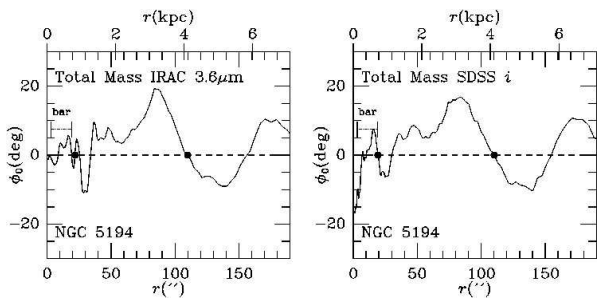}
 \includegraphics{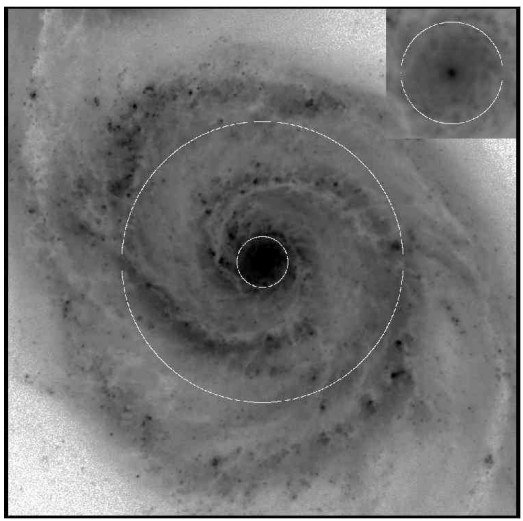}
\caption{{\it Top}: Calculated phase shift vs galaxy radii for NGC 5194
(M51) using the IRAC data as well as SDSS data.  
Two corotation radii (filled circles) are indicated.
{\it Bottom}: CR circles overlaid on the deprojected $g$-band image
of NGC 5194 (M51).
The main frame covers an area 6\rlap{.}$^{\prime}$6 square, while the inset is
0\rlap{.}$^{\prime}$93 square. The units are mag arcsec$^{-2}$.} 
\label{fg:Fig8}
\end{figure}

The potential-density phase shift method was used to derive the main CR
radii for this galaxy (Figure~\ref{fg:Fig8}). By focusing on the area
that just excludes the small companion NGC 5195, which is likely to lie
outside of the M51 galactic plane and thus have minor influence on the
internal dynamics of M51 at the epoch of observation (a conjecture
which is confirmed by our analysis), the phase shift analysis gives two
major CR radii (P/N crossings on the phase shift plot, represented by
solid circles on the overlay image) followed by two
negative-to-positive (N/P) crossing radii (not shown on the overly).  
The latter are
believed to be where the inner modes decouple from the outer modes. The
CR radii, at 21$^{\prime\prime}$$\pm$3$^{\prime\prime}$ and 110$^{\prime\prime}$.
(Table 2), match very well the galaxy morphological features [i.e., the
inner CR circle lies near the end of an inner bar/oval, and the first
N/P crossing circle ($r$=30\arc) is where the two modes decouple].
Also, for the outer mode, the CR circle seems to just bisect the
regions where the star-formation clumps are either concentrated on the
inner edge of the arm, or on the outer edge of the arm -- a strong
indication that this second CR is located right near where the pattern
speed of the wave and the angular speed of the stars match each other
(this can be compared to a similar transition of arm morphology
across CR in the simulated galaxy image of Z96 Fig 3).
This supports the hypothesis that the spiral patterns in this galaxy
are intrinsic modes rather than tidal and transient, and that tidal
perturbation serves to enhance the prominence of the intrinsic mode,
but does not alter its modal shape.

\subsection{NGC 3627 (M66)}

\begin{figure}
\vspace{160pt}
\includegraphics{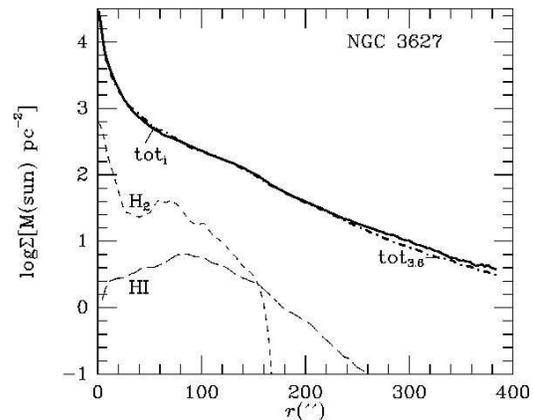}
\caption{Azimuthally-averaged surface mass density profiles of NGC
3627. The layout is the same as in Figure 3.}
\label{fg:Fig9}
\end{figure}

The intermediate-type barred spiral galaxy NGC 3627, of MIR RC3 type
SB(s)b pec (Buta et al. 2010), is a member of the interacting group the
Leo Triplet (the other two members being NGC 3628 and NGC 3623).  It
lies at a distance of approximately 10 Mpc.  The surface densities of
the different mass components for this galaxy derived using IRAC, SDSS,
BIMA SONG and THINGS data are shown in Figure \ref{fg:Fig9}.As for NGC
3351, 4321, and 5194, the total 3.6$\mu$m mass profile and the total
$i$-band mass profile are in good agreement over a wide range of
radius.

Zhang, Wright, and Alexander (1993) observed this galaxy in
high-resolution CO(1-0) and HI, and found that the choice
of an outer CR location at 220\arc coinciding with the outer HI cutoff could account
for many observed morphological features.  Subsequent observations,
e.g. those of Chemin et al. (2003) in the H$\alpha$ line has determined that
that an inner CR enclosing the central bar is a more reasonable choice.
In the following, we use the new
surface density maps to re-evaluate the question of CR determination
for this galaxy, bearing in mind however that the potential-density phase shift
method we use in the current study will give the most reliable CR determination
for potential-density pairs that have achieved
dynamical equilibrium, a condition which is likely to be violated
for this strongly-interacting galaxy that has suffered serious
damage in its outer disk (Zhang et al. 1993).	

\begin{figure}
\vspace{330pt}
\includegraphics{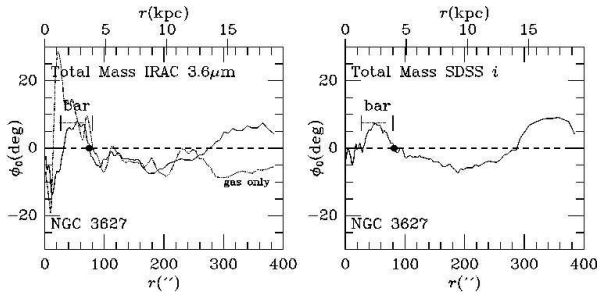}
\includegraphics{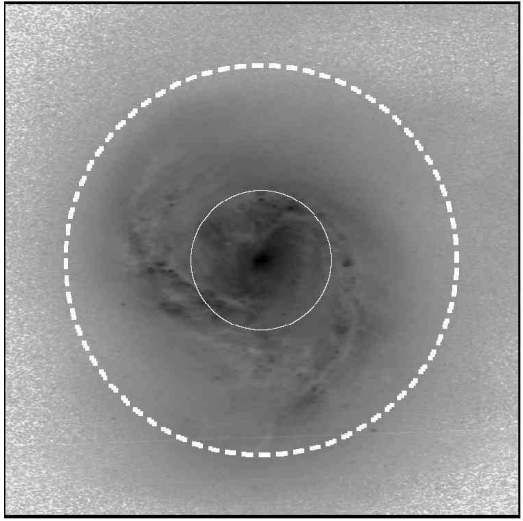}
\caption{{\it Top:} Calculated phase shifts between the stellar mass
and total potential, and the HI gas mass and the total potential,
for NGC 3627. {\it Bottom:}
Overlay of CR circles on the $g$-band image for NGC 3627.
The main frame covers an area 9\rlap{.}$^{\prime}$6 square.
The units are mag arcsec$^{-2}$.}
\label{fg:Fig10}
\end{figure}

In Figure~\ref{fg:Fig10}, top, we show the phase shifts with respect to the
total potential of the total mass distribution, as well as for the total
gas distribution. The total mass phase shifts displays only one inner CR
at 78\arcs 6 $\pm$ 3\arcs 5, which can be compared to the CR radius
obtained by Chemin et al. (2003) of $\sim $70\arc, yet the large-radius
phase distribution shows clear evidence of the disturbance caused by
the interaction, and thus (by implication) a lack of dynamical
equilibrium. The main CR appears to encircle the bar ends in NGC 3627
(Figure \ref{fg:Fig10}, bottom), which would make the galaxy a 
``fast bar" case according to BZ09 (see also Appendix B).

The total gas phase shift shows the presence of a second possible CR
between 220\arc-260\arc, close to the one adopted in Zhang et al.
(1993) of 220\arc.  The radius of this possible CR is ill-defined since
the phase shift barely reaches zero and does not become a clear P/N
crossing.  Figure~\ref{fg:Fig10}, bottom, shows the $g$-band image
overlaid with the two possible CR circles, selecting 220\arc\ for the
outer CR as the minimum likely value. The total mass map shows two
faint outer arms that extend towards the outer CR circle -- in fact these
outer spiral arms extend much further than the impression given in the
$g$-band image, as can be seen in the non-deprojected HI
surface density profile (Zhang et al. 1993, Figure 7; as well as a
similar HI image in the THINGS website).

According to the one-CR view of Chemin et al. (2003), the main spiral
would be bar-driven. However, this view would be contradicted by the
two strong disconnected bow-shocks at the CR circle, indicating the
interaction of two pattern speeds (these disconnected bow shock
segments in fact show up more clearly in the gas surface density map).
Truly bar-driven spirals do not contain these disconnected bow-shock
segments and the spirals would appear as the further continuation of a
skewed bar (see Appendix B).  The outer spiral arms in the case of NGC
3627 clearly do not connect smoothly with the inner bar -- in fact,
there is evidence of two sets of spiral patterns, the inner set with
short arms appears bar-driven whereas the outer set is offset in phase
in the azimuthal direction from the inner one.

With this additional evidence, we propose that prior to
the encounter with NGC 3628, NGC 3627 originally had a two-pattern
structure similar to the CR3 and CR4 regions of NGC 4321.
The interaction tore out a large part of the outer surface density
from NGC 3627, including a segment of the outer spiral arm (Zhang et
al. 1993, Figure 5, clump L).  Therefore the galaxy is evolving
towards a new dynamical equilibrium, with the possibility
of losing the coherent outer spiral pattern eventually.

\subsection{NGC 628}

\begin{figure}
\vspace{160pt}
\includegraphics{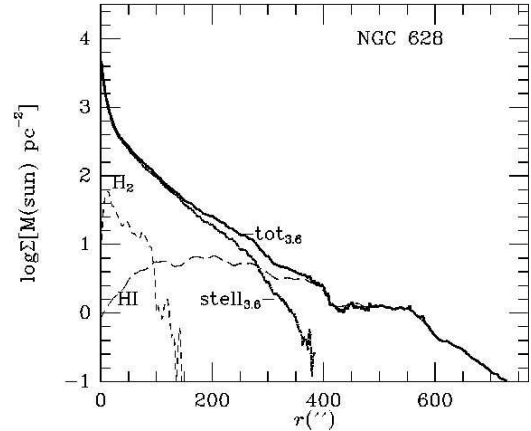}
\caption{Azimuthally-averaged surface mass density profiles of NGC
628. The HI and H$_2$ profiles are shown, and the stellar and
total mass profiles based on a 3.6$\mu$m image.}
\label{fg:Fig11}
\end{figure}

NGC 628 is a late-type spiral of MIR RC3 type SA(s)c (Buta et al. 2010).  
It lies at a distance of 8.2 Mpc.
The surface densities of the different mass components for this galaxy
derived using IRAC, BIMA SONG and THINGS data are shown in Figure
\ref{fg:Fig11}.  For this galaxy, no SDSS data are available.
Unlike the other galaxies in our sample, NGC 628 has considerable HI
gas at large radii, and this accounts for the significant departure
of the total mass profile from the stellar density profile alone.

\begin{figure}
\vspace{310pt}
 \includegraphics{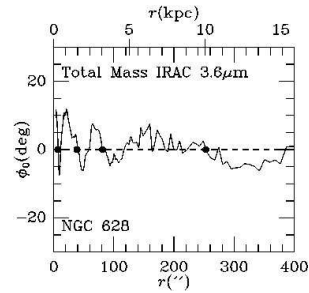}
 \includegraphics{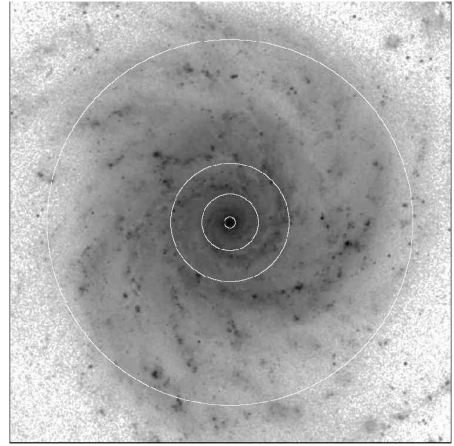}
\caption{{\it Top:} Phase shift between the total mass surface density
and total potential for NGC 628. {\it Bottom:} Deprojected
$B$-band image of NGC 628 overlaid with CR circles.
The main frame covers an area 10\rlap{.}$^{\prime}$1 square.
The units are mag arcsec$^{-2}$.}
\label{fg:Fig12}
\end{figure}

The phase shift plot (Figure~\ref{fg:Fig12}, top) shows
noisier organization compared to all the other
galaxies in this sample, which is consistent with
this galaxy being of very late type. In the
secular evolution picture this corresponds to a
young galaxy still in the process of settling down towards
a dynamical equilibrium state. Nevertheless, there is 
an indication of four P/N crossings which are indicated
by the filled circles in the figure. Table 2 lists the radii
of these crossings. Figure~\ref{fg:Fig12}, bottom shows
the overlay of CR circles on a deprojected $B$-band
image of NGC 628.

\subsection{NGC 4736}

\begin{figure}
\vspace{160pt}
\includegraphics{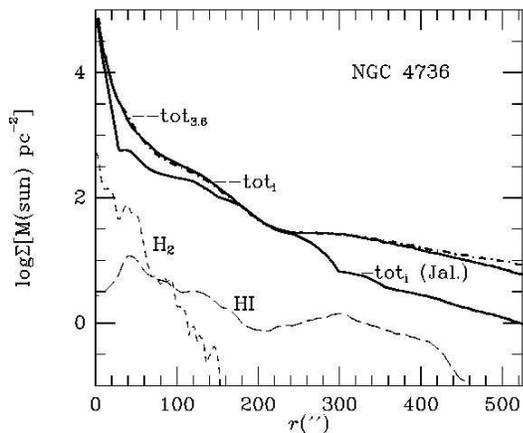}
\caption{Azimuthally-averaged surface mass density profiles of NGC
4736. The layout is the same as in Figure 3, except that graph includes
the profile scaled according to the mass-to-light ratio profile
used by Jalocha et al. (2008).}
\label{fg:Fig13}
\end{figure}

NGC 4736 is an early-type spiral of MIR RC3 type (R)SAB(rl,nr',nl,nb)a
(Buta et al. 2010), implying a galaxy with many distinct features.  It
lies at a distance of about 5 Mpc.  For the derivation of the stellar
mass map, we have used the SDSS $i$-band data and the mass-to-light
profiles obtained through a fitting procedure by Jalocha et al.
(2008).  This is because the usual color-based approach (Bell \& de
Jong 2001) gives a stellar surface density distribution that leads to a
predicted rotation curve much higher than the observed rotation curve
in the central region of the galaxy.  We have not made further
correction to account for the fact that Jalocha et al. used the Cousins
$I$-band data whereas we are using the SDSS $i$-band data, since we
expect the uncertainties in the M/L determination would be larger than
these differences. The surface mass density curves for this galaxy are
shown in Figure~\ref{fg:Fig13}. The tot$_{3.6}$ and tot$_i$ curves were
derived in the same manner as for the other galaxies. The tot$_i$
(Jal.) is based on the purely radius-dependent $M/L$ from Jalocha et
al. (2008).

\begin{figure}
\vspace{310pt}
 \includegraphics{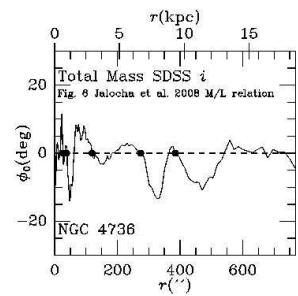}
 \includegraphics{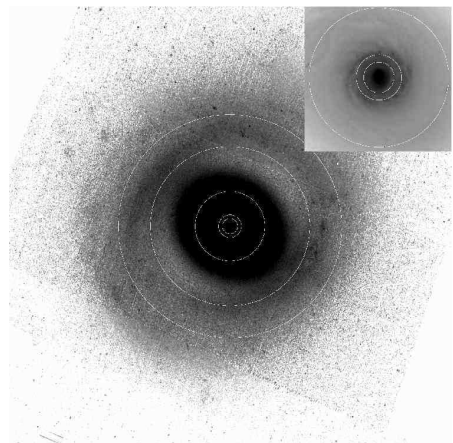}
\caption{{\it Top:} Phase shift between the total mass surface density
and total potential for NGC 4736, derived using the SDSS data.
{\it Bottom:} SDSS $g$-band image of NGC 4736 with the CR circle overlay.
The main frame covers an area 25\rlap{.}$^{\prime}$4 square, while the
inset has an area 4\rlap{.}$^{\prime}$25 square.
The units are mag arcsec$^{-2}$.}
\label{fg:Fig14}
\end{figure}

The phase shift derived using the total mass map and the overlay of CR
circles on the image is given in Figure~\ref{fg:Fig14}. The
shift plot (Figure~\ref{fg:Fig14}, top) shows several well-delineated
P/N crossings (Table 2) which appear to correspond well to the resonant
structures in the image (Figure~\ref{fg:Fig14}, bottom). CR$_1$ could
be associated with an inner nuclear bar, while CR$_2$ is a mode
associated with the bright spiral inner pseudoring. CR$_3$ could be
associated with an intermediate spiral pattern outside the prominent
inner ring. The fact that CR$_4$ passes through the gap between the
inner and outer rings suggests that it is the actual CR of the massive
oval. CR$_5$ may be associated with the outer ring pattern. 

\section{ANALYSES OF THE SAMPLE, AND DISCUSSION}

\subsection{Rotation Curves, Secular Mass Flow Rates, and 
Their Implications on the Compositions of Galactic Dark Matter}

\begin{figure}
\vspace{430pt}
\includegraphics{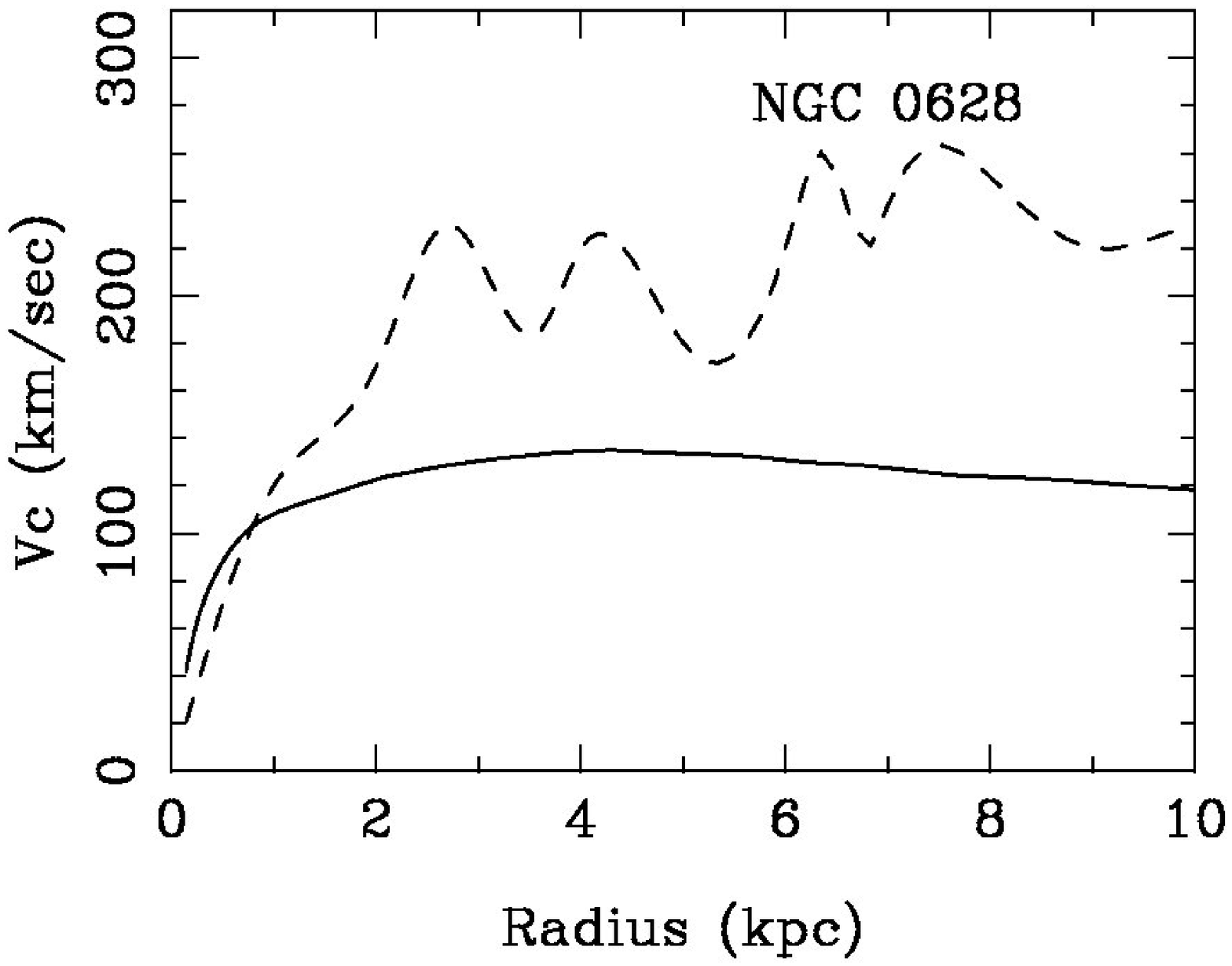}
\includegraphics{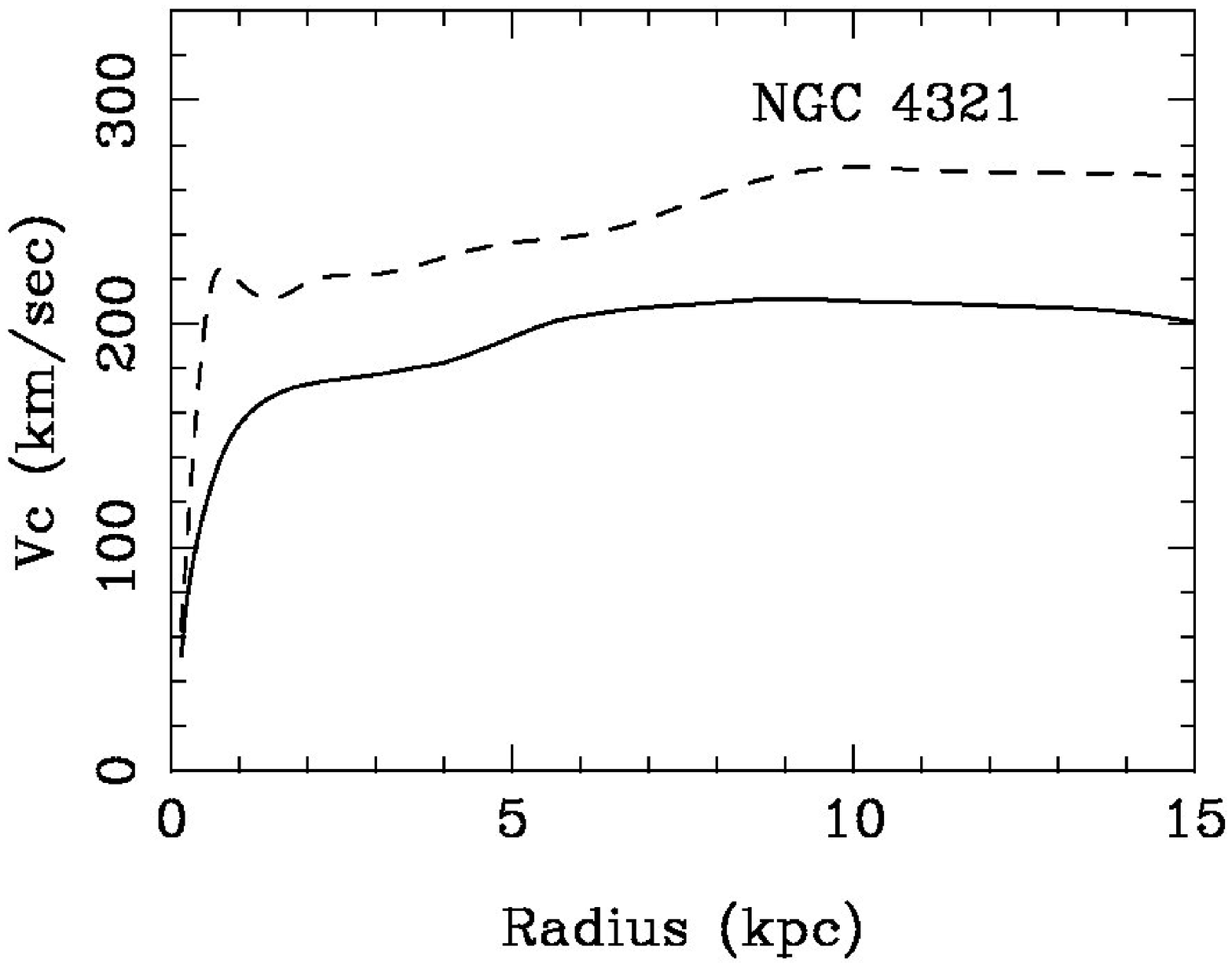}
\includegraphics{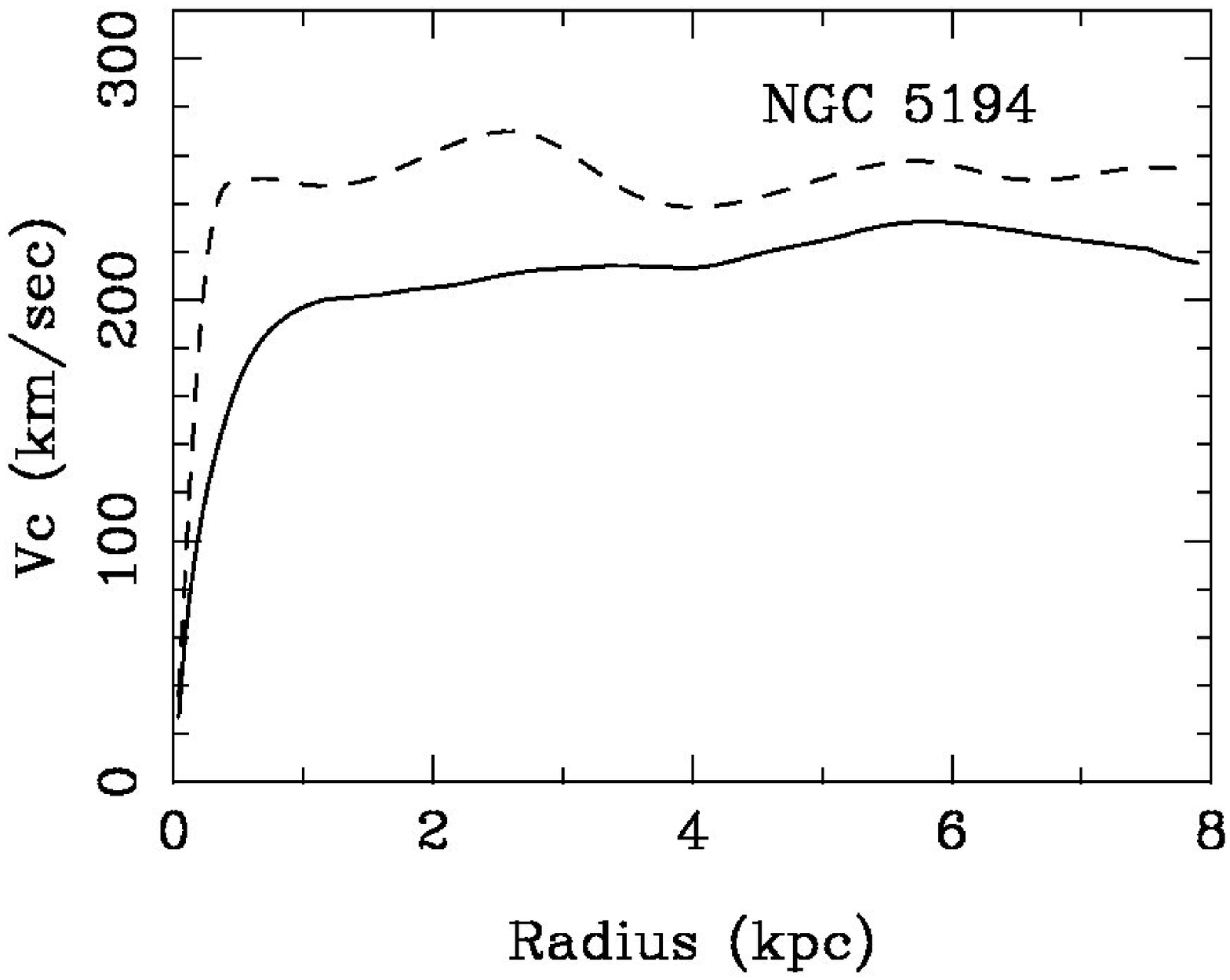}
\includegraphics{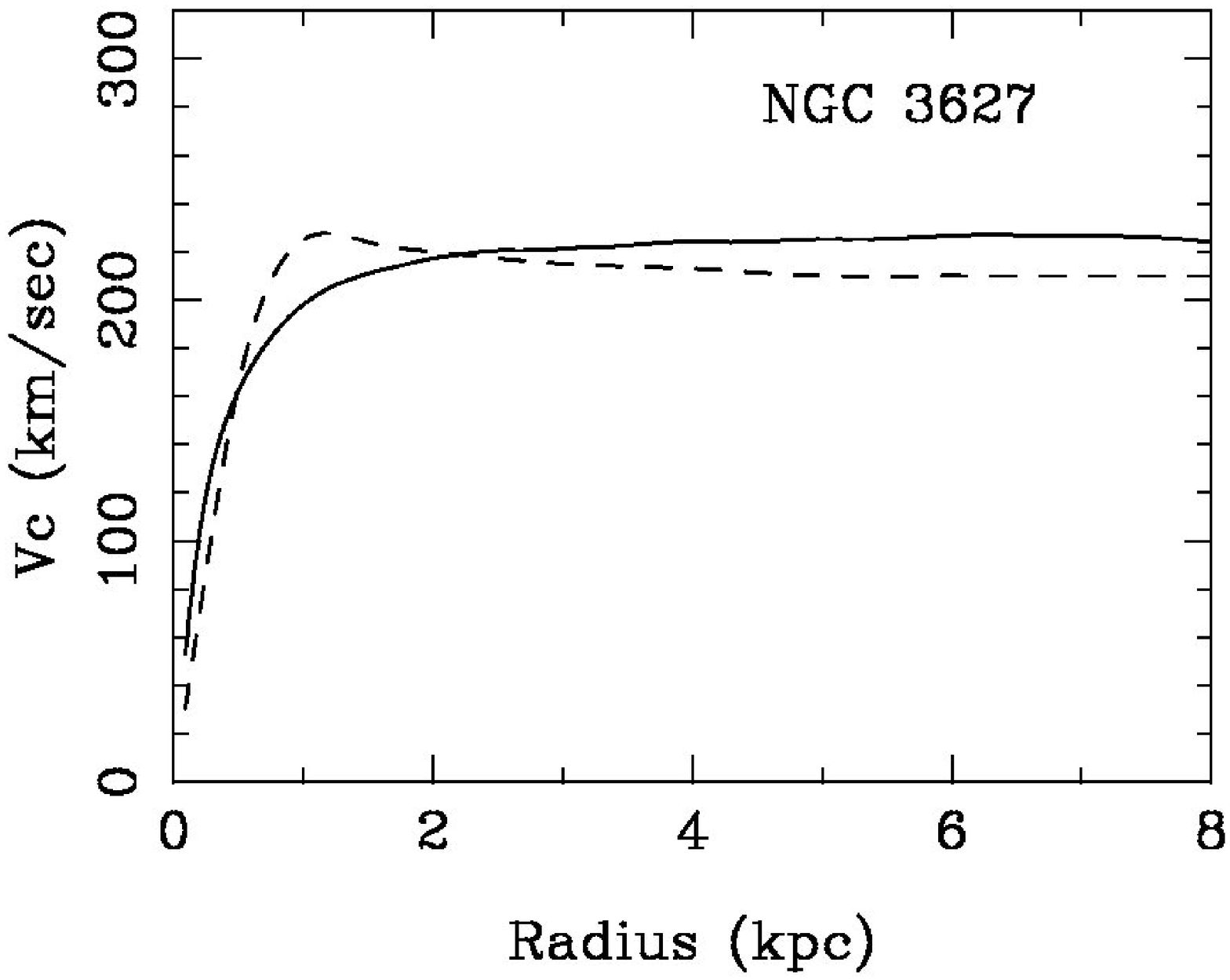}
\includegraphics{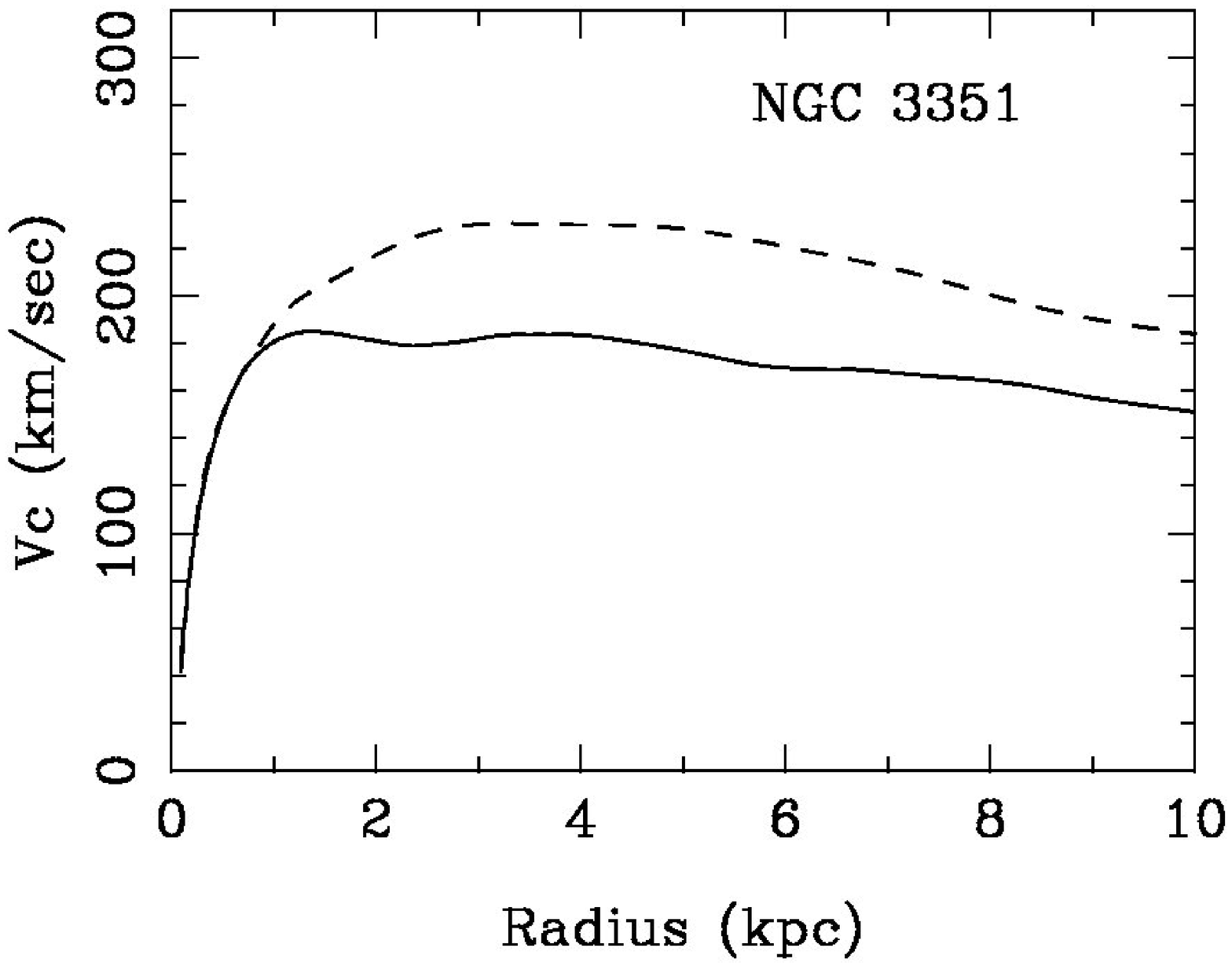}
\includegraphics{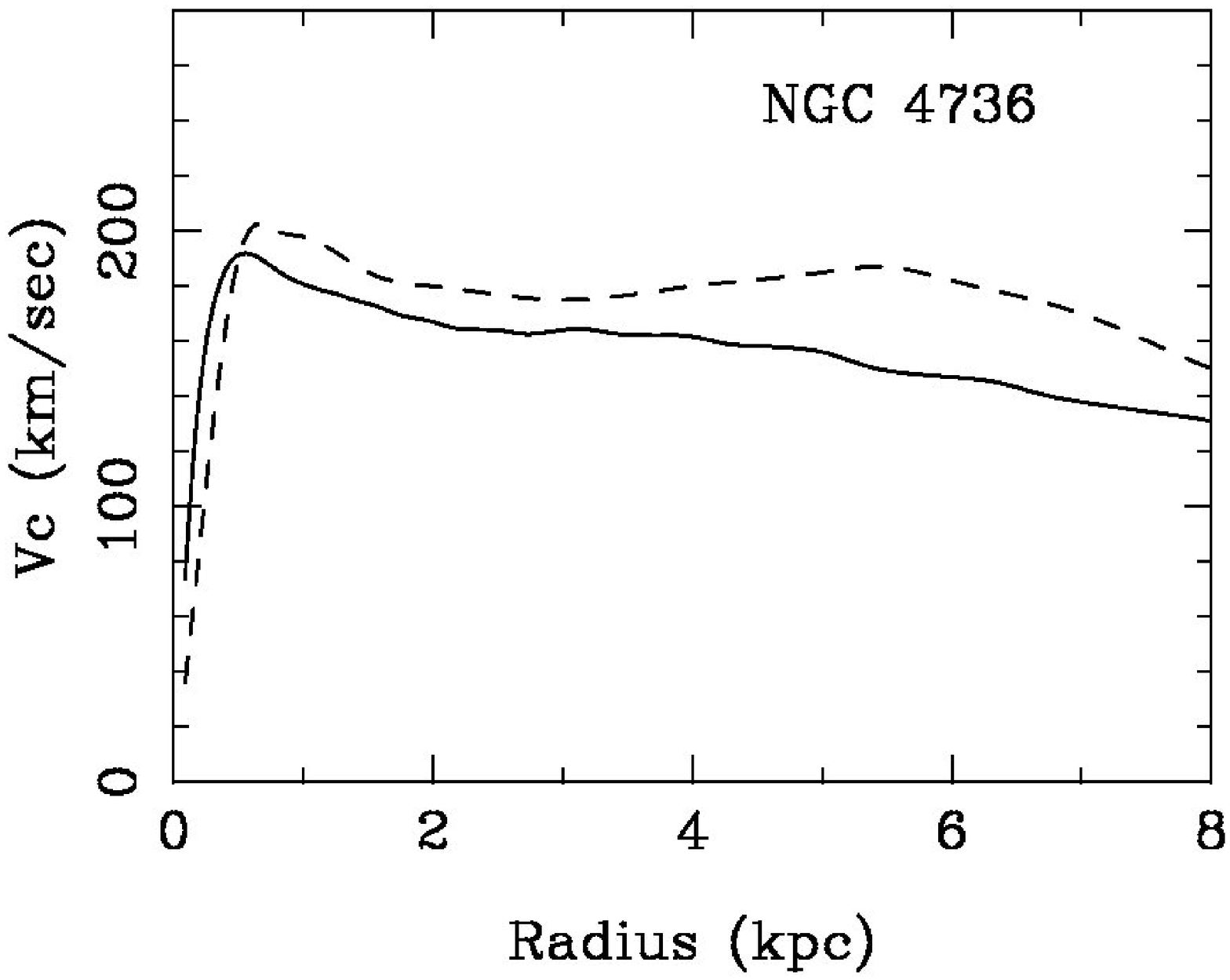}
\caption{Comparison of rotation curves for the six sample galaxies.
{\it Solid lines:} Disk rotation curves inferred from total disk surface density
(stellar plus atomic and molecular hydrogen mass.  Note that very small 
correction to account for helium and metal abundance in the gas mass has not
been made).  The inferred disk rotation curves
were derived using IRAC 3.6 $\mu$m data for NGC 0628, using an average
of IRAC 3.6 $\mu$m and SDSS i-band data for NGC 4321, 3351, 3627, 5194, and
using SDSS i-band data for NGC 4736, plus the atomic and molecular gas
contribution from the VIVA, THINGS and BIMA SONG observations.
{\it Dashed lines:} Observed rotation curves. The sources for these
observations are given in the main text.
}
\label{fg:Fig15}
\end{figure}

\begin{figure}
\vspace{430pt}
\includegraphics{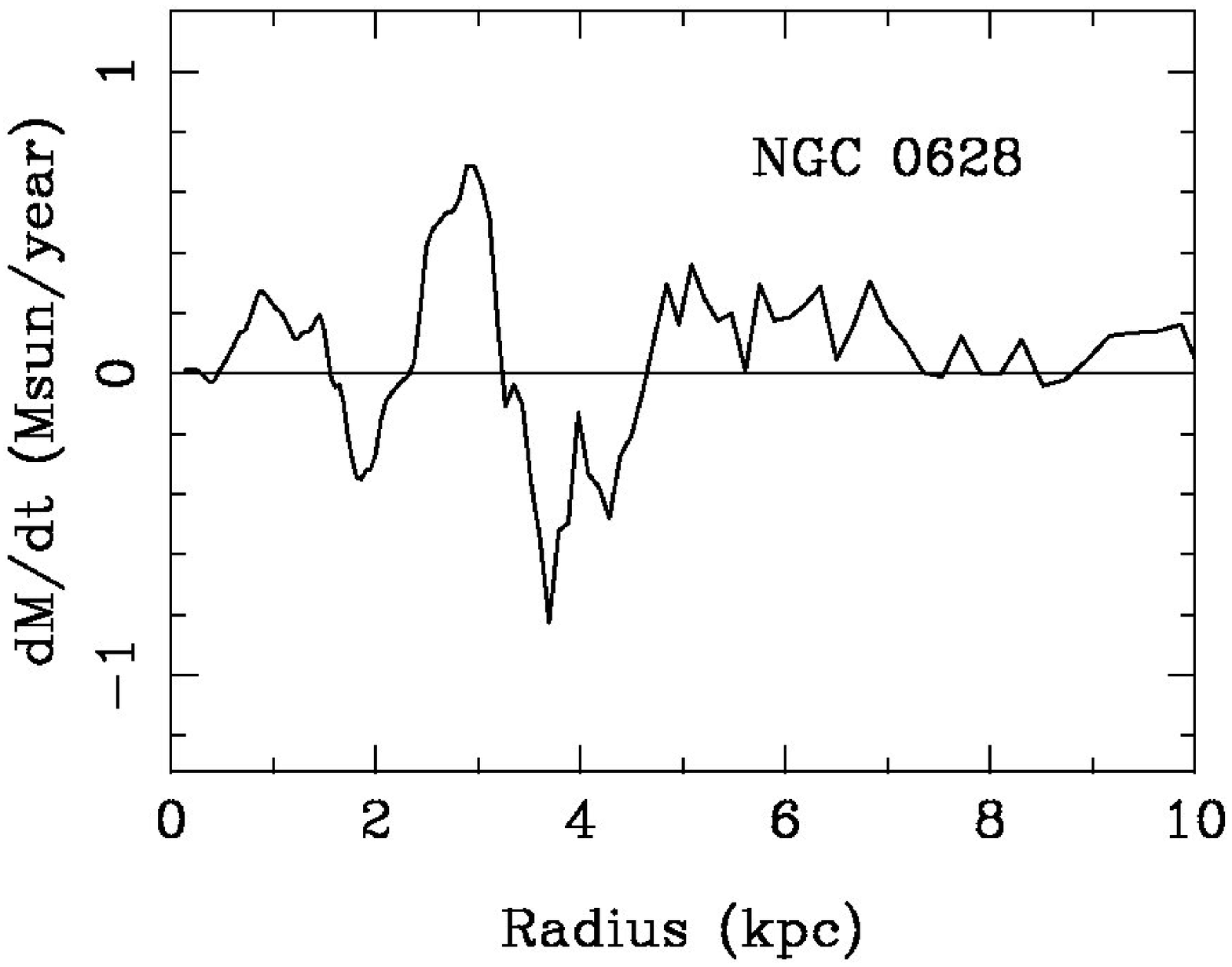}
\includegraphics{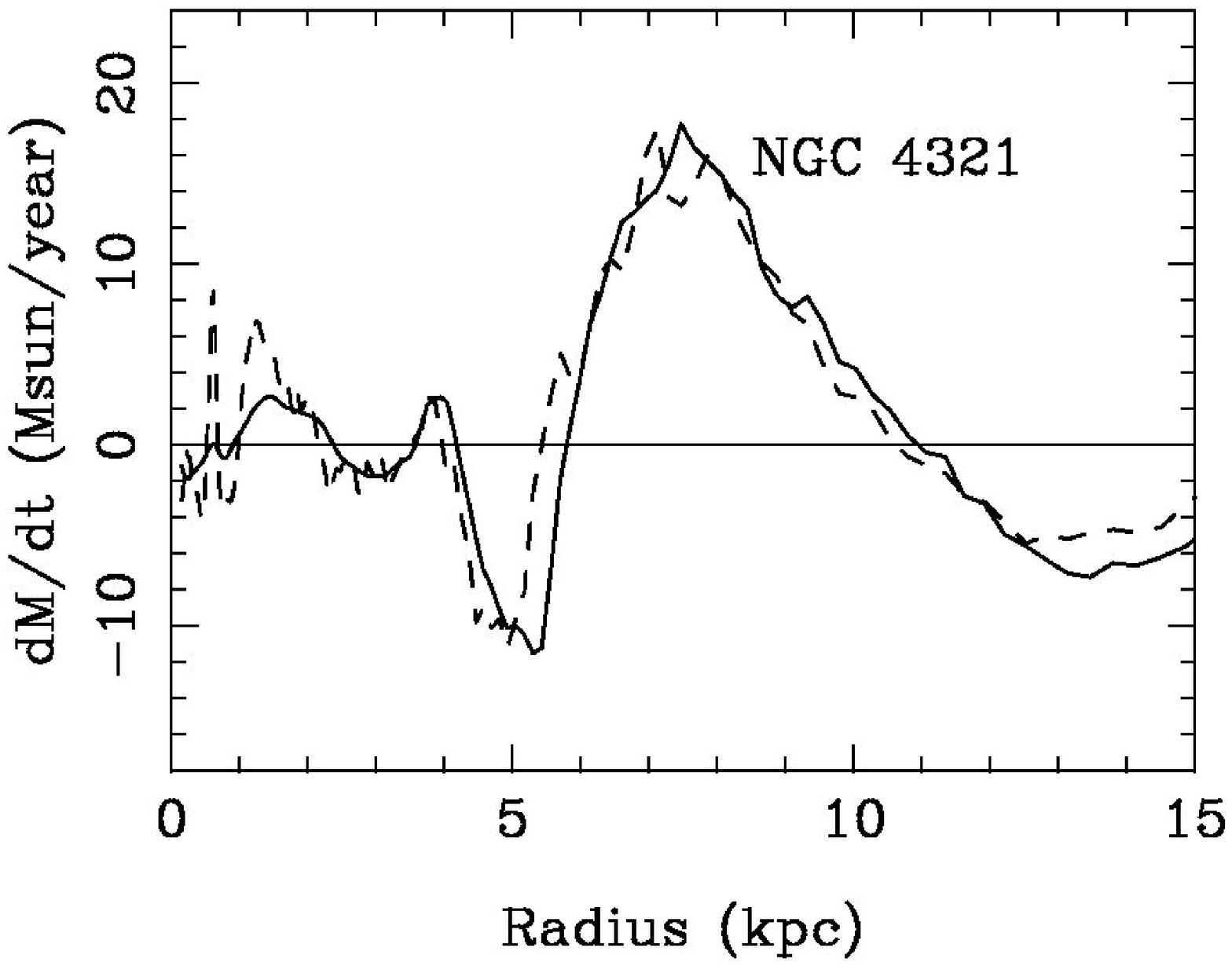}
\includegraphics{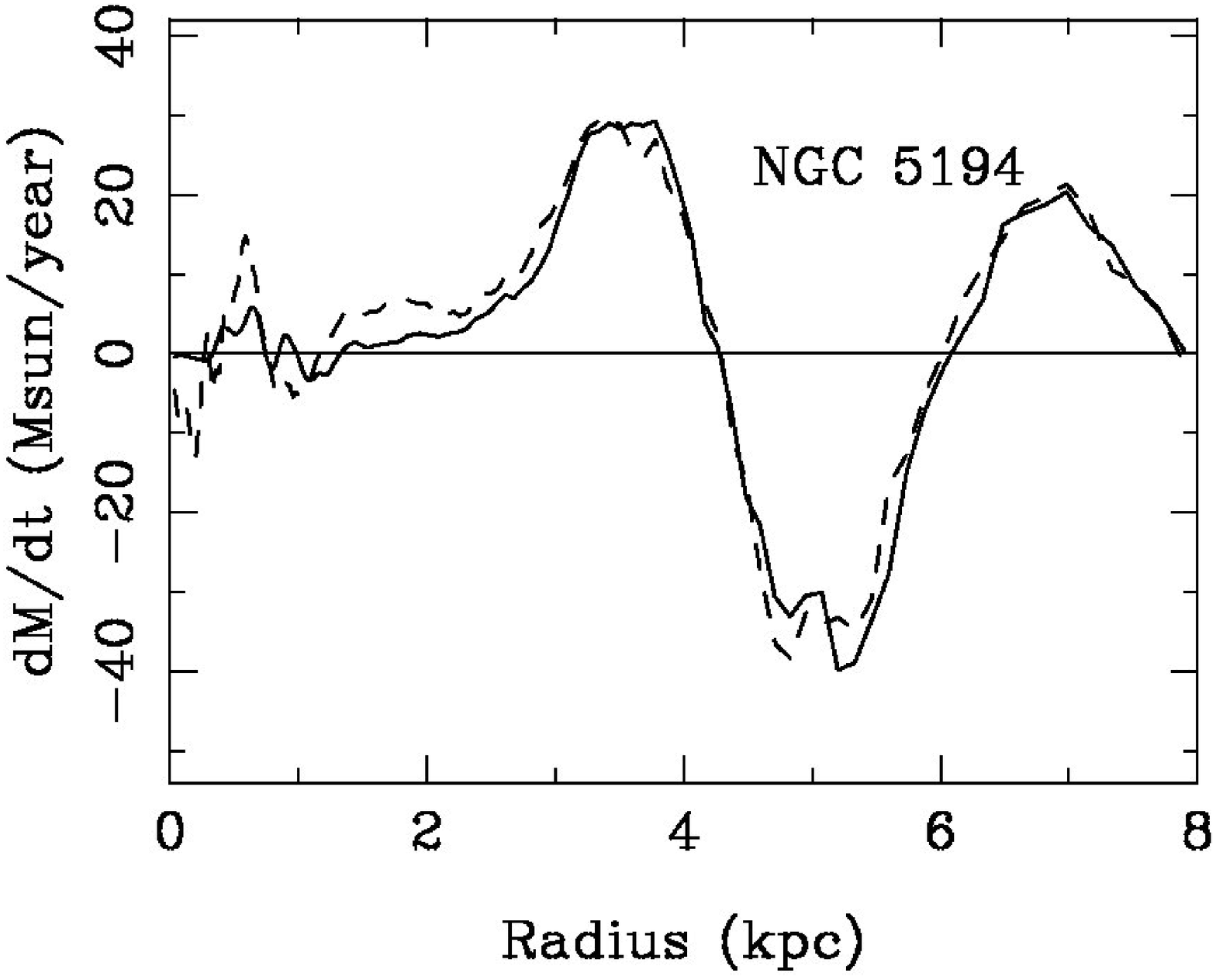}
\includegraphics{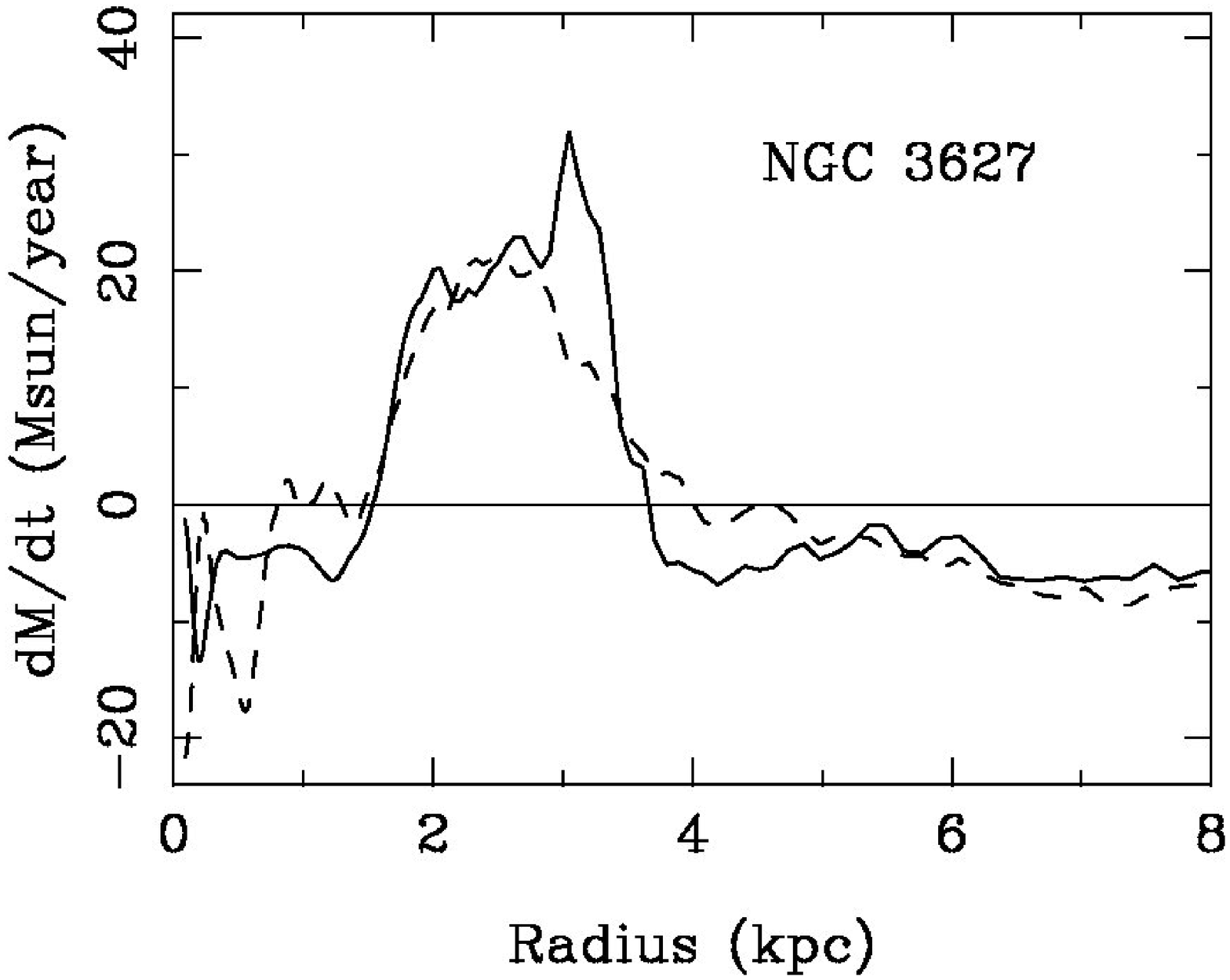}
\includegraphics{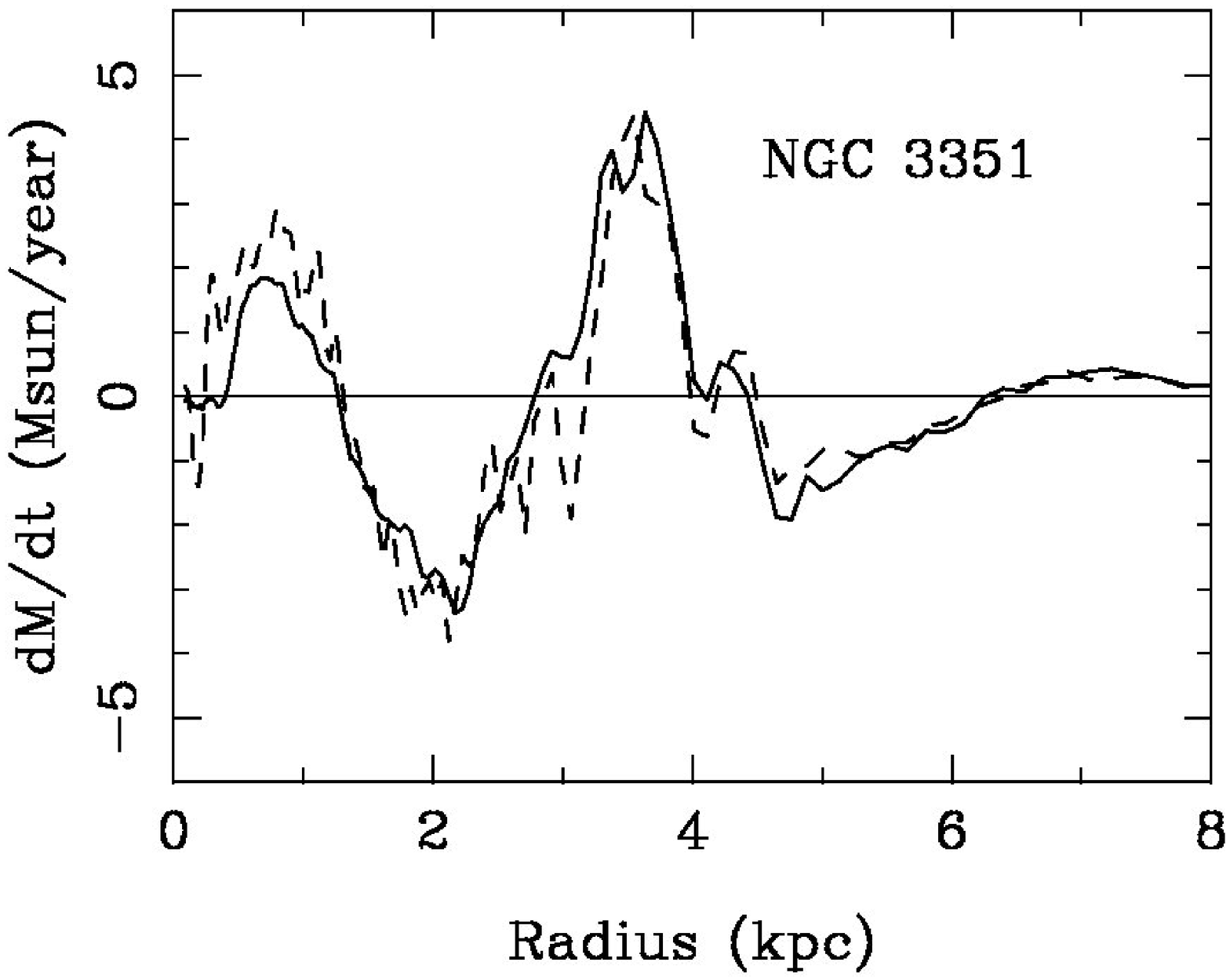}
\includegraphics{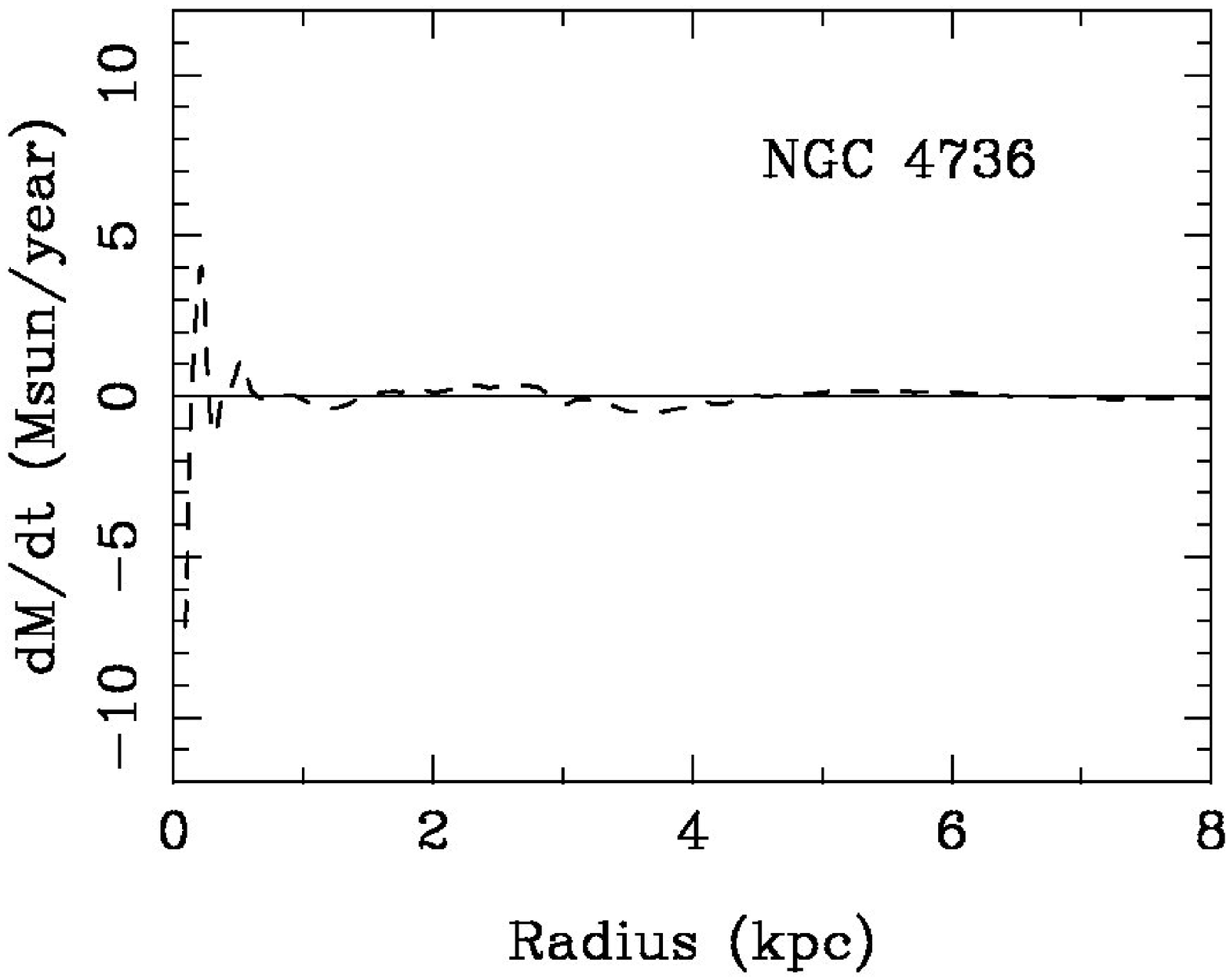}
\caption{Total radial mass flow rate for the six sample galaxies. Positive
portion of the curves indicate inflow, and negative portion of the
curves indicate outflow. The solid lines have stellar mass derived from
IRAC 3.6$\mu$m data, and the dashed lines have stellar mass derived
from SDSS i-band data.  Atomic and molecular gas mass from VIVA, THINGS
and BIMA SONG were added to obtain the total mass maps which were used
to derive these flow rates.}
\label{fg:Fig16}
\end{figure}

In the next group of figures (Figure~\ref{fg:Fig15}
and Figure~\ref{fg:Fig16}), we present the rotation
curves (observed, versus disk-total-mass-derived) as
well as the total mass flow rates for our sample galaxies.
The sequence of arrangement for the different frames,
NGC 0628, NGC 4321, NGC 5194, NGC 3627, NGC 3351, NGC 4736,
from left to right, then top to bottom,
is chosen to be roughly along the Hubble diagram 
from the late to the early types,
in order to reveal any systematic trends along this sequence.
Notice, however, that two of the intermediate types
(NGC 5194 and NGC 3627) are strongly interacting galaxies, and thus
they might deviate from the quiescent evolution trends.
Note also that since there is usually a limited radial range
for the availability of observed rotation curves, our subsequent
plots in this section (Figures \ref{fg:Fig15}-\ref{fg:Fig21}) 
will be displayed with a smaller radial range than
their corresponding phase shift plots in the last section.

The sources of the observed rotation curves are as follows:
N0628: from Nicol (2006)'s SINGS sample follow-up H-alpha observations.
N4321: From the CO/H-alpha compiled rotation curve of Sofue et al. (1999).
N5194: From Sofue (1996) compiled rotation curve.
N3627: Inner part from Zhang et al. (1993)'s
CO 1-0 observation (rescaled to the current distance value, and
changed to use terminal velocity instead of peak velocity of
the molecular contours), outer part from Chemin et al. (2003)'s
H-alpha observation.
N3351: Inner part from Devereux, Kenney, \& Young (1992)'s CO 1-0 observation, 
outer part from Buta (1988) H-alpha observation.
N4736: From Jalocha et al. (2008), which was adapted 
from Sofue et al. (1999) compilation.
From Figure~\ref{fg:Fig15}, we observe that the contribution
of the disk baryonic matter (excluding the contribution from
Helium and heavy elements) to the total rotation curve
increases as the galaxy's Hubble type changes from the late 
to the early as the galaxy evolves along the Hubble sequence.  
For early-type galaxies such as NGC 4736, the entire rotation
curve may be accounted for by disk baryonic matter (Jalocha et al. 2008).
Note that the close match between the observed and disk-inferred
rotation curves for NGC 3627, which is unusual because of its
intermediate Hubble type, maybe be a result of the close
encounter with NGC 3628 which likely to have striped a large
portion of its halo.

We have verified that the mass flow rates calculated
according to the prescriptions in \S2 is not sensitively
dependent on the choice of scale height in the
potential calculation (i.e. a change of
Hz from 12.6" to 3.8" in the case of NGC 4321 changed
the scale of mass flow by only 10\% or less for the
entire radial range considered).
From Figure~\ref{fg:Fig16}, we see that the mass flow
rates for the various Hubble types range from a few solar
masses per year to about a few tens of solar masses a year,
except for the very late type galaxy NGC 628 which is still
in the process of coordinating a significant
galaxy-wide mass flow pattern. The intermediate-type
galaxies appear to have the largest mass flow rates (the one
galaxy, NGC 1530 studied in ZB07, which has mass flow
rates on the order of 100 solar mass a year, is also
an intermediate-type galaxy), whereas 
for the early-type galaxies the mass flow is more concentrated
to the central region.  Although both inflow and outflow
of mass are present at a given instant across a galaxy's
radial range, due to the need to support the
structure of density wave modes which have alternating
positive and negative angular momentum density outside
and inside CR, respectively (this inflow and outflow of
mass for a density wave mode was also previously confirmed
in the N-body simulations of DT94 and Z96), there is a general trend
for the mass to concentrate into the inner disk with time, together
with the build-up of an extended outer envelope,
consistent with the direction of entropy evolution.
This trend will be confirmed next in Figure~\ref{fg:Fig17}, which shows
that for late and intermediate-type galaxies the gravitational
torque couple (and so is advective torque couple which has similar
shape) has mostly positive values in the inner disk,
which shows that the angular momentum is consistently
being channeled out of the galaxy. There is also a trend
there that the mass flow and angular momentum channeling
activity progressively moves from the outer disk to the
inner disk as a galaxy evolves from the late to the early
Hubble types. 

Note that the radial mass flow rates presented
in these curves are actually the lower bounds on the actual
radial mass flow rates due to nonaxisymmetric instabilities in 
physical galaxies, since these instabilities are expected
to result in skewed distributions in parts of the thick-disk
and halo as well, and thus result in the radial redistribution
of the halo dark matter along with the luminous disk matter
(with the composition of the dark component likely to be made of baryons too,
see discussions that follow). The nonaxisymmetric instabilities
in the spheroidal components are reflected in the twisted
isophotes often observed in disky elliptical galaxies.  These
twisted distributions in the spheroidal components
will lead to the same potential-density
phase shift and collective secular mass redistribution as
we discussed for the disk component.
Therefore, the average mass flow rate
is sufficient for the build-up of Milky Way type Bulge
in a Hubble type, and for some of the strongly interacting
or else high-wave-amplitude isolated galaxies (like NGC 1530),
evolution towards disky ellipticals within a Hubble time
is also entirely possible.

This picture of the gradual build-up of the Hubble sequence by
the radial accretion of the baryonic matter, however,
poses a significant challenge to the currently popular LCDM paradigm:
The magnitudes of the mass flow rates
appear large enough to be able to effect a significant Hubble type 
transformation during the past Hubble time (i.e. 10 solar masses
a year corresponds to $10^{11}$ solar mass during the past
$10^{10}$ years, enough to build up a massive central bulge
in a Milky-Way type galaxy).
The mass flow rates are especially large for galaxies 
experiencing strong tidal perturbations (i.e. NGC 3627 and NGC 5194),
due to the large-amplitude density wave patterns excited
during these strong interactions.  The interaction-enhanced
secular evolution effect can thus
underlie the well known morphological Butcher-Oemler effect
in clusters (Butcher \& Oemler 1978) which transform late
type disks to early-type spirals and S0s.
The effect is also likely responsible for producing the morphology-density
relation in different galaxy environments (Dressler 1980).

If the galaxy rotation curve (RC) and mass flow rates indicate that 
along the Hubble sequence from late to early, more and more of the RC 
is contributed by luminous baryonic matter as a result of secular mass accretion,
then where did the non-baryonic cold dark matter (CDM)
go as this evolution proceeds?  The CDM could not simply be
displaced by the baryonic matter, since a well-known result
of the LCDM numerical simulations is the so-called
adiabatic compression of the dark matter halos (see, e.g. Sellwood
\& McGaugh 2005 and the references therein), i.e. as the baryons
become more centrally-concentrated, the dark matter particles
should be dragged into the central potential wells thus formed,
since the dark matter particles are cold as hypothesized.

To resolve this paradox, in fact the most natural solution is
to hypothesize that the galactic dark matter is mostly
made of dark baryons.  This baryonic dark matter
could be in the form of brown dwarfs and other non-luminous 
Massive Compact Halo Objects (MACHOs) as revealed by microlensing 
observations (Alcock et al. 1997; Koopmans \& De Bruyn 2000).
Some may argue that the current microlensing surveys have not
uncovered enough dark matter to account for all that is needed
to reproduce the galactic rotation curve, and that the extrapolated
initial mass function (IMF) for star formation in the galactic plane does
not seem to produce enough mass to close the gap either.  We comment that
the current microlensing observations may still be incomplete,
and part of the discrepancy from the IMF argument may be due to the fact that
IMF in different environments may be significantly different (i.e.,
in the disk plane the star formation is caused by spiral density
wave shocks, and thus produce a family of stars that weigh heavily
towards high mass stars.  The mass formed in the quiescent environment
of the halo and outer galaxy may on the other hand weigh
heavily towards low mass stars and sub-stellar objects.

This proposal for the form of galactic dark matter
not only makes the secular evolution picture
more consistent (since the brown dwarfs have smaller mass, they
can be displaced to the outer galaxy as the galaxy establishes
new dynamical equilibrium -- this is a kind of diffusion or
dynamical friction effect), but it also makes other known paradoxes
in the LCDM paradigm, such as the core-cusp controversy of the
LSB and dwarf galaxies (de Blok 2010), as well as the difficulty of forming
galaxies with realistic rotation curves in the LCDM simulations
whenever the baryonic matter fraction is not high enough (White 2009),
be easily solved.  This solution would make the Hubble sequence galaxy dynamics
all the more transparent:  The early-type galaxies have cuspy
mass distribution because the baryons there are more concentrated
through secular evolution.  In certain early-galaxies, such as NGC 4736,
it has been found that the contribution of dark matter to the
total mass distribution is very small (Jalocha et al. 2008).
The late-type galaxies, on the other hand, are dark-matter dominated
but were often found to have core-shaped mass distribution contrary to
the LCDM prediction.  In our current picture this can be understood
easily since in late-type galaxies secular evolution has just been
launched, and not enough central flowing of matter has been
accomplished.  The secular evolution scenario can naturally account
for many other observed regularities of galactic structure
and kinematics along the Hubble sequence, such as the galaxy
scaling relations and the existence of a Universal Rotation Curve
trend (Zhang 2004, 2008), which were previously difficult to account
for in the LCDM framework.  

It is a well-known fact that a fraction of the baryonic matter 
has to be dark, from the discrepancy between the predicted 
cosmic baryon fraction through Big Bang nucleosynthesis,
and the local observed values 
(Persic and Salucci 1992; Fukugita, Hogan, \& Peebles 1998; McGaugh 
et al. 1010).  If a fraction of the baryons can be dark, then more of it potentially
can be dark as well.  This proposal for the composition of dark matter
to be made of mostly dark baryons works well for many known dark matter
observations such as the colliding Bullet Cluster (Clowe et al. 2006) and 
the Train-Wreck Cluster (Mahdavi et al. 2007), which are sometimes difficult
to reconcile under the standard LCDM paradigm (i.e. in the Bullet
Cluster this dark matter seems to cluster with the stars in the outer
two humps and not with the gas that is in the middle of the collision
reminant, and yet in the Train-Wreck cluster the dark matter clusters
with both the stars in the outer two humps as well as the central gas clump).
In our new picture, since the baryonic 
dark matter consisting of MACHO-type objects
that have a mass spectrum that can go from sub-stellar objects all the
way to dust particles, it can display a more or less degree of
dissipation as its detailed composition entails.  In the Bullet cluster,
the mass spectrum of the baryonic dark matter probably lean more towards
slightly sub-stellar, whereas in the Train-Wreck cluster a more continuous
spectrum from the barely sub-stellar to the super-dusty were all present,
and the collision dynamics serves as a graded sieve to separate and arrange these
mass components along a continuous path.  

This proposal also easily eliminated the frustration of the non-detection of CDM
particles, as well as some obvious challenging consequences we have to face if
the CDM particles are in our midst.  If the CDM particles gravitationally cluster 
with the baryons as hypothesized, and are mixed with the baryons in a ratio of
5 to 1, how come we could not feel their 
presence and detect their dynamical consequences
say in the earth's atmosphere, which is gravitationally bound to
the earth itself.  And the increasingly-strong evidence
of the down-sizing trend for structure formation 
(Cimatti et al. 2006; Nelan et al. 2005)
is another serious challenge to the bottom-up mass
assembly scenario of the LCDM paradigm, and the same trend is much
more easily understood in the secular evolution picture.

We comment here also that our preference for MACHOs as candidates for a
significant fraction of the galactic dark matter, as opposed to exotic
forms of non-luminous gaseous baryons that reside in the galaxy disk
(e.g. Pfenniger \& Combes 1994), is due first of all, to the trend of
properties seen along the Hubble sequence, i.e.  the late-type disk
galaxies have the highest dark matter contribution to their RCs, and
yet their (luminous) disks are the lightest.  Therefore, if the
hypothesized non-luminous disk-gas baryons come in fixed fraction with
the luminous disk-gas baryons, this trend of spheroidal-to-disk mass
ratio along the Hubble sequence would be hard to account for.
Secondly, from the modal theory of Bertin et al. (1989), density wave
morphology along the Hubble sequence is linked to a galaxy's basic
state properties.  The fact that late-type galaxies have open and
multi-armed spiral patterns require that their parent basic state is
such that most of the dynamical mass resides in a spherical halo, which
is naturally consistent with the the dark matter being a MACHO type of
halo dark matter, and is inconsistent with it being cold and
non-luminous disk gaseous matter.

The unlikeliness of the galactic dark matter being made of hot gas,
on the other hand, is due to the fact that hot gas usually has a
very diffused and extended distribution, and its role is very
minor in the secular transformation of Hubble types which
requires mass to be accreted into the central region of a galaxy
on the time scale of a fraction of a Hubble time.  Hot gas will
first need to dissipate its energy and settle onto the disk
for it to participate in the global-instability-facilitated
secular evolution, and even after that it is a long way from
the outer galaxy to the central region.  Even with the facilitation
of disk global instabilities, we know the secular orbital decay
rate is only a few kpc per Hubble time (Z99).

The observations of the Bullet Cluster are also inconsistent with the
dark matter being made of a mostly dissipative gas component
(either cold or hot gas).  

This choice on the nature of galactic dark matter
does leave us with the demanding task of rethinking
the cosmic evolution history given by the concordance LCDM
cosmological model.  But, it is a reasonable assumption that
we are in better understanding of local physics than we are of the distant
universe, and the uncertainties in the details of the models
of the universe certainly do not exclude the possibility
of baryons as the major contributor of the
dark matter (for example, a slight variation with time of 
the fundamental constants,
such as the gravitational constant G, could lead to a different
prediction on the expansion rate in the early universe, and thus
different predictions on the results of primordial nucleosynthesis that
would make it possible to accommodate a larger present-day baryon fraction;
Likewise, the relative strengths of the different peaks on the CMB
angular spectrum are not uniquely fitted by the current concordance
parameter set -- and it is well known that the acoustic peaks in the
CMB angular spectrum were first predicted in purely barynoic
cosmological theories [Sakharov 1965, Sunyaev \& Zeldovich 1969;
Peebles \& Yu 1970]).  And the local physics, as revealed by observations of
the physical properties of galaxies and by the secular evolution
scenario for galaxy morphological transformation, seems to strongly
suggest that our current fundamental cosmological theories about 
structure formation is in need of major revision.

\subsection{Relative Contributions from
Gravitational and Advective Torque Couples, and 
Relations to Other Studies of Symmetry Breaking
and Collective Effects}

\begin{figure}
\vspace{410pt}
\includegraphics{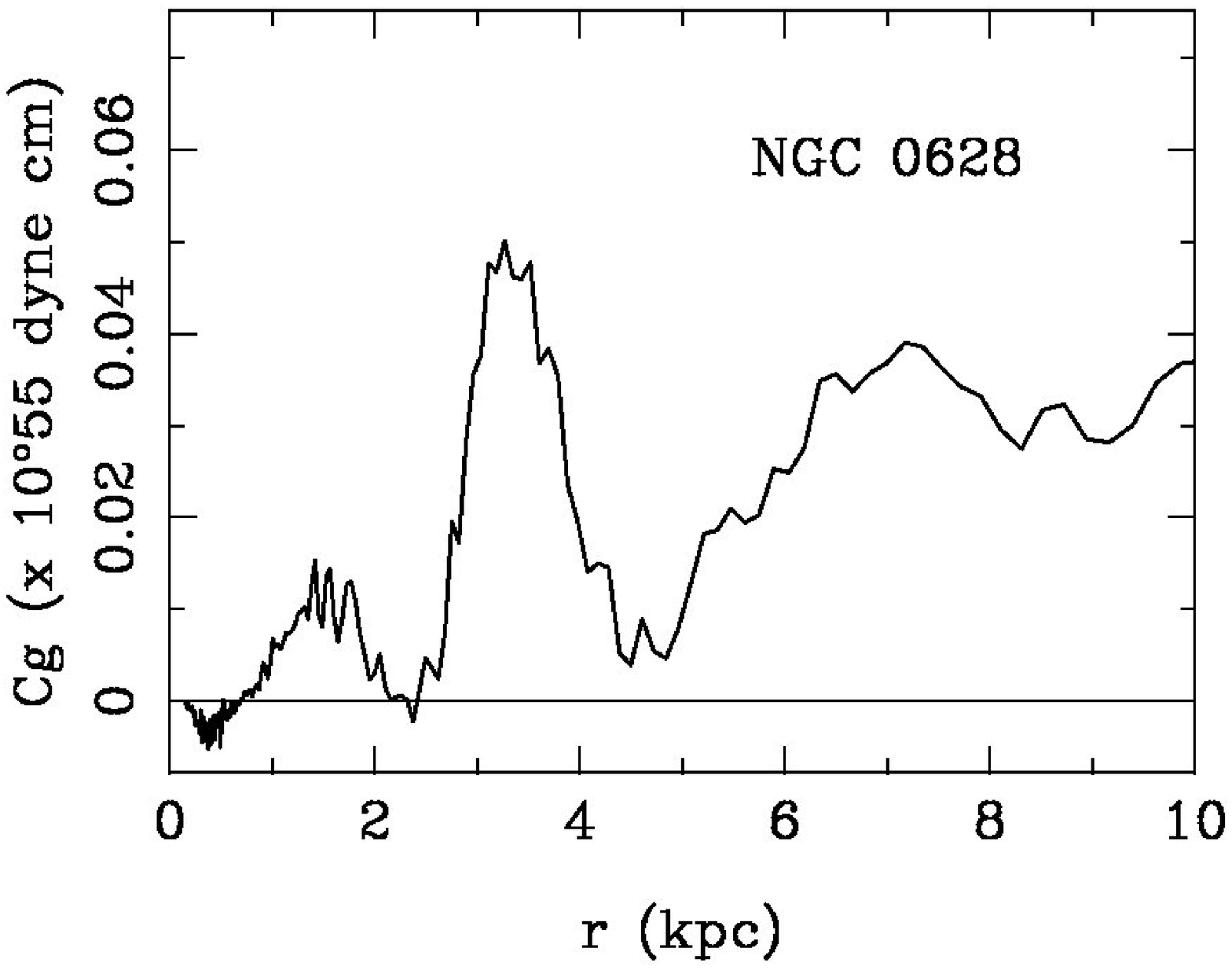}
\includegraphics{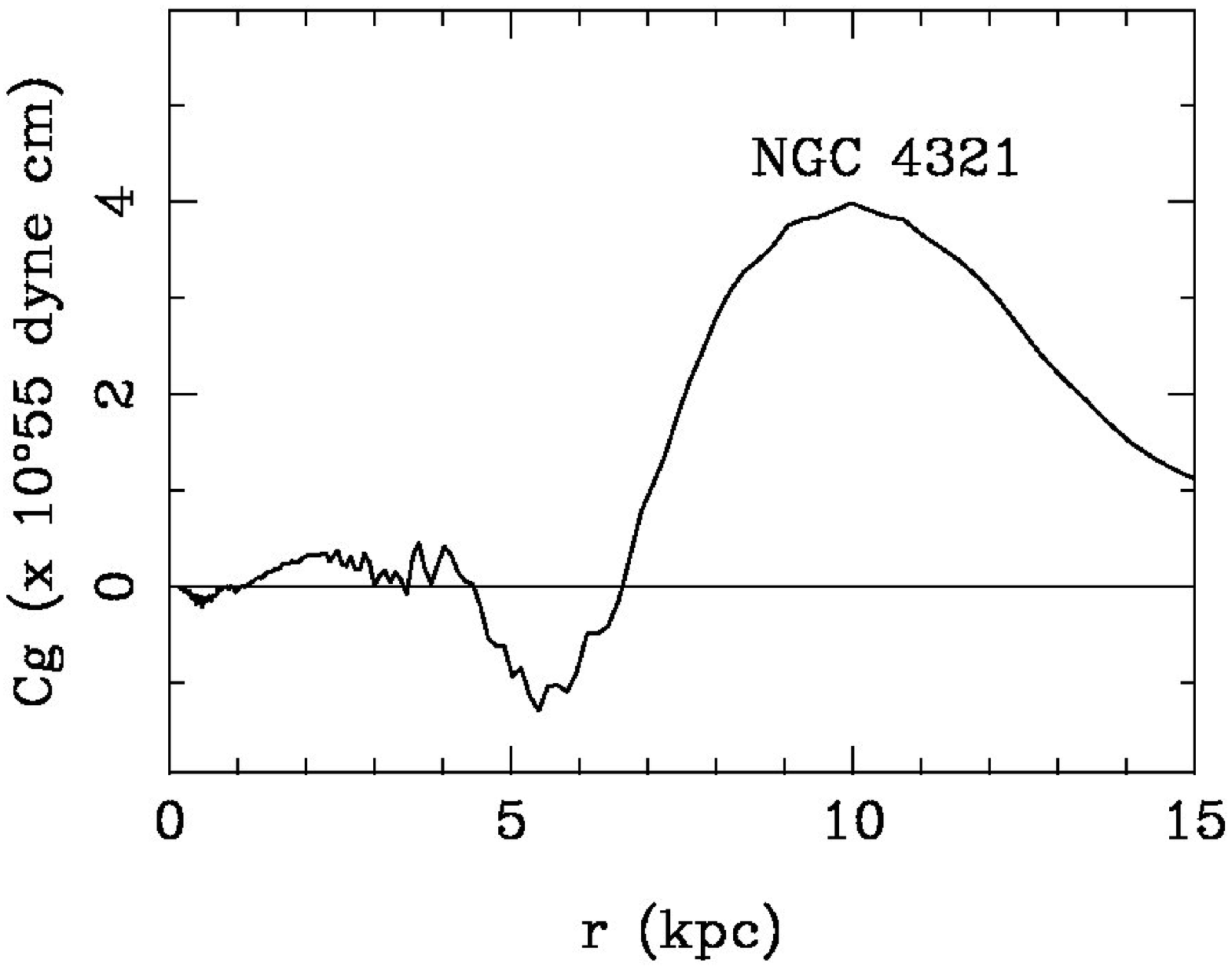}
\includegraphics{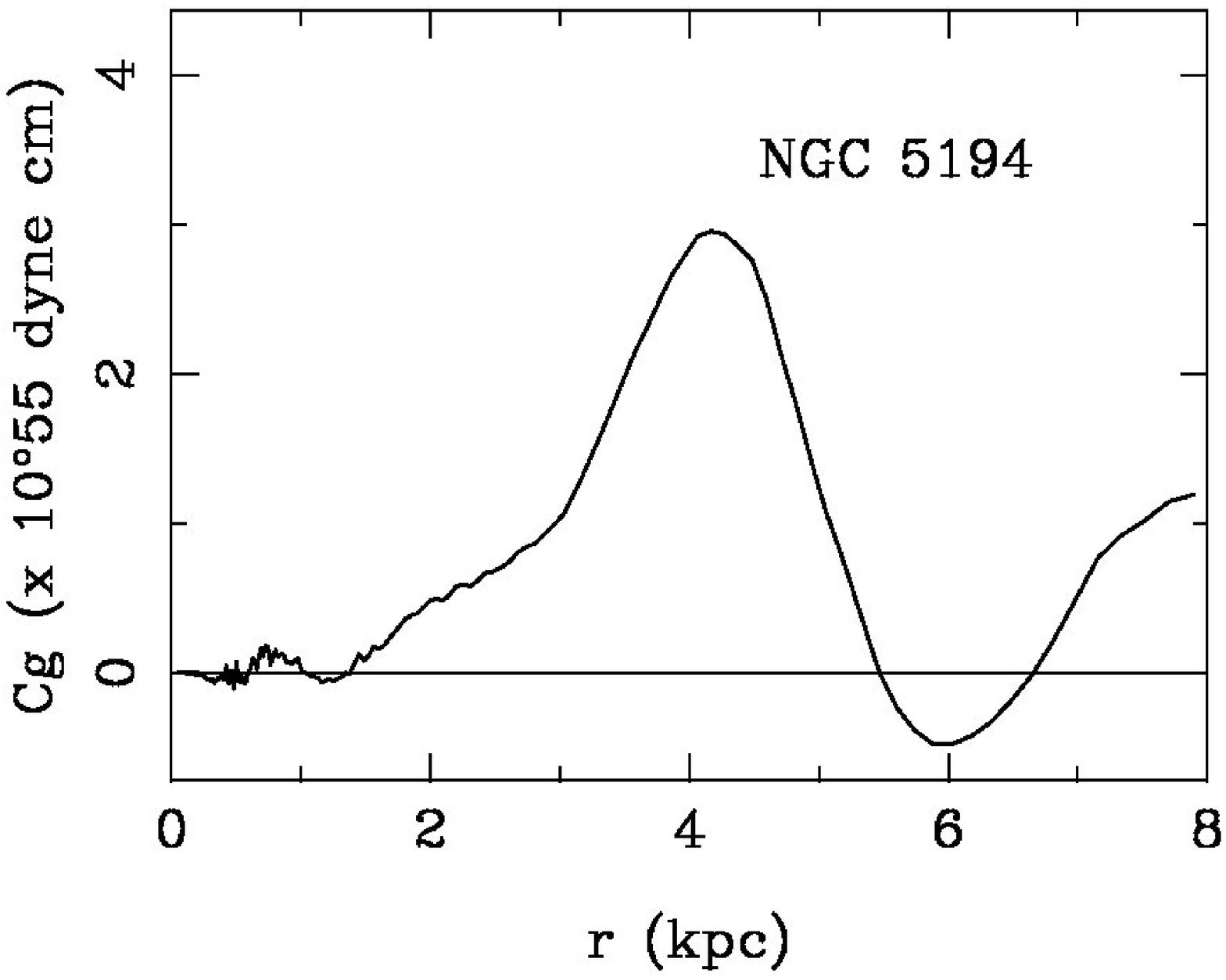}
\includegraphics{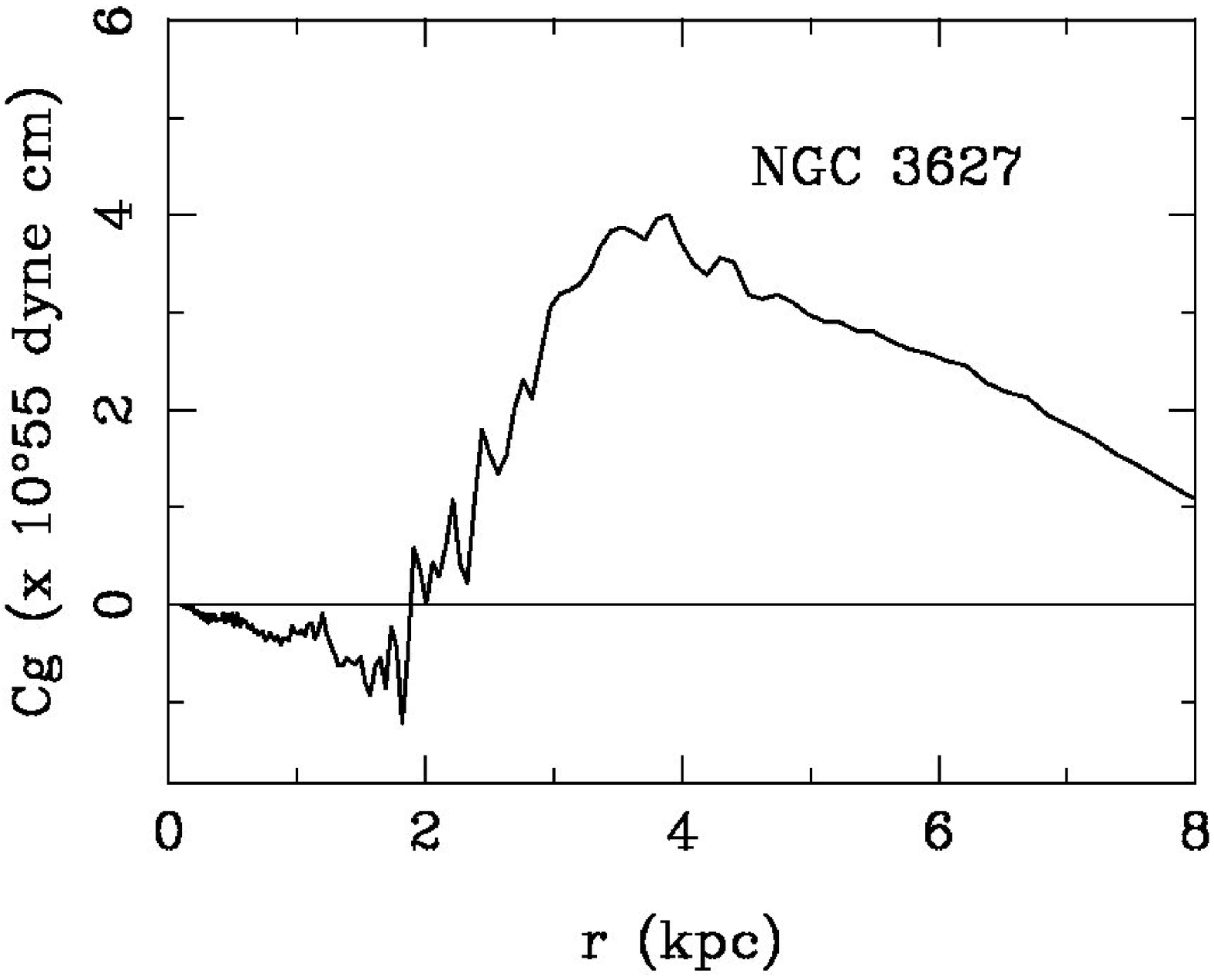}
\includegraphics{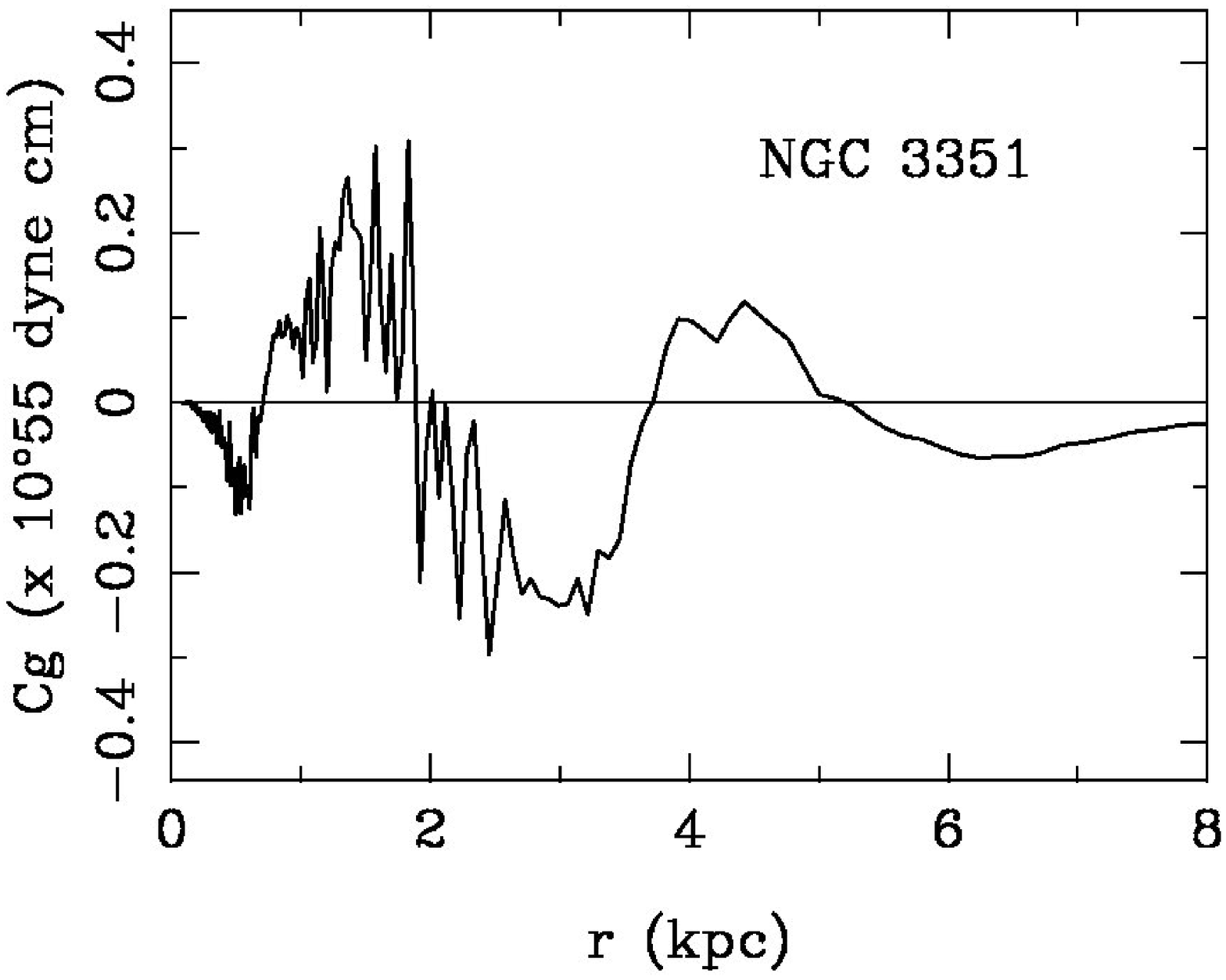}
\includegraphics{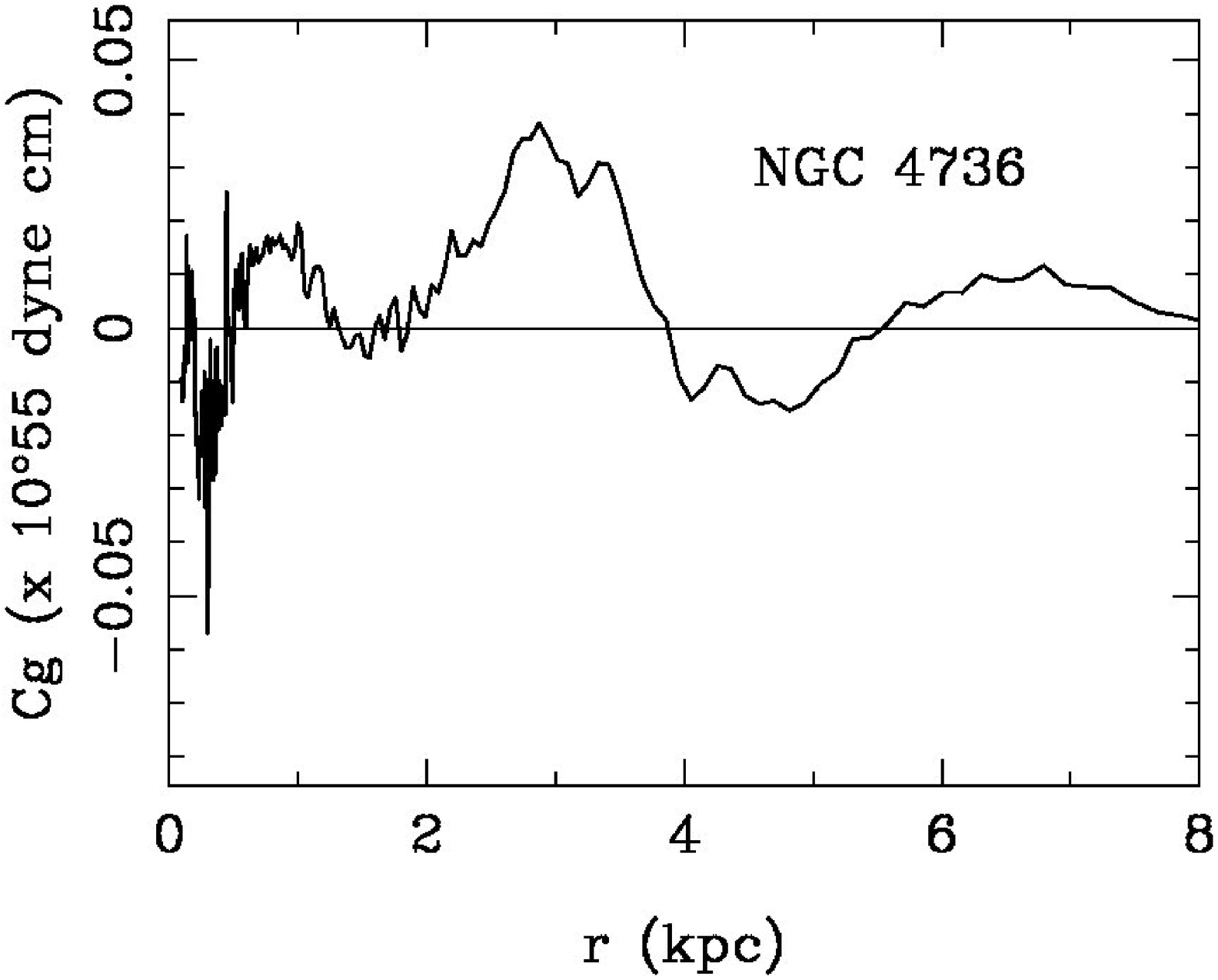}
\caption{Calculated gravitational torque coupling for the six sample
galaxies, using IRAC 3.6 $\mu$m data for NGC 0628,
an average of IRAC 3.6 $\mu$m and SDSS i-band data for NGC 4321, 3351,
5194, 3627, and SDSS i-band data for NGC 4736, to these the
usual gas maps were added.}
\label{fg:Fig17}
\end{figure}

In Figure~\ref{fg:Fig17}, we plot the calculated gravitational
torque couple for the six sample galaxies.  First we discuss the
result for NGC 4321.  This
torque result is very similar in shape to the one calculated
for the same galaxy by Gnedin et al. (1995), though the scale factor
is more than a factor of 10 smaller than obtained in their paper.
Part of the difference can be accounted for from the difference
in galaxy images and in the galaxy parameters used between these two studies.
We have tried to switche to use an R-band image as in Gnedin et al.,
and rescaled the galaxy parameters to be in agreement
with what they used. However, after these adjustments the resulting scale
is still smaller by a factor of $\sim 5$ from that in the Gnedin et al. (1995).
The same amount of magnitude difference is recently found by Foyle et al. (2010)
as well when they try to reproduce the Gnedin et al. result.  
It is possible that there was an internal error in the
Gnedin et al. calculation.

As we see from a close inspection of Figure~\ref{fg:Fig17},
for most of the galaxies in the sample, especially
those of late to intermediate Hubble types, there is a main peak of the
gravitational torque couple, which is centered
in the mid disk.  Thus the implied mass inflow/outflow
should be focused around the mid disk near the peak location 
of the main bell curve.
This location, however, is expected to change with time 
for a given galaxy as secular
evolution proceeds, since the basic state configuration and
the resulting modal configuration will both evolve due to the
redistribution of the disk mass.  In fact, by examine this
small sample of six galaxies, we can already see a trend that the
main peak of the $C_g$ curve moves from the outer region of the galaxy
(as for NGC 0628) to the mid-disk (as for NGC 4321, 5194 and 3627),
and then onward to the central region (NGC 3351 and 4736).

The existence of the multiple humps of the gravitational torque
couple (as well as advective and total torque couple, since these latter
two are found to be of similar shape to the gravitational torque couple
for self-sustained modes) in a single galaxy,
and by implication multiple nested modes of differing pattern speeds, 
shows that the secular mass flow process is not a one-way street.  
During the lifetime of a galaxy a given radial range will experience 
inflows and outflows of different magnitudes, but there is an overall
trend of the gradual central concentrating of matter together
with the building up of an extended envelop, consistent
with the entropy evolution direction of the self-gravitating
systems.  Galaxies accomplish this in a well-organized fashion,
by first employing a dominant mode across the galaxy's
outer disk, and subsequently with more of the modal activity moving
into the central region of a galaxy, and the single dominant
mode also bifurcates into multiple nested modes.

\begin{figure}
\vspace{410pt}
\includegraphics{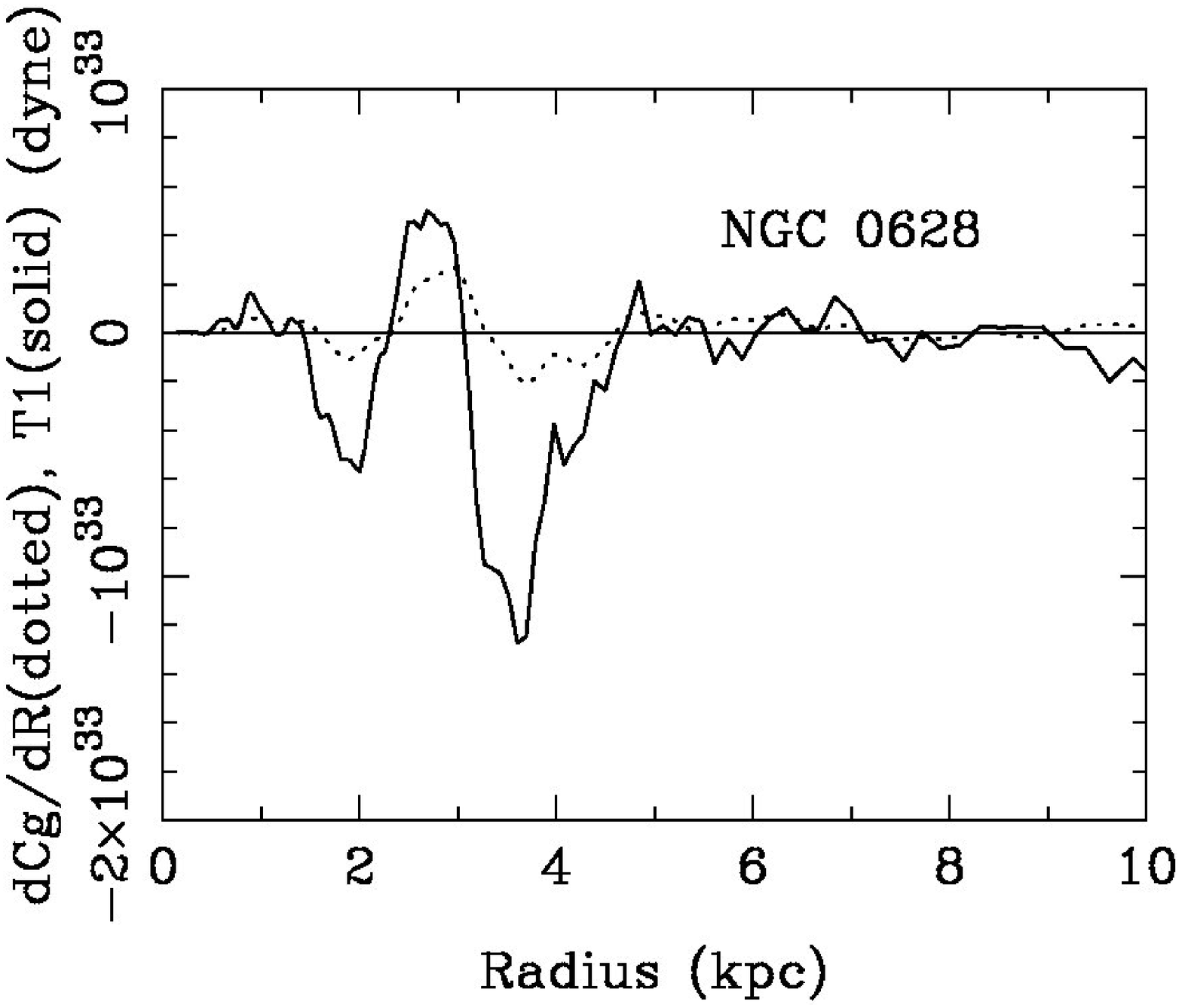}
\includegraphics{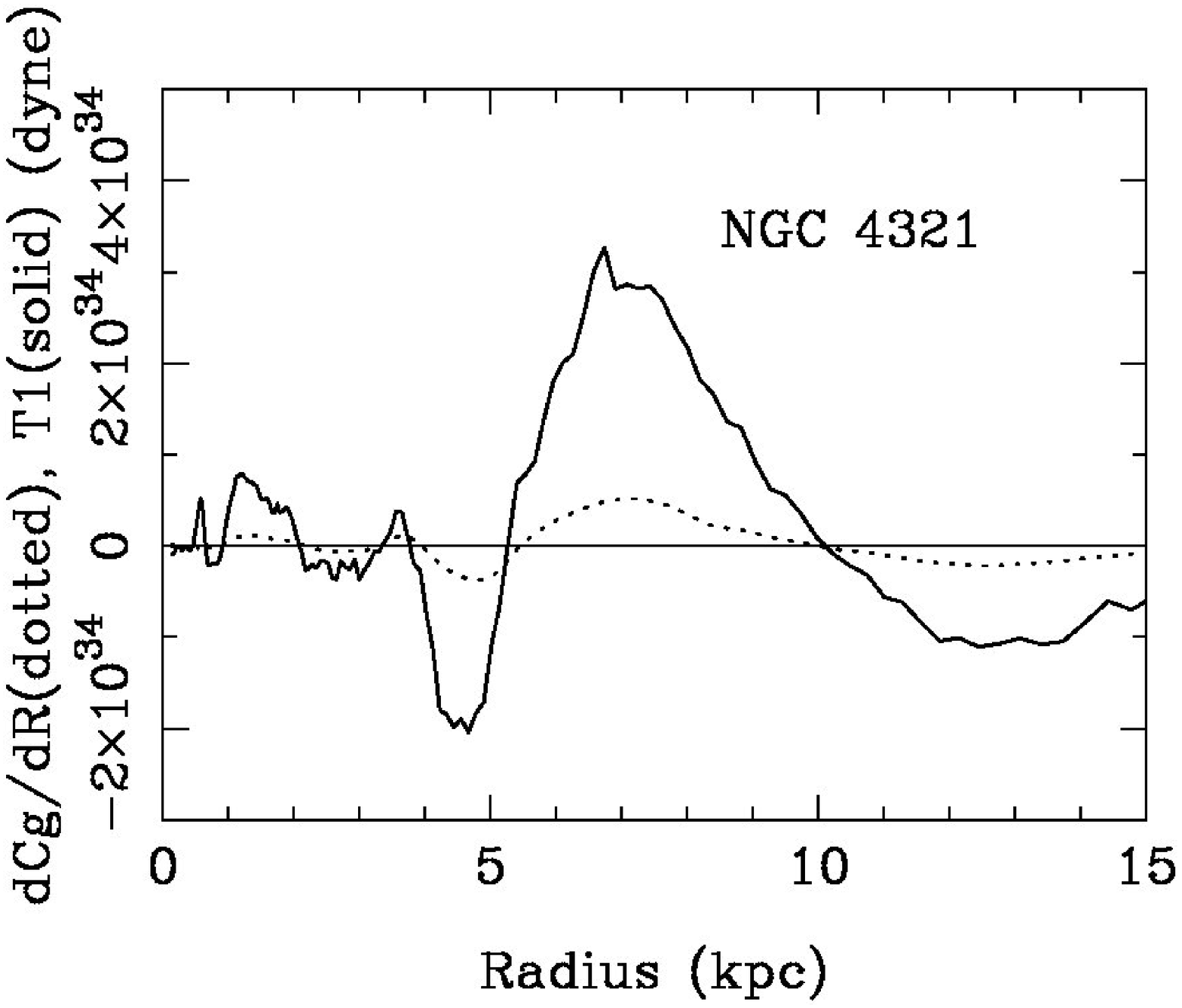}
\includegraphics{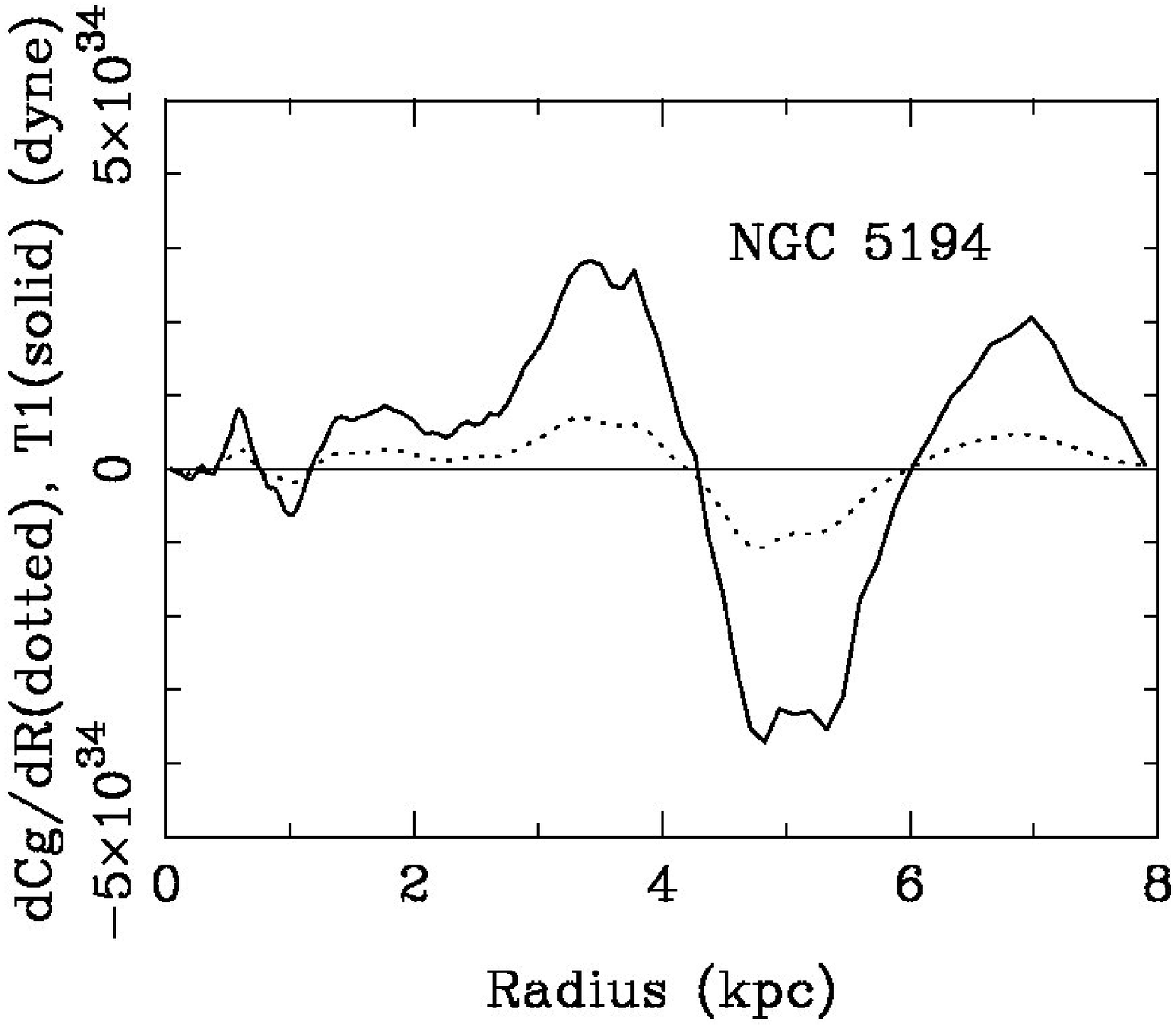}
\includegraphics{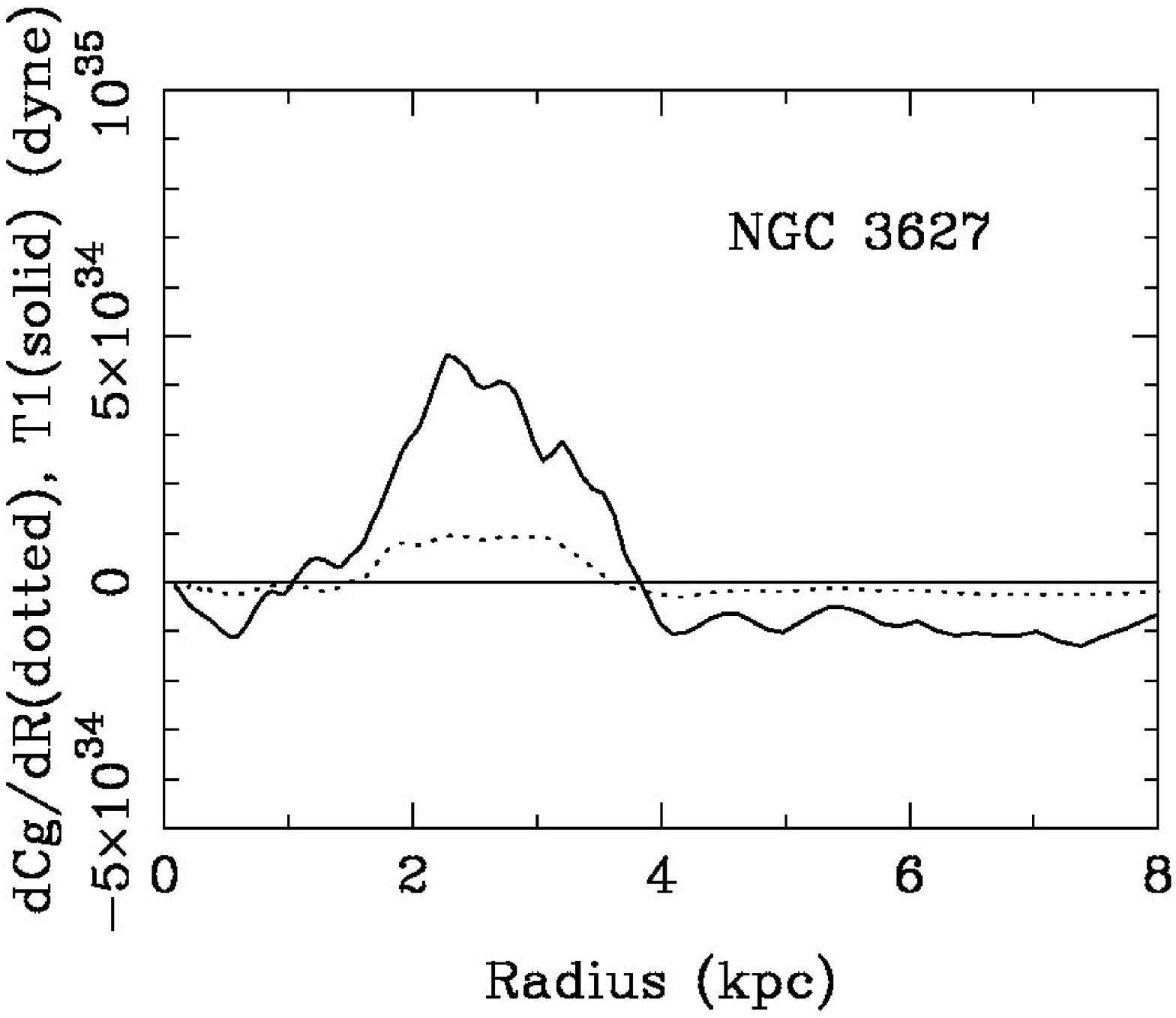}
\includegraphics{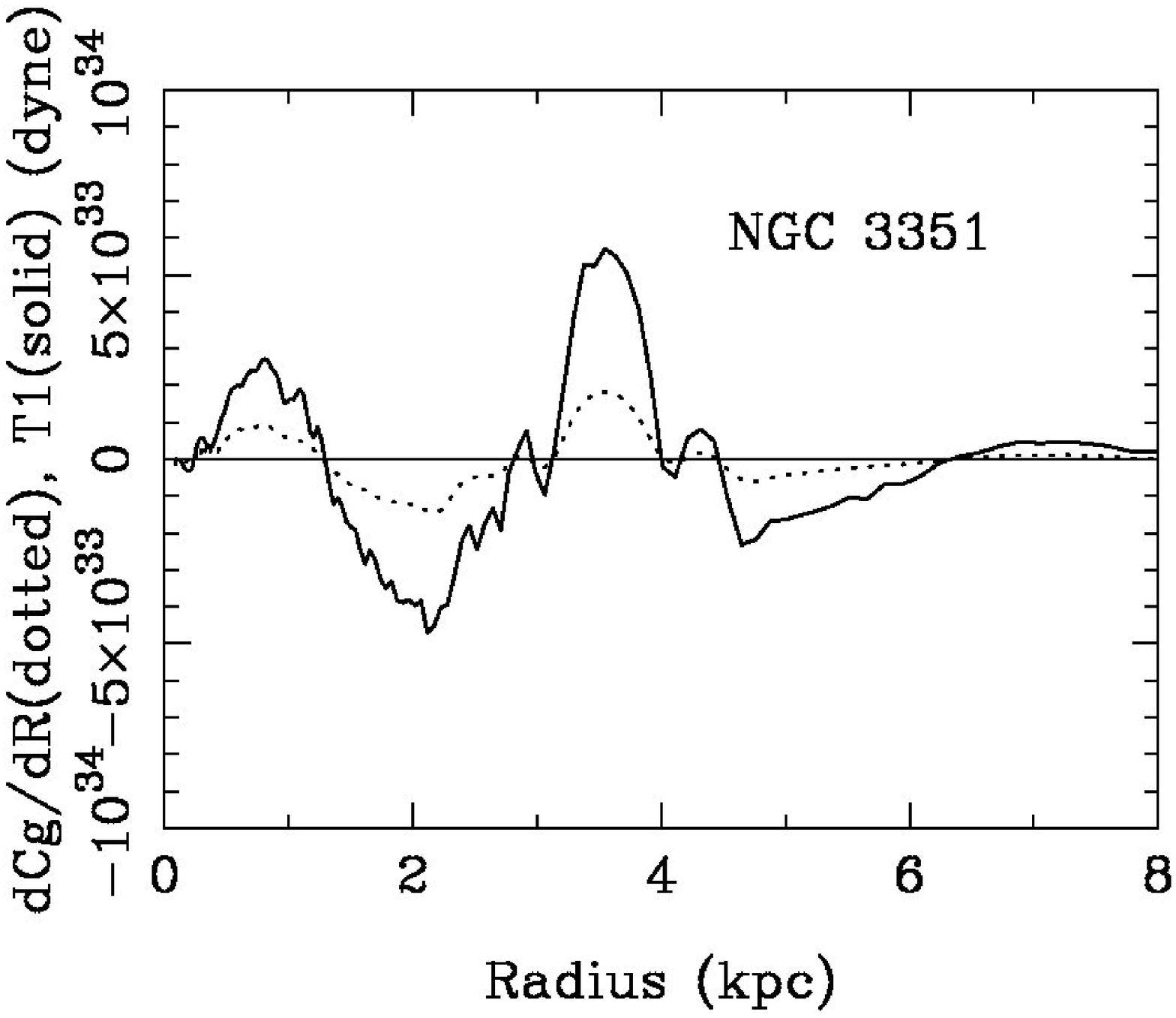}
\includegraphics{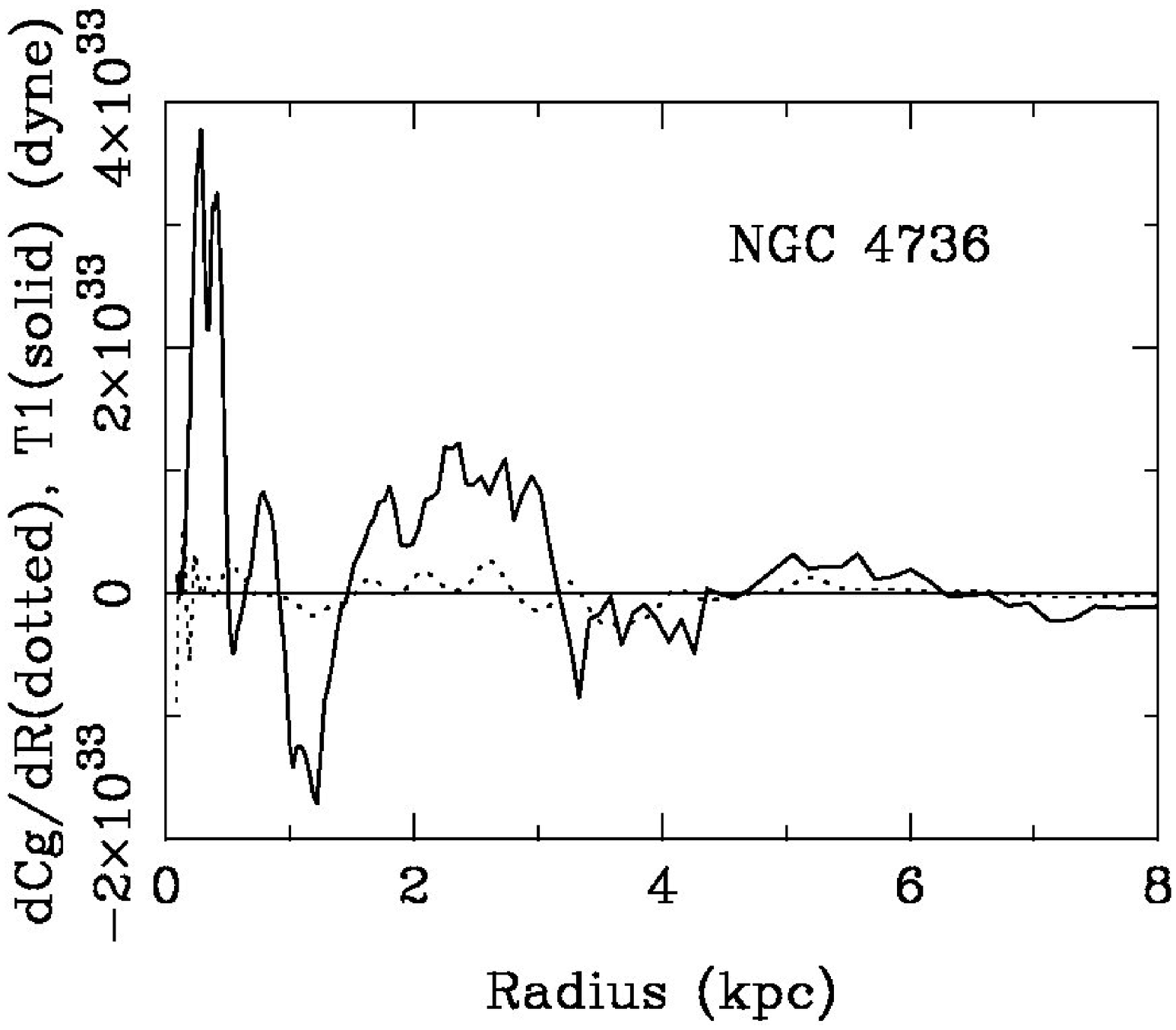}
\caption{Radial gradient of gravitational torque coupling integral compared
with the volume torque integral for the six sample galaxies.  For
NGC 0628, used IRAC 3.6 $\mu$m data; for NGC 4321, 3351, 3627, 5194,
used the average of IRAC 3.6 $\mu$m and SDSS i-band data; for NGC 4736,
used SDSS i-band data; to these the usual gas maps were added.}
\label{fg:Fig18}
\end{figure}

In Figure \ref{fg:Fig18}, we show the calculated
radial gradient of gravitational torque coupling integral 
$dC_g(R)/dR$ as compared to the volume torque integral $T_1(R)$.  
There is about a factor of 3-4 difference between the
$T_1(R)$ and $d C_g(R)/dR$ for most of the galaxies in the sample
(for NGC 4736, this ratio in the inner region is as high as 8),
indicating that the remainder, which is contributed by the gradient
of the advective torque coupling, $d C_a(R)/dR$,
is in the same sense but much greater in value than $dC_g(R)/dR$.
Note that this difference between the volume torque integral 
and the gradient of the (surface) gravitational torque couple
is only expected in the new theory: in the traditional theory 
of LBK these two are supposed to be equal to each other 
(see Appendix A2 of Z98).  

Furthermore, if the LBK theory is used literally,
one should not expect any mass flow rate at all over most
of the galactic radiii in the quasi-steady state, 
except at isolated resonance radii
(since the total angular momentum flux, $C_a(R)+C_g(R)$,
is expected to be constant independent of radius in the LBK theory).
The existence of a mass flux across the entire galactic disk is 
thus also contrary to the LBK's original expectations,
especially since there is strong evidence
that the patterns in these same galaxies were steady
modes (i.e., the fact we can use the potential-density
phase shifts to successfully locate resonances in ZB07 and BZ09
shows that these patterns were steady modes).

We mention here that we have repeated the calculations for 
$dC_g(R)/dR$ and $T_1(R)$ using different scale height values.
Even though both quantities are affected by a particular
choice of the scale height, the relative ratios
between the two quantities appear little affected. 
This is due to the fact that the potential
that is being affected by the scale height choice enters into both
the $C_g$ and the $T_1(R)$ results, so their ratio is insensitive
to the choice of scale height.  So the conclusion of the
significant difference between these two quantities at
the nonlinear regime of the wave mode is robust.

The above result tells us that
the correct treatment of gravitational many-body systems containing 
self-organized global patterns, such as density wave modes in disk galaxies, 
requires a re-examination of classical dynamical approaches and assumptions.  
Our experience so far has shown that entirely new qualitative and 
quantitative results can emerge from the collective mutual
interactions of the many particles in a complex dynamical system 
when the system is undergoing a dynamical instability or nonequilibrium 
phase transition.  Formerly sacred laws 
(such as the differential form of the Poisson equation) can
break down at the crest of collisionless shocks (Z98), and new meta-laws 
(such as the equality of the volume torque integral with the derivative 
of the sum of gravitational and advective surface torque coupling integrals) 
appear as emergent laws.  Such emergent behavior is the low-energy 
Newtonian dynamical analogy of high energy physics' spontaneous breaking 
of gauge symmetry, a well known pathway for forming new meta laws when 
traversing the hierarchy of organizations.

In this context, we would like to elucidate further the relation of 
galactic collective effects with similar phenomena in other classical
and quantum systems.  A terminology that is often used to describe this class
of phenomena is {\em anomaly}.  Wikipedia's definition for this term 
goes as follows: ``In quantum physics an anomaly or quantum anomaly is 
the failure of a symmetry of a theory's classical action 
to be a symmetry of any regularization of the full 
quantum theory. In classical physics an anomaly is the 
failure of a symmetry to be restored in the limit in which 
the symmetry-breaking parameter goes to zero. Perhaps 
the first known anomaly was the dissipative anomaly 
in turbulence: time-reversibility remains broken (and 
energy dissipation rate finite) at the limit of vanishing viscosity''.

From the above definition, we see that the existence of anomaly is 
associated with the symmetry breaking of the parent theory which is not
a smooth continuation (or logical extension) of the underlying
more symmetric theory (that was one of the reasons why the collective
effects in the galactic density wave disk were never rigorously ``derived''
from the classical action-angle-based mean-field theory, and why
so many ``experts'' of the old school have failed to realize these effects
for many decades).  In fluid turbulence 
(Firsch 1995 and the references therein), 
the amount of turbulence viscosity is not determined by microscopic
viscosity of the fluid medium, but by energy injection and cascade
from the largest scales.  This feature has an
analogous situation in galactic disks possessing
density wave modes -- the effective gravitational viscosity
was determined solely by the modal parameters and not by the
molecular viscosity in the underlying medium (a formal
analogy of the density wave disk with a viscous accretion disk,
including the derivation of effective gravitational viscosity
parameters, was given in Z99, \S2.1.1).  It is also for this
reason that the traditional distinction between the roles of
stars and gas in the context of density-wave induced mass redistribution
is so unnatural: In reality the viscosity in both mediums are
dominated by the collective gravitational viscosity whose amount
is determined by the self-consistency requirement of the large
scale pattern.

Another characteristic associated with anomaly or symmetry breaking
is the existence of singularity and closure relations. It is well-known
that one-dimensional Burgers equation in fluid dynamics spontaneous
develops a singularity in finite time if we evolve the solution
from an initial singularity-free distribution (Burgers 1948).  
Similar situation in three-dimensional turbulence was analyzed by 
Onsager (1949).  The analogy in galactic dynamics is that the
Eulerian fluid description of galactic density waves 
(in the cylindrical or disk geometry) will spontaneously
develop singularity at the wave crest in the nonlinear regime, if no
viscosity is introduced.  This spontaneous-steepening tendency of
the nonlinear density wave is the basis for the formation of 
spiral collisionless shocks (\S3.2 of Z96 and the references therein).  
The gravitational viscosity induced by the collisionless shock
offsets the steepening tendency in the azimuthal direction, 
and the balance of the two leads to the closure relation, i.e. the volume
torque relation which relates the total torque couple to the local
angular momentum exchange rate between the basic state and the
wave mode. The analogy of this phenomenon in quantum mechanics
include the formation of axial anomaly (Schwinger 1951), 
whose description can be formally likened to the description of two-dimension
turbulence (Polyakov 1993; Eyink \& Sreenivasan 2006).
The singularities formed in nonequilibrium phase transitions
separate the different hierarchies of organization (as in spontaneous breaking
of gauge symmetry), allowing older laws to cease to be valid and
new laws (or meta-laws, as in our case the volume-torque closure
relation) to take over as the emergent governing relations.
Pfenniger (2009) also brought attention to the relevance of fluid
turbulence analyses to the studies of galactic dynamics.

\subsection{The Relative Contributions of Stellar and Gaseous Mass Flows}

In the past few decades, secular evolution, bulge growth, and the evolution
along the Hubble sequence were studied mainly within the framework of
the secular redistribution of the interstellar medium under the influence
of a barred potential, and the resulting growth of the so-called
``pseudo bulges'' (KK04 and the references therein). KK04 showed, however,
that through gas accretion alone one could at most 
build late-type pseudobulges, but not intermediate and early type bulges, 
since the observed gas inflow rates and nuclear
star-formation rates are both insufficient to account for the formation
of earlier-type bulges through secular inflow of gas and the subsequent
conversion of gas to stars during a Hubble time.  Yet observationally,
early type galaxies (including disky-Es) are known to form a continuum 
with the intermediate and late types in terms of the structural parameters and 
kinematics (Jablonka, Gorgas, \& Goudfrooij 2002; Franx 1993), and thus 
their formation mechanisms are also expected to be similar.

With the recognition of the role of collective effects in the 
secular mass redistribution process, the study of secular 
morphological transformation of galaxies should now give 
equal emphasis on the role of stellar and gaseous mass redistribution.  Due
to their different intrinsic characteristics (radial surface density
distribution, compressibility, star-formation correlation, and dissipation
capability), stars and gas do play somewhat different roles in the secular
evolution process.  In this subsection we illustrate
with our sample galaxies some of the specifics of these roles.

\begin{figure}
\vspace{430pt}
\includegraphics{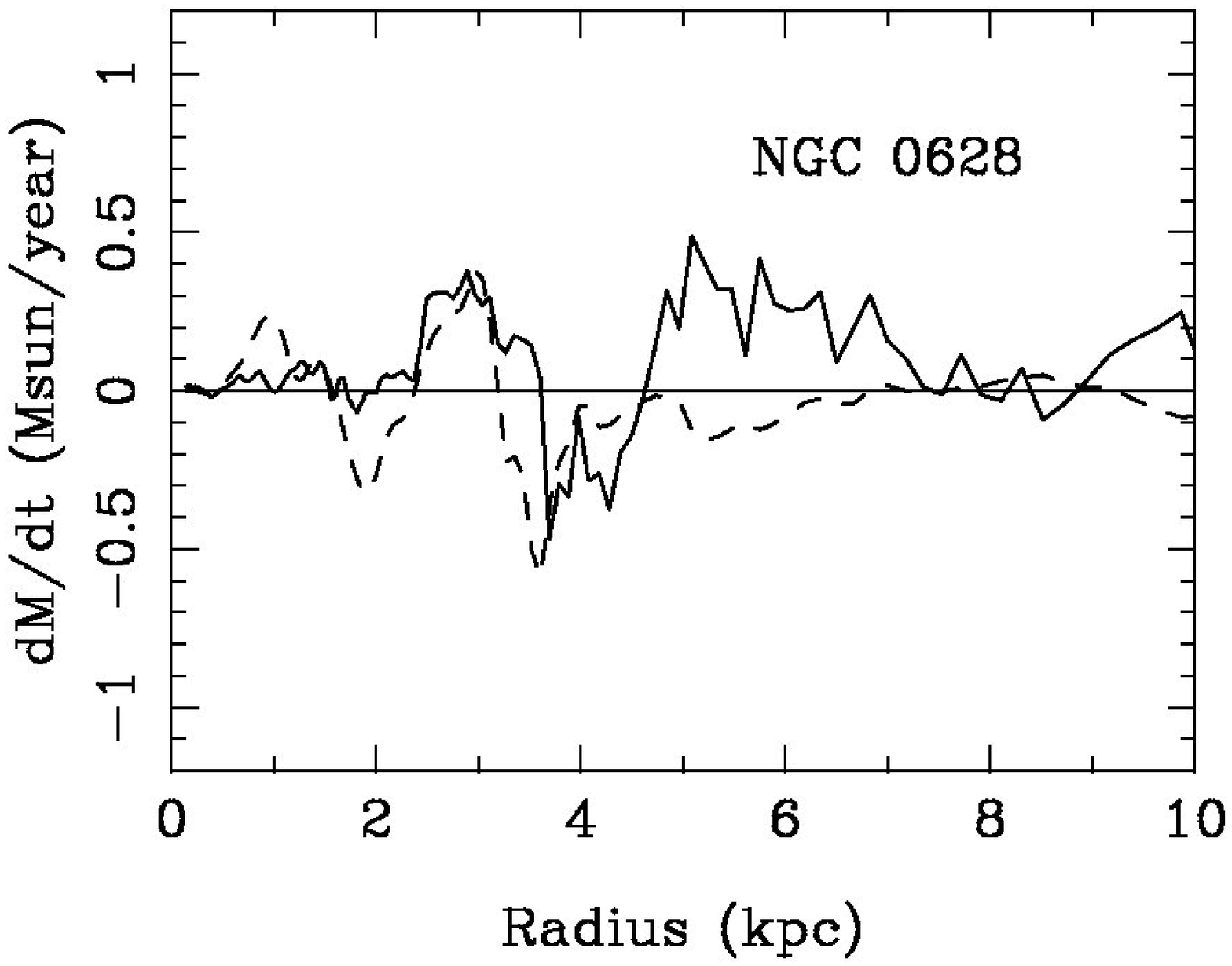}
\includegraphics{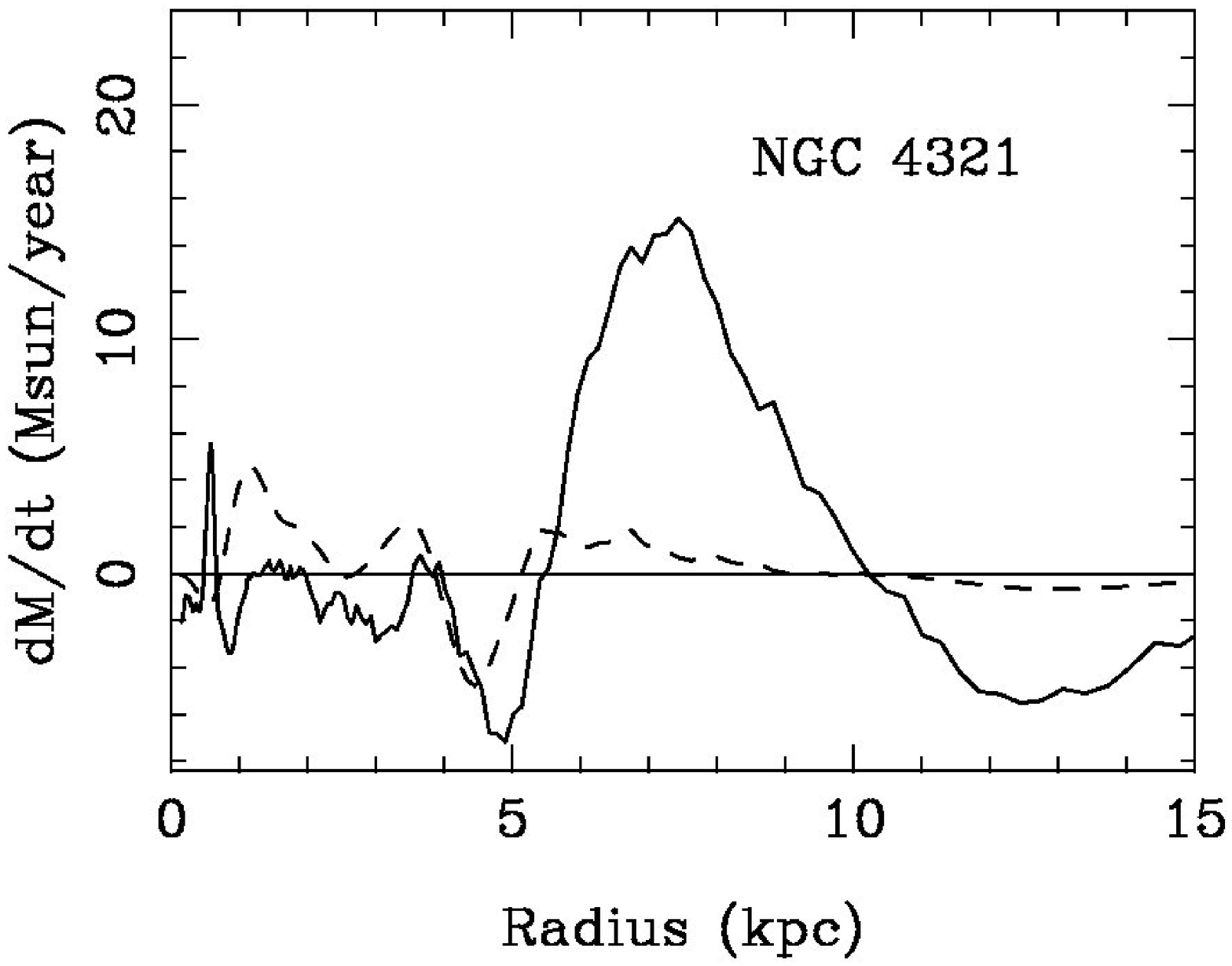}
\includegraphics{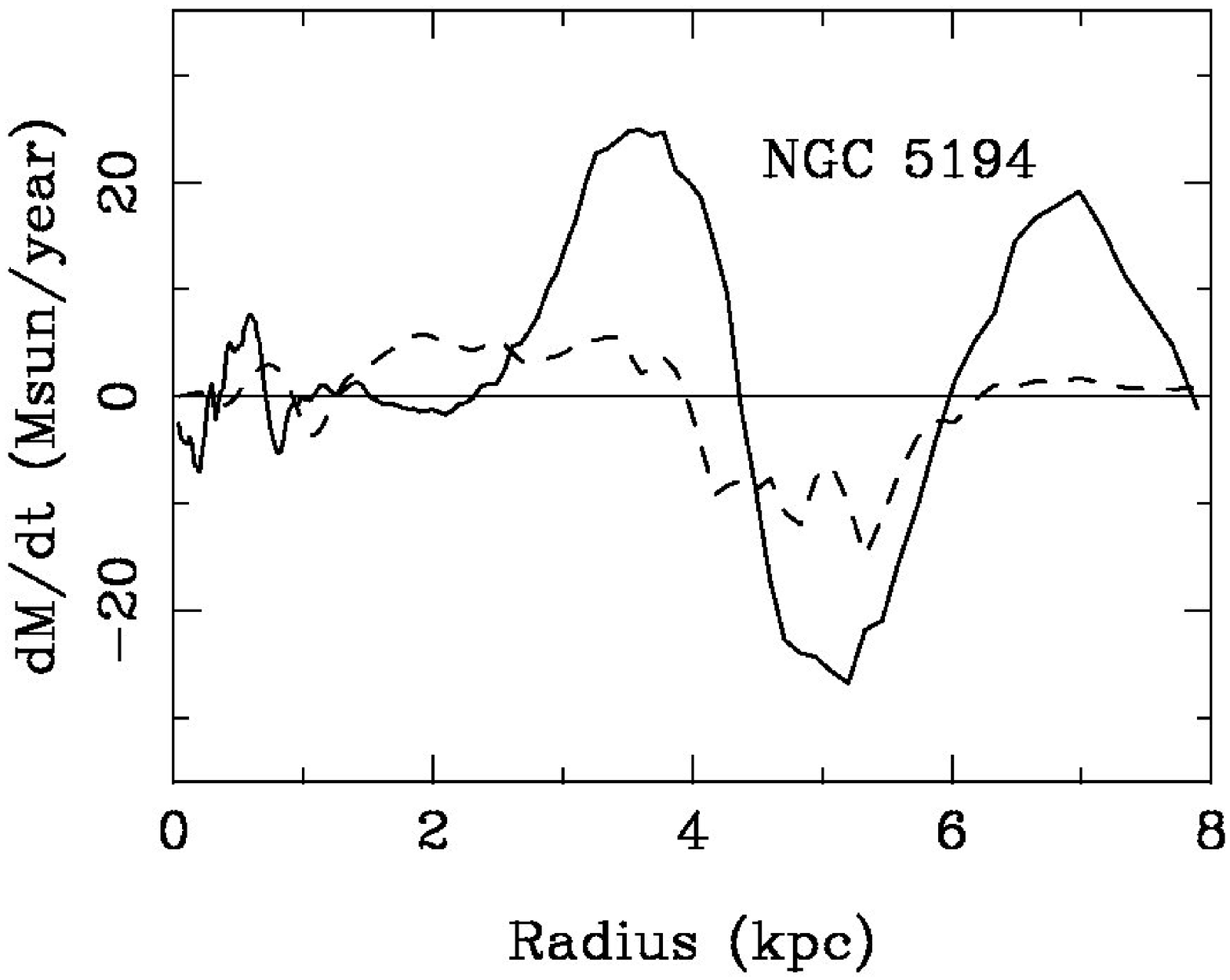}
\includegraphics{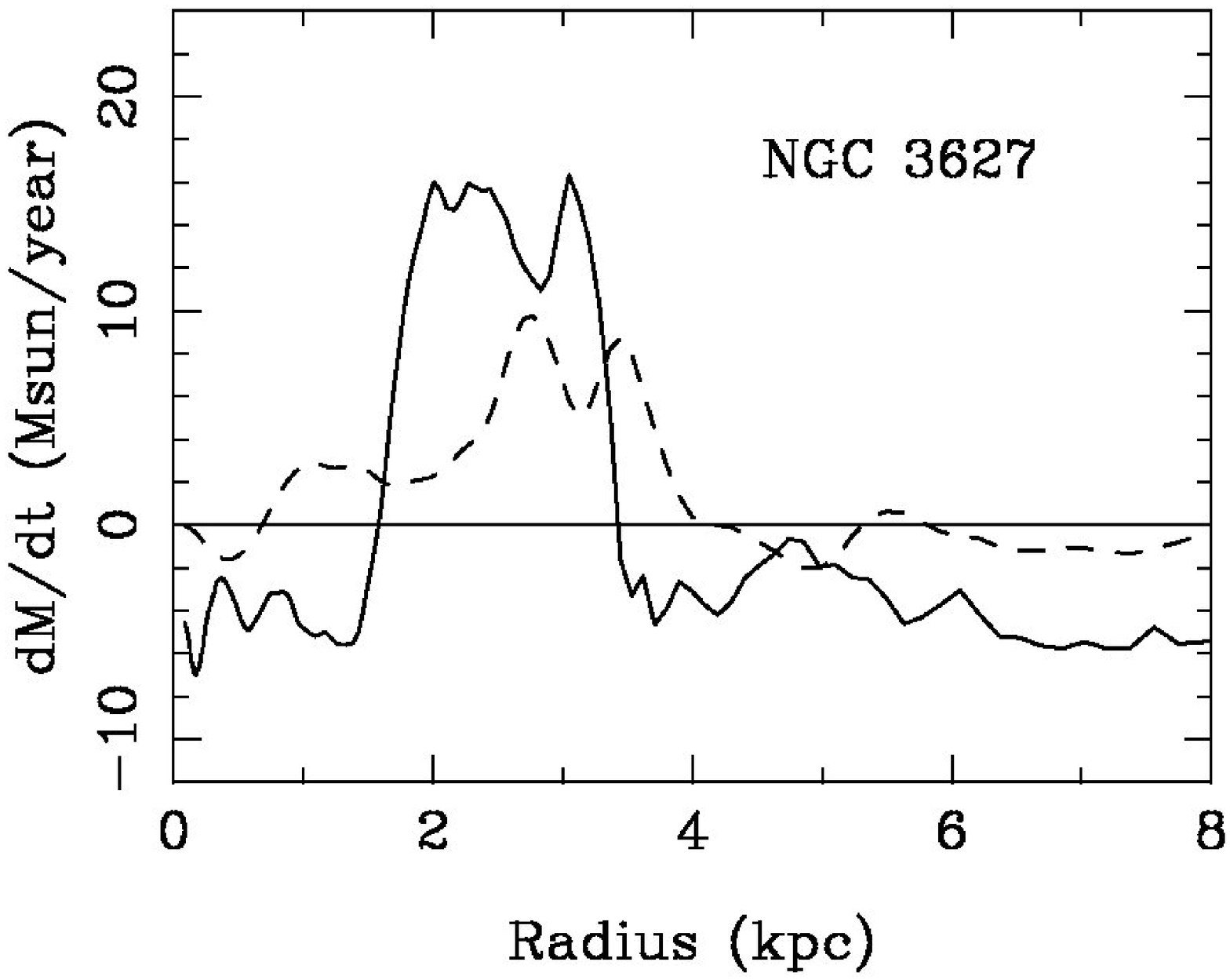}
\includegraphics{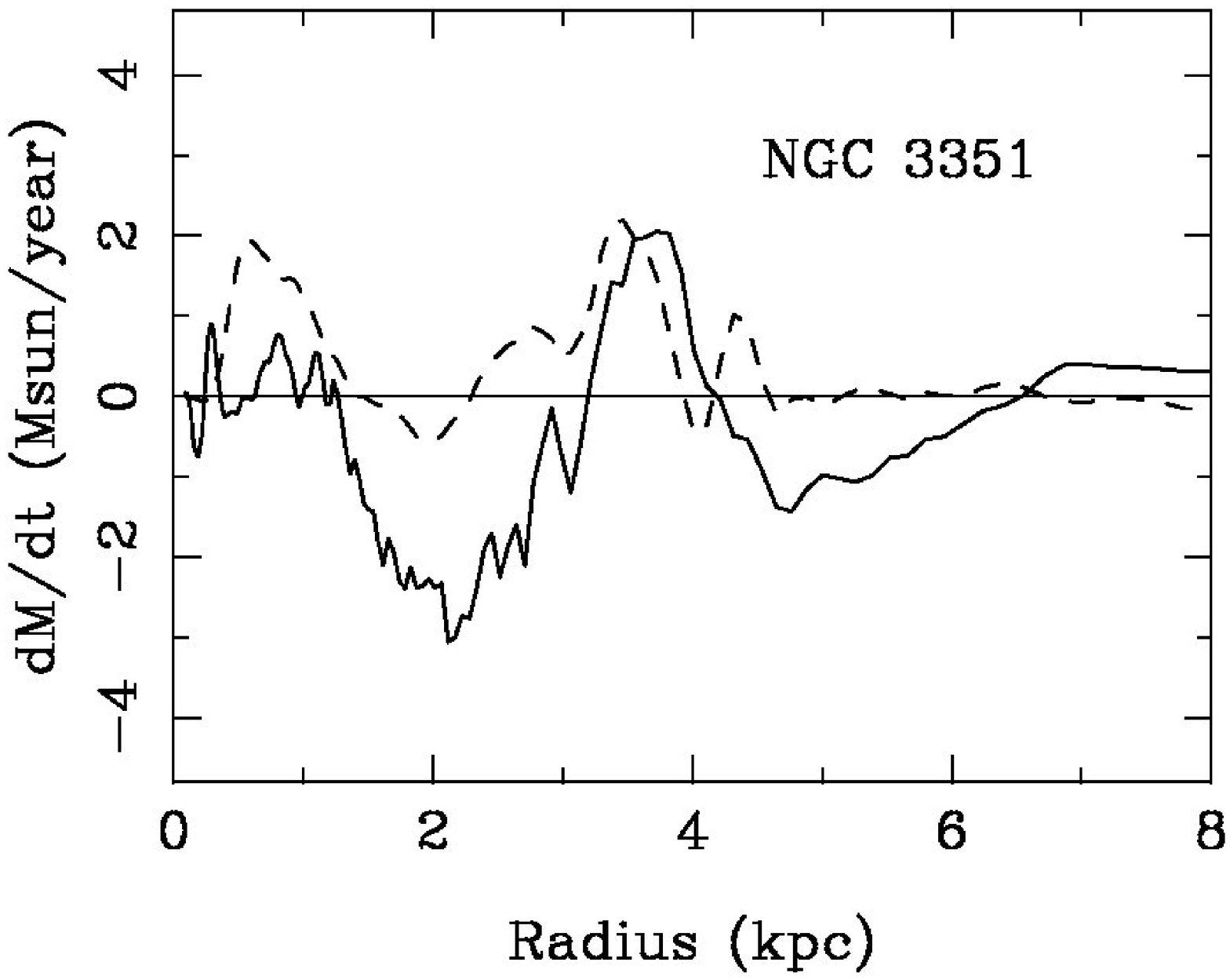}
\includegraphics{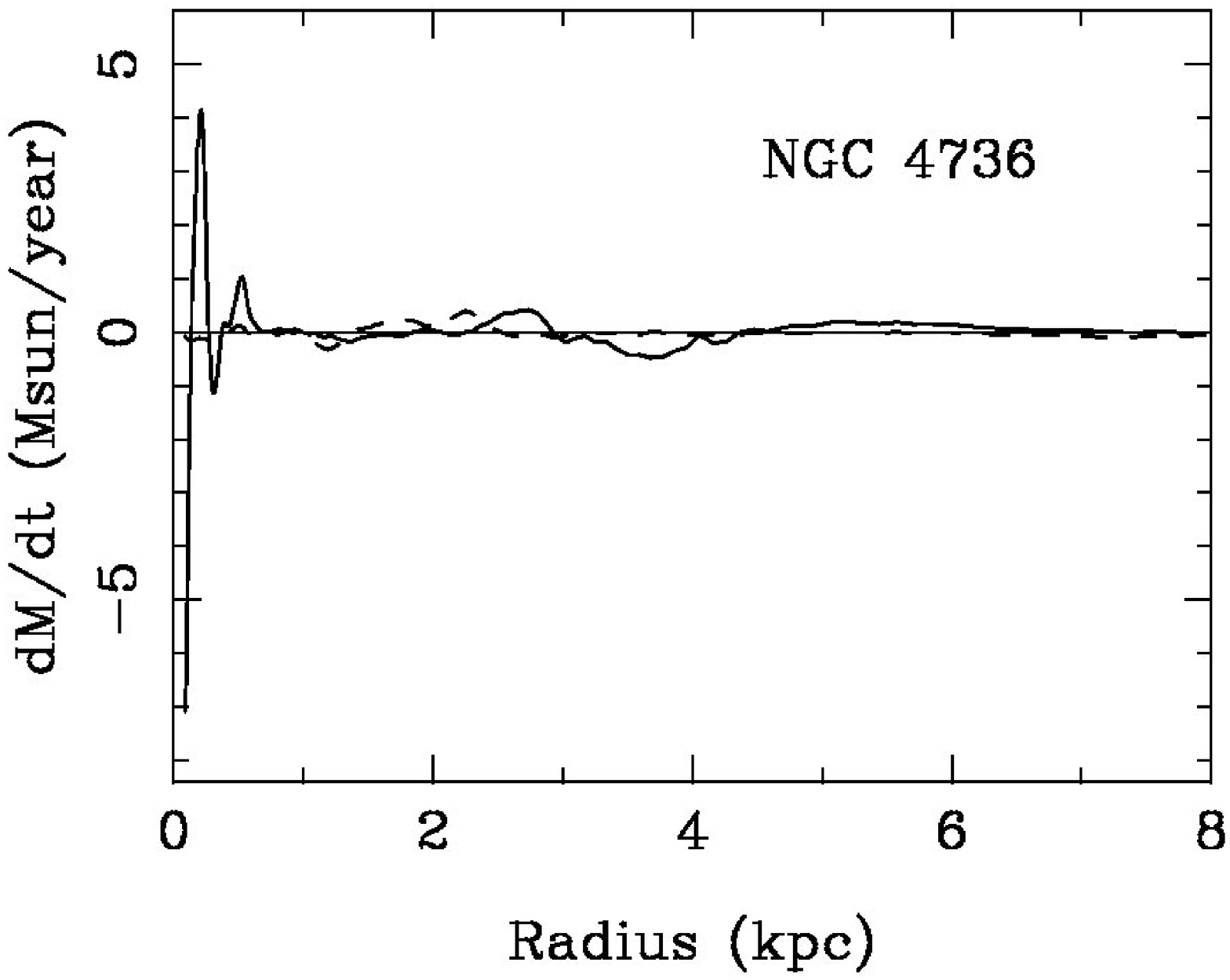}
\caption{Stellar and gaseous mass flow rates for the six sample galaxies,
calculated from the IRAC 3.6 $\mu$m for NGC 0628,
an average of IRAC 3.6 $\mu$m and SDSS i-band for NGC 4321, 5194, 3627,
3351, and SDSS i-band for NGC 4736, plus VIVA, THINGS and BIMA SONG data.
The total potentials used for these
calculations were the same as previously derived using IRAC
and/or SDSS for the stellar contributions, with appropriate averaging,
plus the gas contributions.}
\label{fg:Fig19}
\end{figure}

\begin{figure}
\vspace{430pt}
\includegraphics{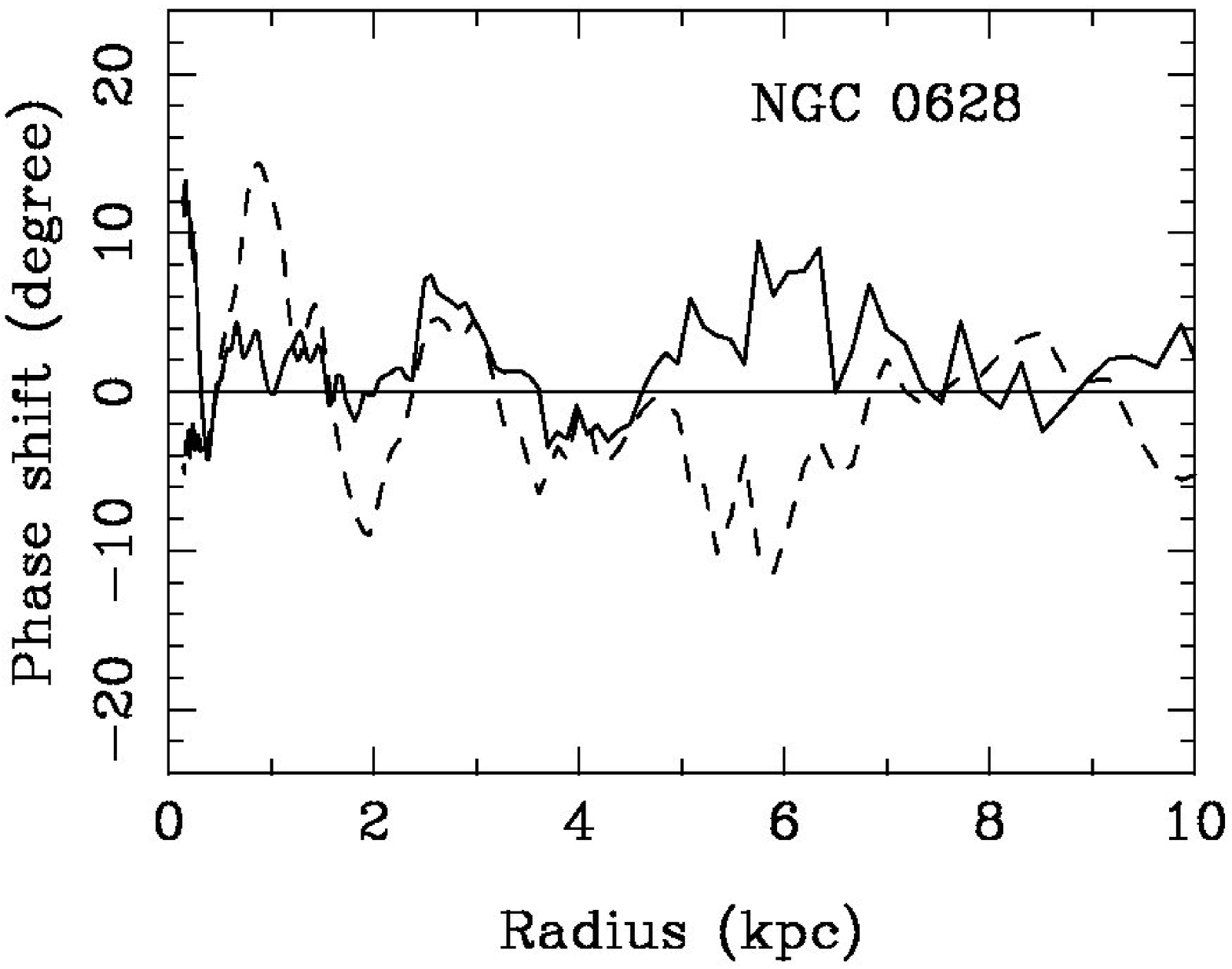}
\includegraphics{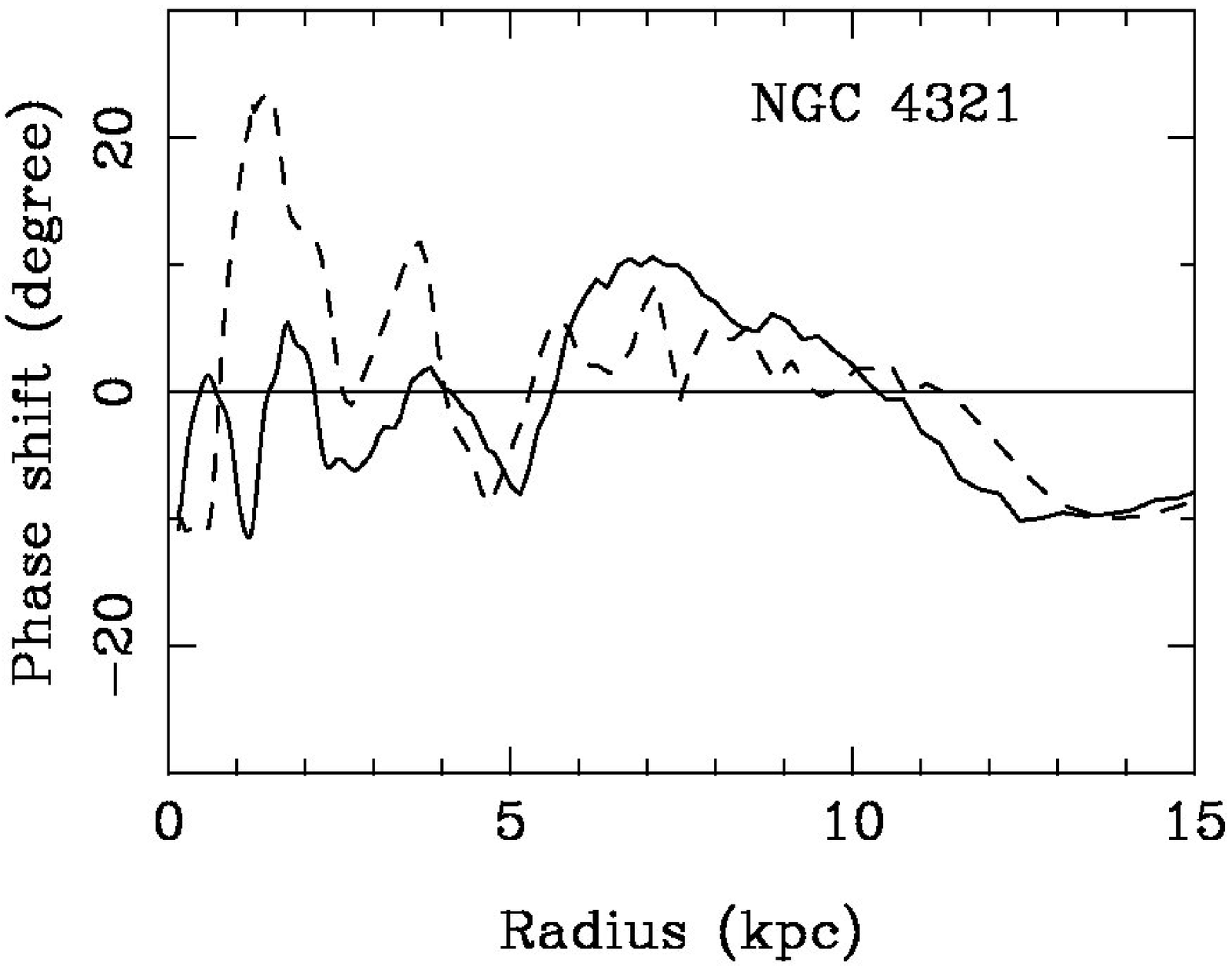}
\includegraphics{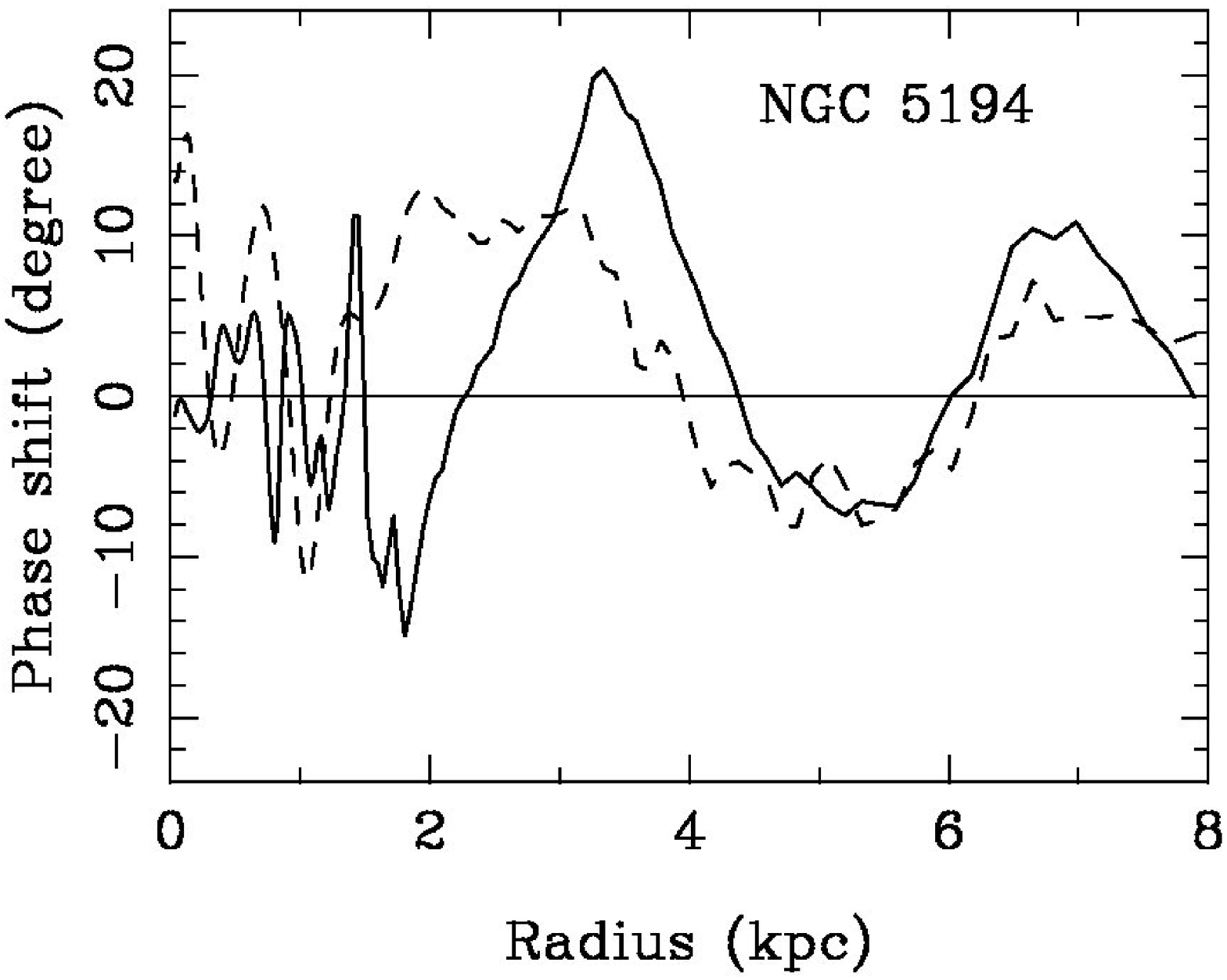}
\includegraphics{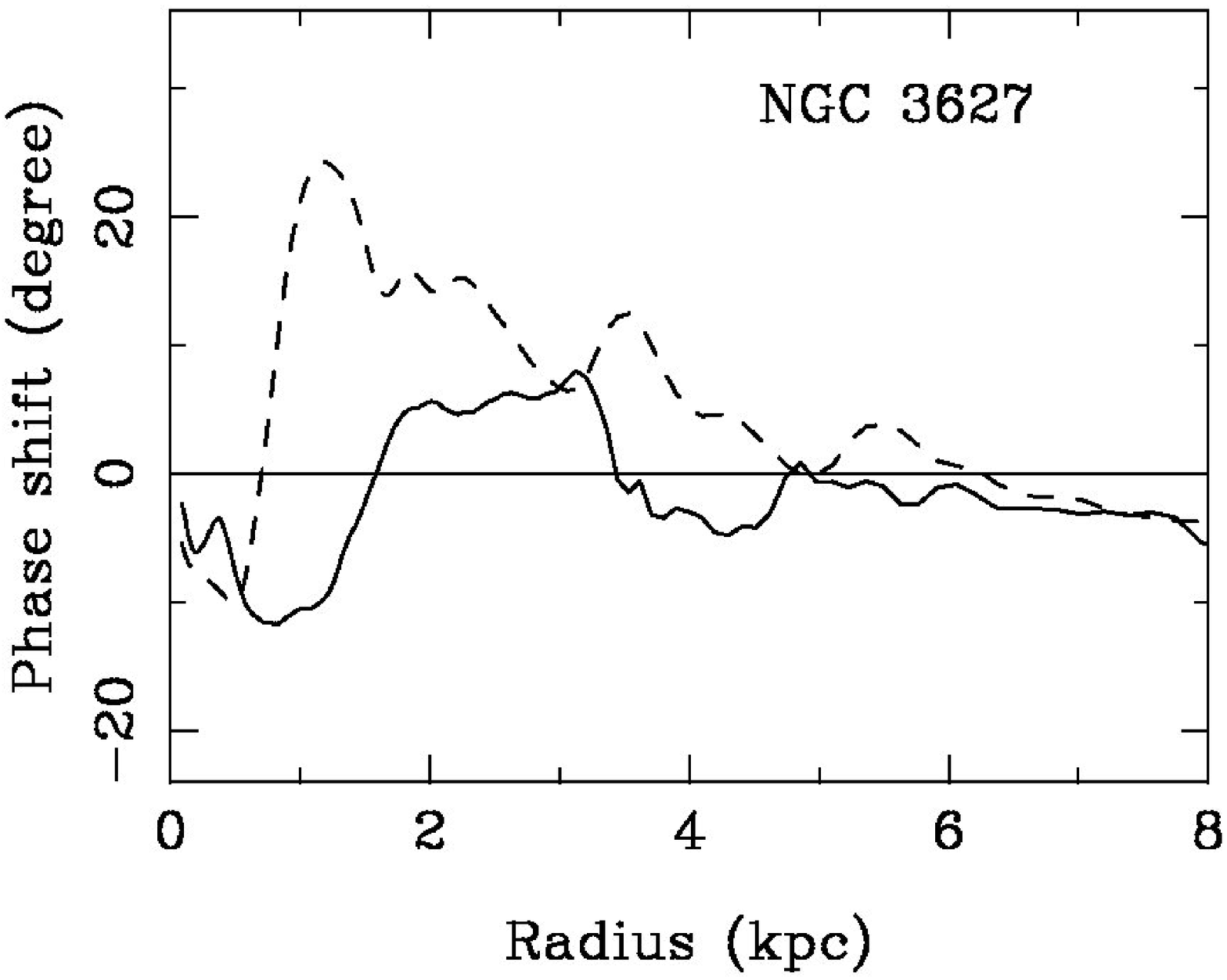}
\includegraphics{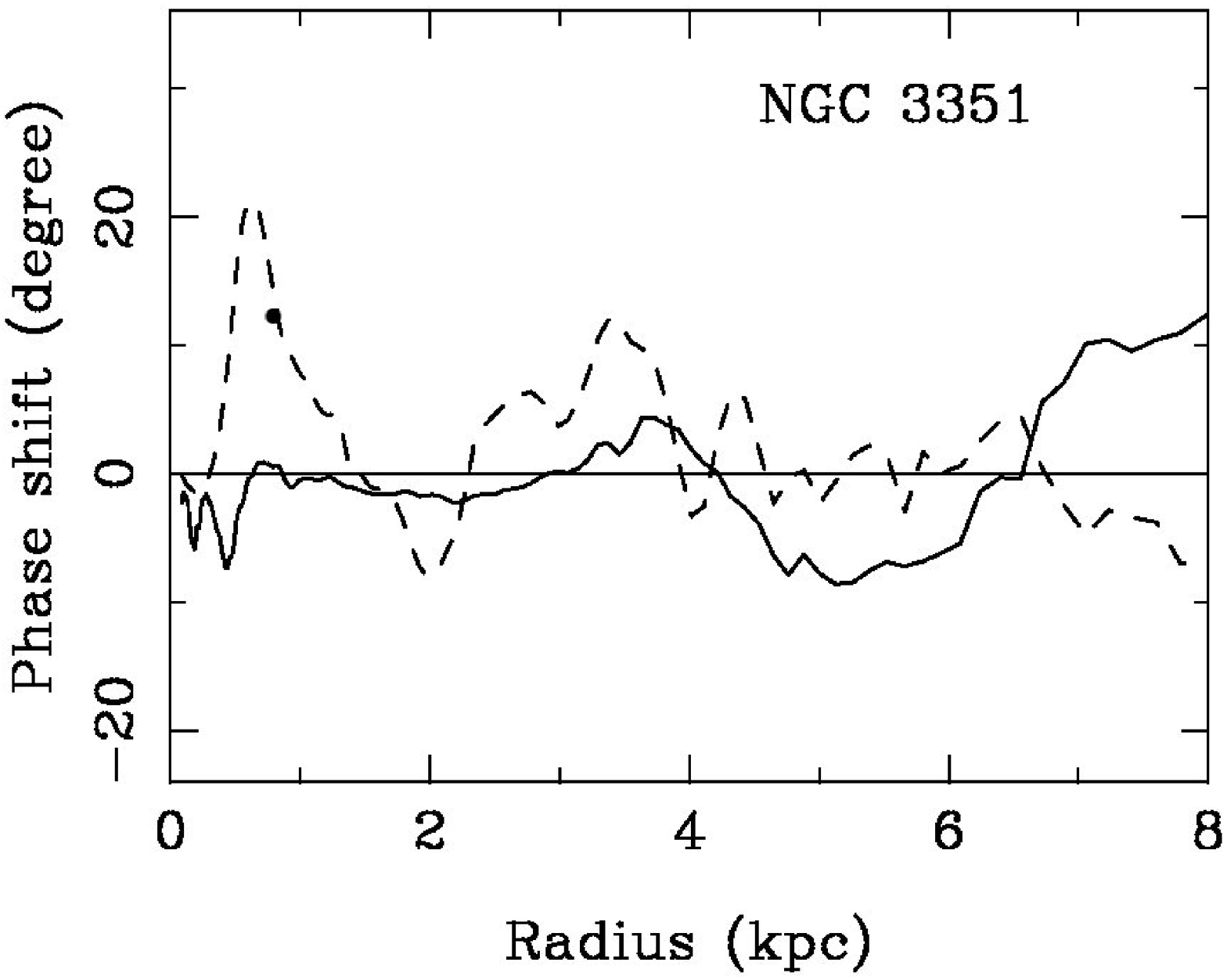}
\includegraphics{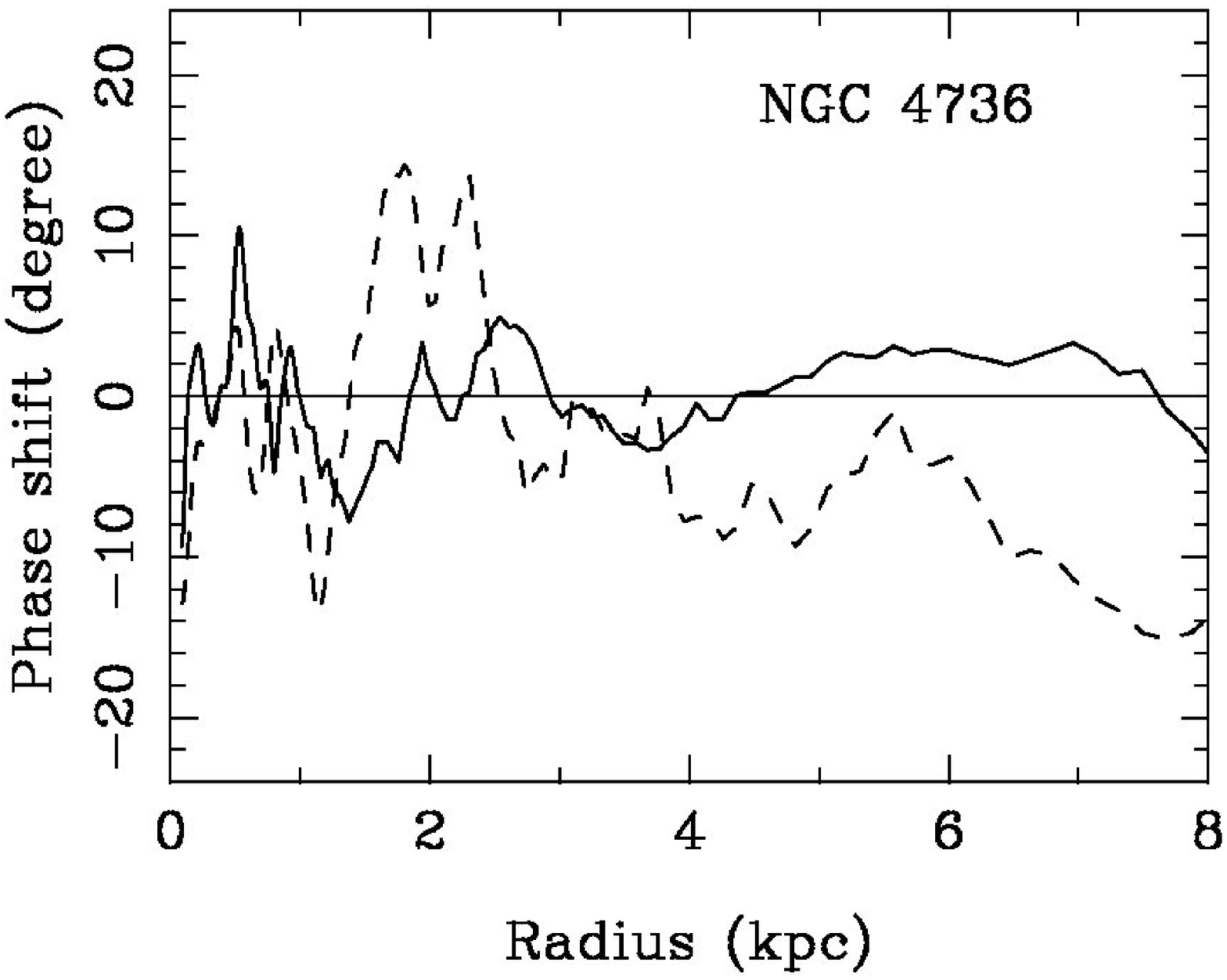}
\caption{Stellar and gaseous phase shift with respect to total
potential for the six galaxies. {\it Solid lines:} Stellar phase shifts.
{\it Dashed lines:} Gaseous phase shifts.
The stellar maps were derived using the IRAC 3.6 $\mu$m data
for NGC 0628, using an average of IRAC and SDSS data for NGC 4321, 3351, 3627,
5194, and using SDSS i-band data for NGC 4637. 
The maps were from VIVA, THINGS and BIMA SONG observations.
The total potentials used for these
calculations were the same as previously derived using IRAC
and/or SDSS for the stellar contributions, with approproate averaging,
plus the gas contributions.}
\label{fg:Fig20}
\end{figure}

In Figure~\ref{fg:Fig19} and \ref{fg:Fig20}, we
present the comparison of stellar and gaseous (HI plus H$_2$)
mass flow rates, as well as their respective phase shifts with respect to the
total (star plus gas) potential. The phase shift plots tell us
the relative efficiency of the stellar and gaseous mass accretion
processes, since (from Z98, equation 25)

\begin{equation}
{\overline{ {{dL} \over {dt}}}} (R)
= {1 \over 2} F^2 v_c^2 \tan i \sin(m \phi_0) \Sigma_0
\label{eq:e}
,
\end{equation}
where $F^2 \equiv F_{\Sigma} F_{\cal{V}}$ is the product
of the fractional density and potential wave amplitudes,
and m is the number of spiral arms (usually taken to be 2).
Therefore, from equation (\ref{eq:eq4}), (\ref{eq:eq5}) in the current paper,
\begin{equation}
\sin(m \phi_0)_i
={{1} \over {\pi v_c R F_i^2}}{ {{dM_i } \over {dt}} \over {(\Sigma_0)_i}}
\end{equation}
where the subscript i represent stars or gas, respectively.
So for similar wave amplitudes between stars and gas, the mass component
that has higher phase shift will have higher mass flow rate per
unit surface density.

For galaxy NGC 0628,
the stellar and gaseous accretion efficiencies are similar,
as are the respective total mass accretion rates, since
this is a late-type spiral galaxy and is gas rich.  The alignment
of the stellar and gaseous phase shifts are not consistent,
expecially for the outer disk, indicating that the galaxy is 
yet to evolve into a state of dynamical 
equilibrium.

NGC 4321 is relatively quiescent and of intermediate Hubble type, 
From Figure~\ref{fg:Fig20}, we see that for much of the central region
(except the very center) the gaseous phase shift with respect 
to their common potential is much larger than the stellar phase shift,
indicating that gas leads in phase in this region compared to stars, 
revealing the higher dissipation rate and thus mass redistribution 
efficient of the gas compared to stars for the central region of this galaxy.
The values of phase shifts of stars and gas are comparable for
the outer region indicating that the two mass components have 
similar mass-redistribution efficiency there.
The shapes of the positive and negative humps of phase shifts for stars
and gas have more similar radial distributions for this galaxy
compared to NGC 0628, especially for the outer region,
indicating a higher degree of dynamical equilibrium.
The overall contribution of the stars to mass redistribution,
however, is much higher for stars than for gas in the outer region
(Figure~\ref{fg:Fig19}),
because of the higher overall stellar surface density there .

Next the stellar and gaseous mass flow rates and phase shifts
for NGC 5194 (M51) are presented. For this galaxy, 
unlike for NGC 4321, the stellar and gaseous phase shifts 
have significantly different radial distribution, even though
both galaxies were of intermediate Hubble type.
This indicates the non-dynamical-equilibrium state of M51
due to the tidal pull of the companion,
and the inevitable evolution towards forming a new set of
nested resonances, with the gas playing a leading role in
seeking the new dynamical equilibrium because of its
more dissipative nature, and the stellar component lagging
somewhat behind in this action. But at every moment of this
re-establishment of the dynamical equilibrium the overall
density (i.e. the sum total of stellar and gaseous) still
has a much more coherent phase shift distribution with respect
to the total potential than each component considered separately
(i.e. compared to Figure~\ref{fg:Fig8}).  The overall mass flow rate of stars is
much higher than gas for this galaxy.

Next the stellar and gaseous mass flow rates and phase shifts
for NGC 3627 are presented. For this galaxy,
even though the phase shift in the central region shows
that gas has a higher accretion efficiency than stars,
the overall accretion rate of stars much exceeds those of gas.
The second CR location is not shown here because of more
limited radial range plotted, but is present in the total
gas phase shift curve when examined further outward.

For the earlier-type galaxy NGC 3351, we
see that the central region gas-star relative
phase shift for this galaxy is even more severe than for NGC 4321.
This is likely due to the fact that the straight bar potential
in the central region of this galaxy is mostly contributed by stars,
which has small phase shift with respect to the total potential,
and gas thus contributes a much larger phase shift (through its
dissipation in the bar potential and the phase offset of its
density peak from the stellar density peak) to
the overall potential-density phase shift. The overall
mass flow rates are contributed similarly by stars and gas for this galaxy.
Note that the somewhat chaotic appearance of phase shifts in the outer region
of this galaxy is due to the low surface density there,
and thus noise begins to dominate.

For NGC 4736, gas leads stars in part of the radial range, but the
overall contribution to mass flow is mostly due to stars, especially
for the central region,
due to the fact that in this early-type galaxy the stellar
surface density much exceeds that of gas.
Also once again the low surface density in the outer disk
of this early type galaxy leads to the more chaotic phase shift
distribution there.

We see from this set of plots that some of the old myths, such as that 
the gas always torques stars inside and outside CR in the right
sense, and from the change of sign of the relative phase between
stellar and gaseous density distributions one can tell the CR location, is only
true in small number of instances.  To obtain a reliable CR
estimate, one really needs to use the total density (star plus gas)
and total potential, and calculation of the phase shift zero crossings
between these two components to determine the CR locations.  In the
absence of the gas surface densities, stellar surface density alone
and the potential calculated from it (as is done in ZB07), come as
the next best compromise.  Both of these approaches are much more reliable than
if using the phase shift between the stellar density and gas density
distributions.  We also see that in the majority of the galaxies
in the local universe, secular mass flow is dominated by the
stellar mass redistribution rather than by gas redistribution,
unlike what has been emphasized by many of the earlier works
on secular evolution.

We comment here that even for the gas accretion in disk galaxies,
the mechanism responsible for its viscosity is still the
collective gravitational instabilities (which manifests as the
phase shift between the gas density and total potential), rather 
than the microscopic gaseous viscosity, which was long known to
be inadequate both for the accretion phenomenon needed
to form young stars, and for the accretion phenomenon
observed in the gaseous disks of galaxies -- thus the
well-known need for anomalous viscosity in generalized accretion disks
(Lin \& Pringle 1987).
Z99 showed that the large-scale density wave-induced
gravitational viscosity is likely to be the source of anomalous
viscosity in both the stellar and gaseous viscous accretion disks 
of galactic and stellar types.

\subsection{The Relative Contributions of Atomic and Molecular Mass Flows}

\begin{figure}
\vspace{430pt}
\includegraphics{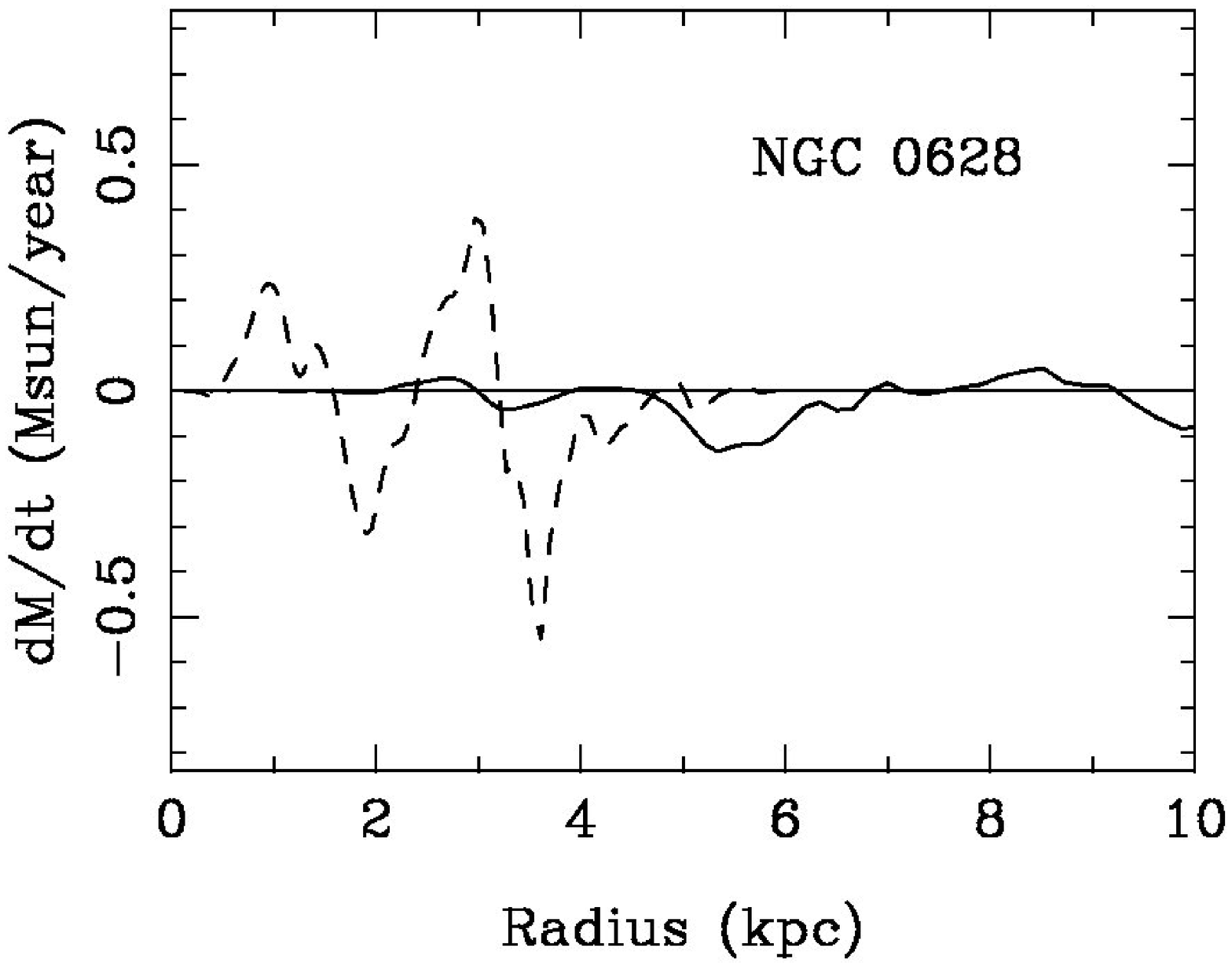}
\includegraphics{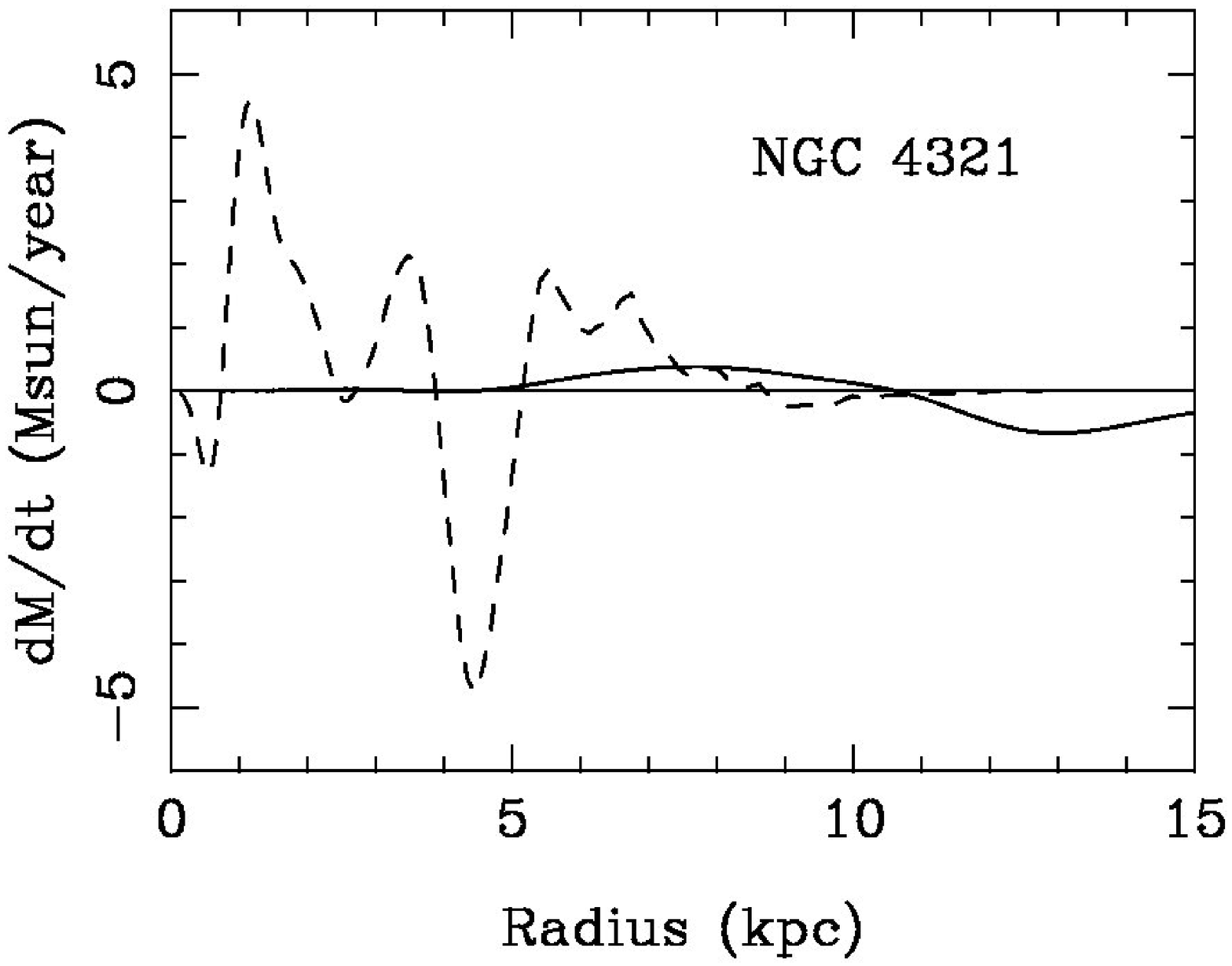}
\includegraphics{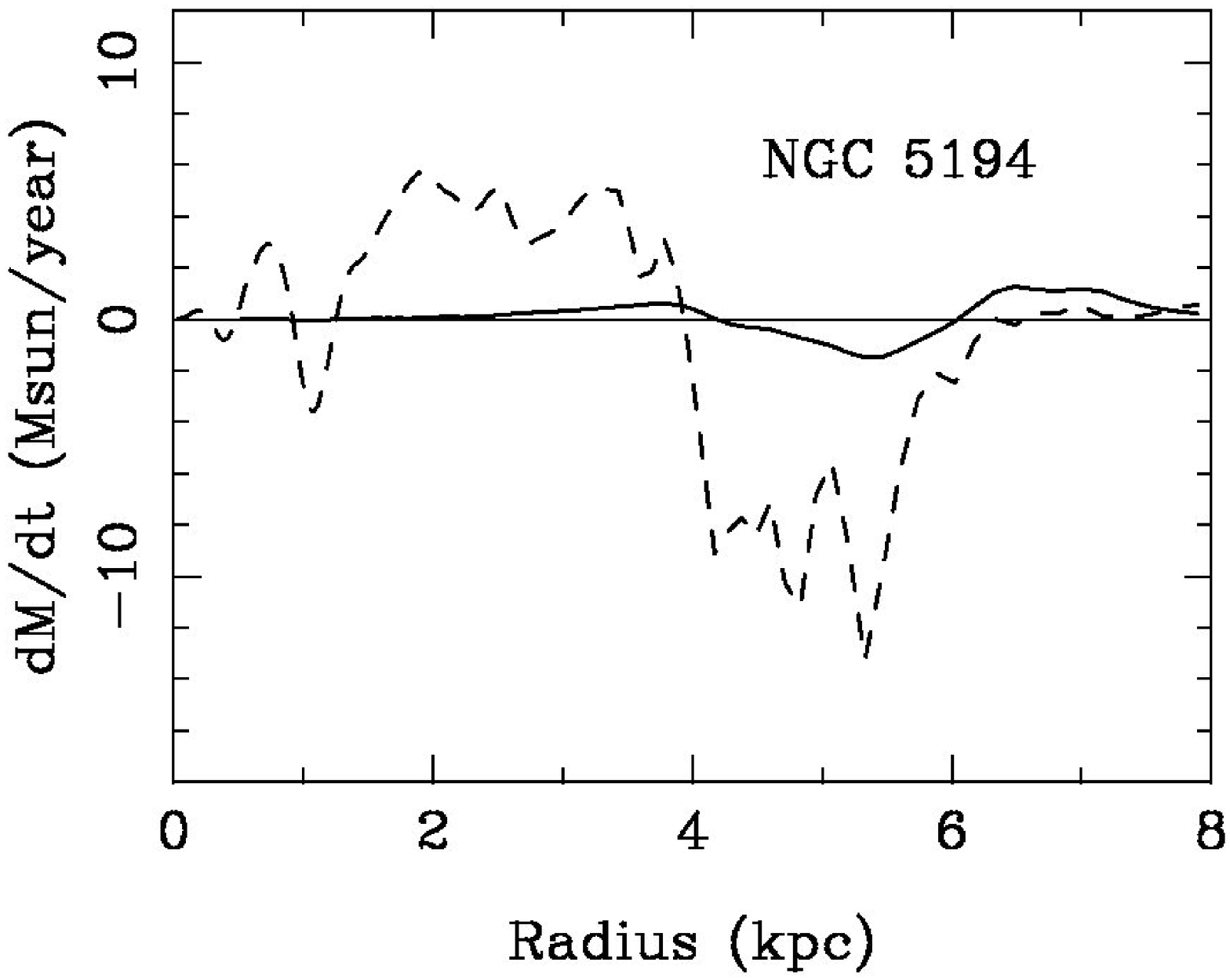}
\includegraphics{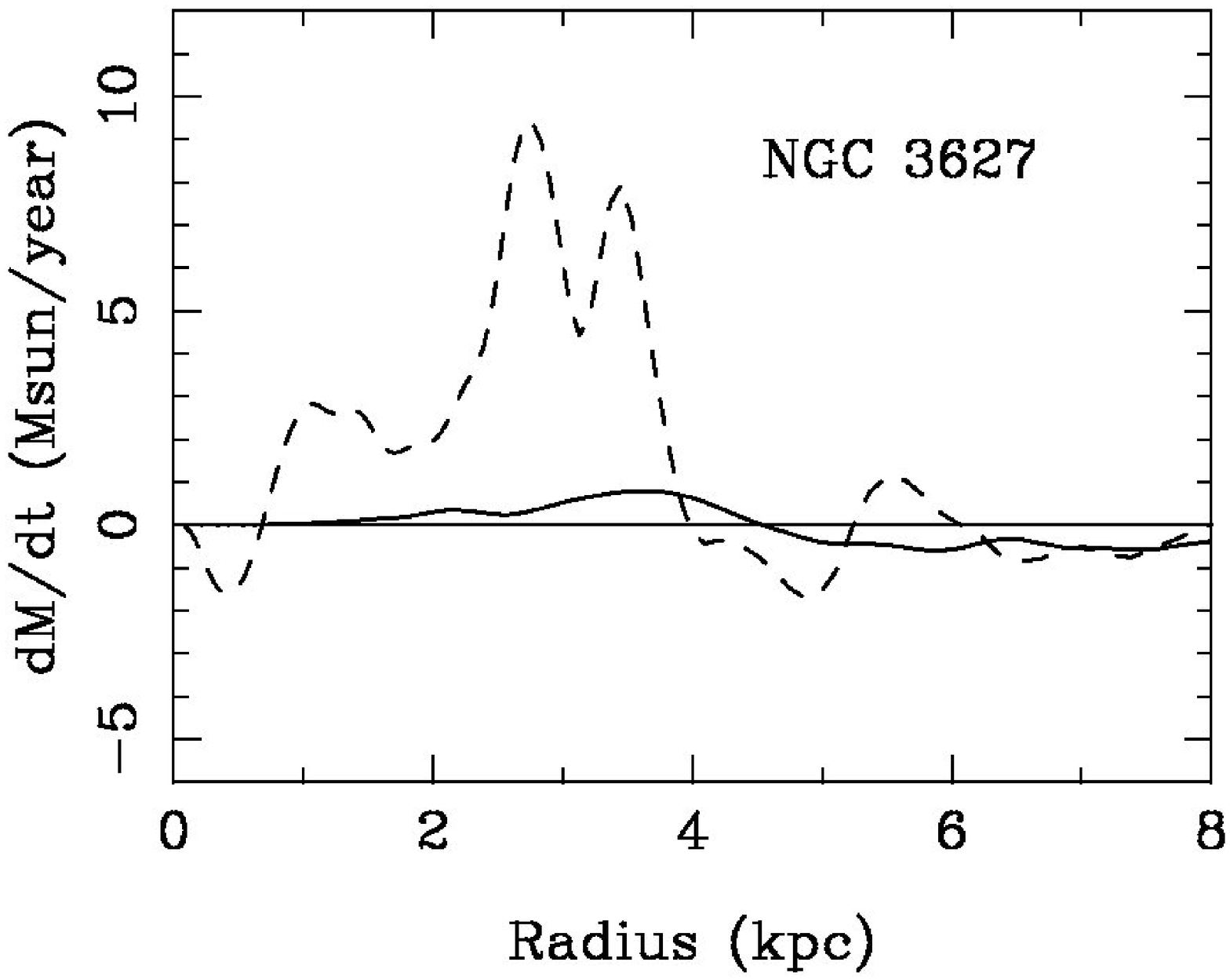}
\includegraphics{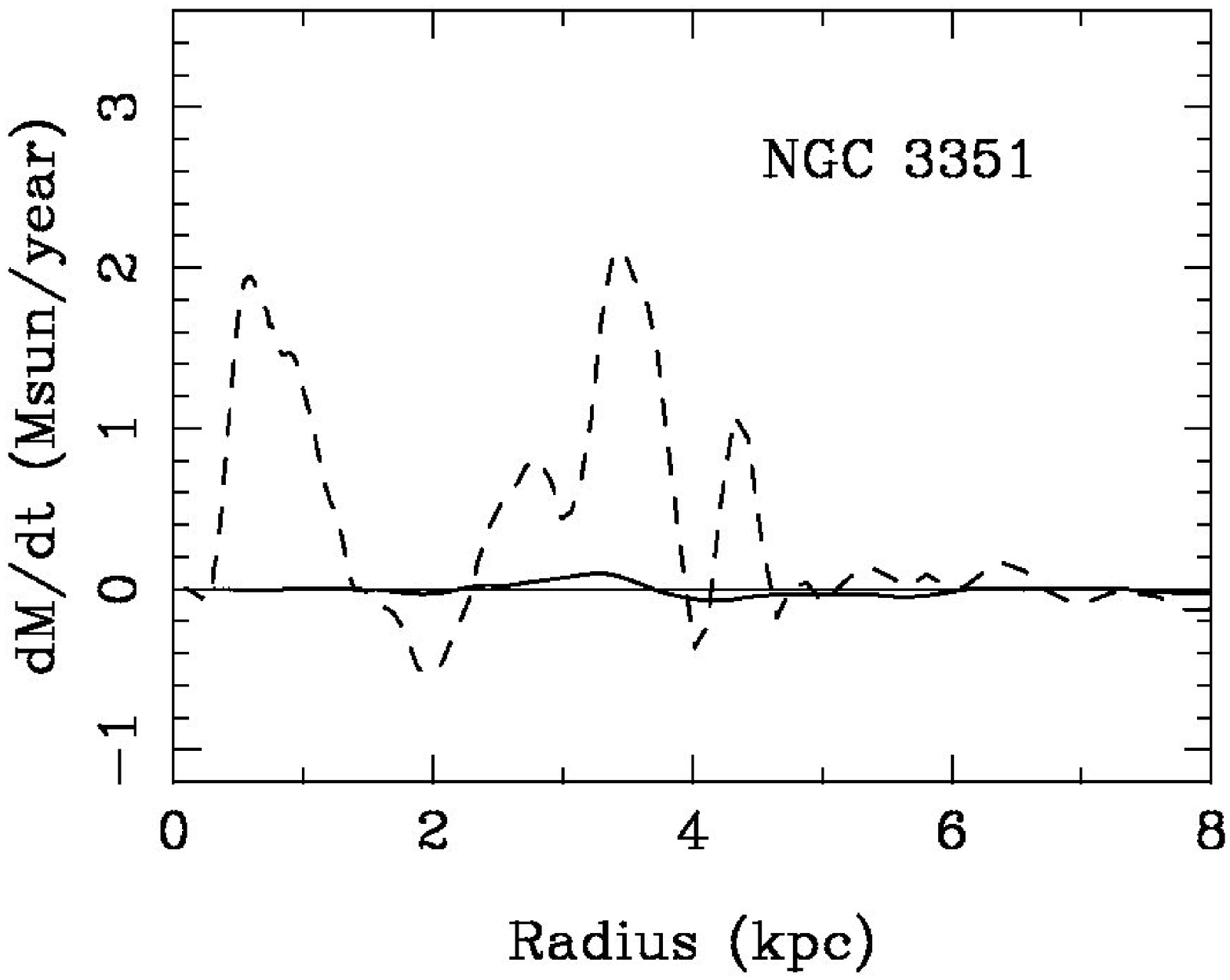}
\includegraphics{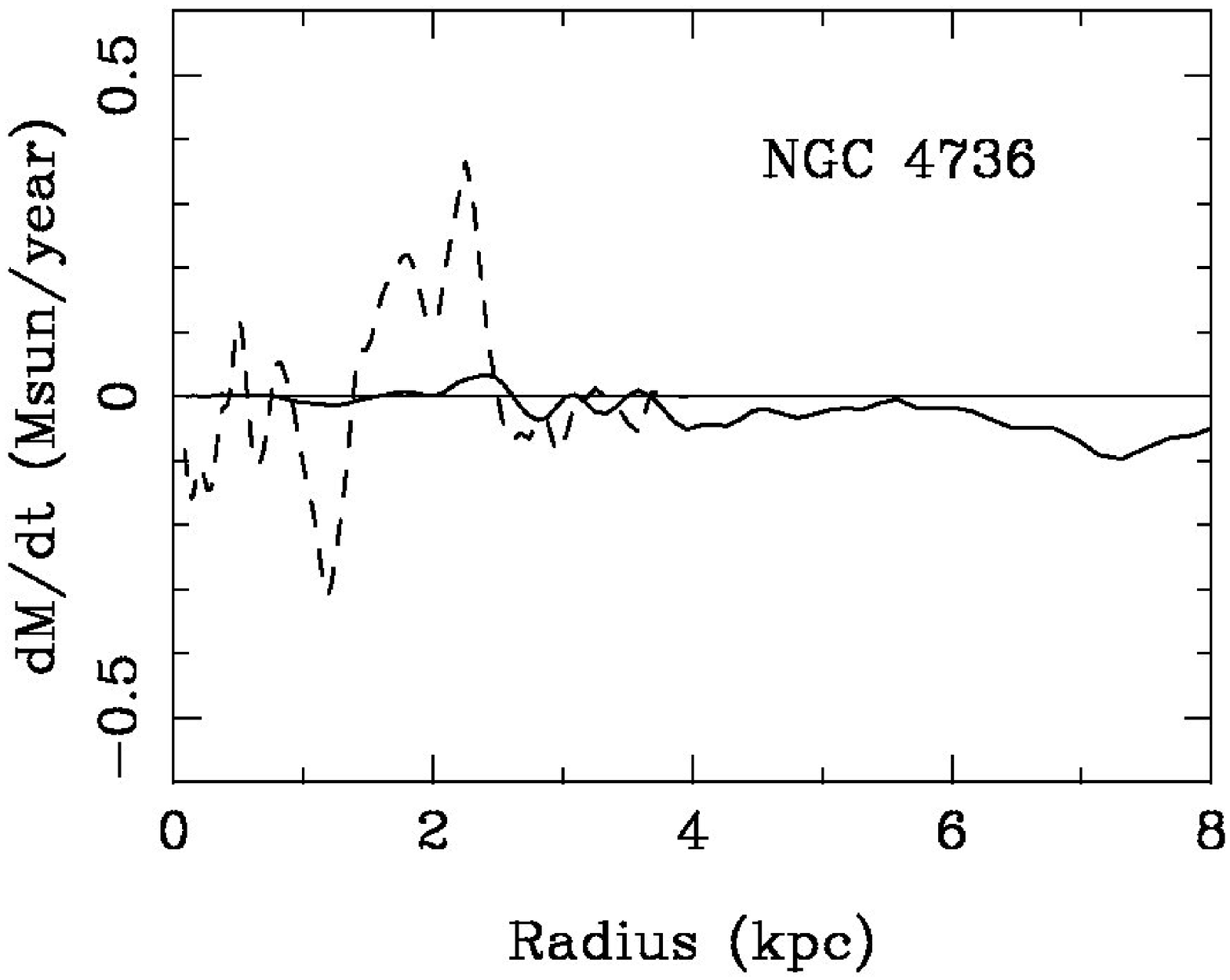}
\caption{HI and $H_2$ mass flow rates for the six sample galaxies. 
{\it Solid lines:} HI mass flow rates. 
{\it Dashed lines:} H$_2$ mass flow rates.
The HI mass flow rates were derived using mass maps from VIVA and THINGS
observations, and H$_2$ flow rates were derived using mass maps
from BIMA SONG observations. 
The total potentials used for these
calculations were the same as previously derived using IRAC
and/or SDSS for the stellar contributions, with approproate averaging,
plus the gas contributions.}
\label{fg:Fig21}
\end{figure}

In Figures~\ref{fg:Fig21},
we present the comparison of $HI$ and $H_2$ mass flow rates for our
six galaxies.  It is seen that for all five galaxies the $H_2$
mass flow follows more closely the stellar mass flow distributions,
whereas the HI has a more smooth and gradual distribution.  This
is likely to be a result of the process of the formation of molecular
gas (and molecular clouds and complexes as well)
at the spiral arms due to the density wave shock, and their subsequent
dissociation.  The HI gas distribution, on the other hand, follows a more
quiescent dynamics, even though a mild correspondence to the density
wave patterns can be discerned.
We emphasize that these plots should not be read as that
the HI gas is less responsive to the gravitational perturbation of
the density wave than H$_2$, but rather that when the response of
the gas happens in the spiral shock, HI will be converted to H$_2$ and
thus will show up as $H_2$ in the response.

\subsection{Implications on the Origin of Density Wave Patterns
in Galaxies}

Our current work has another important implication on the
continuing debate regarding the modal-versus-transient nature
of density wave patterns in disk galaxies (see, e.g., BT08 
and the references therein). A recent
contribution to this debate is Sellwood (2011), 
where N-body simulation results
of spiral patterns based on the linear modal theory of
 Bertin et al. (1989), as well as on previous modal 
simulations by Donner \& Thomasson (1994, hereafter DT94) 
and by Z96, Z98 are given.  Sellwood (2011) concluded that 
the spiral patterns in his simulations are all transient 
waves rather than modes.  While the cause for disagreement with
the Bertin et al. theory may be due to the simplifying assumptions
made in that theory, which are up to these authors to respond,
in what follows we present our own response specific to 
Sellwood's comments regarding his repeat simulations of the DT94 and Z96,Z98
results (some of these text were extracted from earlier 
email exchanges between the first author and J. Sellwood).

\begin{itemize}
\item There is a significant distinction between the repeat simulations
Sellwood conducted using the basic state specifications of
DT94 and Z96, Z98, and those simulations he had done in the past 
for genuine transient spirals (see, e.g., Sellwood 2008 and 
the references therein).
Almost all of his past simulations of transient spirals were
carried out using basic states that were {\em over-stable} to the
formation of spiral modes, and therefore the spiral patterns
emerged in those disks were indeed transient wave trains.  For
barred galaxy simulations, such as in Sparke \& Sellwood (1987),
{\em unstable} basic states were on the other hand used, and these
bar patterns formed were genuine bar modes.  BZ09
had commented on the inconsistency of Sellwood's past practice
in this regard, and pointed out that in realistic galaxies,
such as the more than 150 OSUBGS galaxies analyzed by us
in the same paper, bars and spirals often co-exist in the same
galaxy as nested patterns, it is not possible to specify the portion of
the disk for bars as unstable to modal formation, and
specify the portion of the disk for spirals as overstable for modal formation.
In his 2011 paper, when Sellwood repeated the older
simulations of DT94 and Z96, Z98, he had on the other hand used our 
basic state specification that is unstable for spiral mode formation --
i.e. this is the first time Sellwood has simulated a galaxy
disk that allowed the formation of unstable {\em spiral modes}.  
\item In Sellwood (2011)'s new simulations, he was able to reproduce 
the gross morphological features in our previous simulations.  There is 
little question that the numerical codes we each used (in DT94, Z96, Z98, 
and Sellwood 2011) behaved similarly, apart from the larger number of 
particles used in the new simulations.  The mere fact that the derived
modal characteristics are nearly identical in both the old and the new simulations,
shows that we indeed are all simulating {\bf modes}, not transient waves.  This
is like someone plucking a guitar string: if the string length is held the
same, no matter how much difference there is in the applied force, the
fundamental mode excited has to be the same (i.e. the guitar will sound
with the same basic tone).  The same thing is true for disk galaxies:
despite the unavoidable differences in the noise and random-variable 
performances of our various codes, the modal characteristics
stay the same since the basic state characteristics (i.e. the 
radial distributions of surface density, random velocity and the 
rotation curve) are specified to be the same.  
\item Regarding the generating mechanisms of galactic density wave patterns, 
our view is that these are modes which are the fundamental 
instabilities of the basic state of the disk, whereas Sellwood thinks that 
they are transient waves regenerated on short time scales
(Note that even Sellwood himself, during our private communication, 
had acknowledged that the mere fact that the power spectrum of the 
dominant spiral structure is coherent and spans much of the galactic radii
means that its generating mechanism cannot be totally random.  Yet he
did not offer an alternative mechanism to the modal picture). 
In all other naturally occurring resonant systems, such
as atmospheric convection (Benard instability), microwave resonant cavities, 
etc., unstable modes inevitably emerge if the boundary conditions allow them.  
In galaxies, the boundary condition naturally favors the appearance of modes: 
i.e., the disk would reach a marginally stable state for axisymmetric 
instabilities due to the self-regulating balance of gravitational and 
pressure forces.  Then, when the disk is marginally stable to axisymmetric 
instabilities, it is unstable for bi-symmetric instabilities 
(spiral and bar modes).  So for disk galaxies to spontaneously form 
these density wave modes is almost an unavoidable outcome.  That is why 
these patterns appear to be so predominant in disk systems.
\item Sellwood (2011) has pointed out that besides the 
dominant mode in the inner disk, which has its initial CR at 30 and 
OLR at 42 in Z96 and DT94's simulations, and which is 1/10 of these 
values in Sellwood's new simulation because of the different normalizations, 
there are spurious modes present at the same time. We agree with this
conclusion of his.  In fact, Figure 5 of DT94 
showed other power spectrum peaks besides the main one (XZ has independently
confirmed this but did not present these results in the published papers), 
similar to what Sellwood has shown in his Figure 8 (though Sellwood identified
a total of four modes, i.e., three spurious modes -- a difference that is
likely to be due to the difference in the detailed realization of our codes).  
However, these spurious power spectrum peaks (in the old as well as the new
simulations) are all located in the outer region of the disk, where the surface 
density is extremely low, and the outer disk was not the main focus of 
our previous studies.  Nonetheless, Figure 4 of Z98 (reproduced here
as the top plot in Figure 1 of the current paper) did indeed show the 
presence of the first spurious edge mode as the second small
bell hump.  Z98 has limited the
plotting region in that figure to R=70, therefore the third peak in 
DT94's Figure 5 in the extreme outer region of the disk does not show up.
Due to the partial inter-penetration of the dominant and the spurious modes, 
beating and fluctuation of modal amplitudes do happen as Sellwood (2011) had
commented.  This had shown up in both Z98 and DT94's simulations 
(for example see Figure 2 in Z98 paper, or Figure 3 in DT94).  
So the recent simulations of Sellwood had not shown any
qualitatively different results from our previous studies.
Furthermore, the values of the power spectrum intensity in 
both DT94 and Sellwood (2011) was normalized to the surface density
at each radius, and thus the outer peaks have much exaggerated
amplitudes relative to the inner one, due to the low surface
density in the outer disk, as Sellwood himself had also acknowledged in the
same paper.  In actuality the absolute amplitudes of these outer modes 
in the relevant simulations are very
small in comparison with the absolute amplitude of the inner mode,
which is precisely why the dominant spiral feature 
remains coherence for many pattern rotation periods in the Z98 simulation,
with the spiral pattern lasting a significant fraction of a Hubble time.
In physical galaxies (rather than this particular set of N-body
simulations), the phenemenon of nested modes are of course common
(see, e.g., ZB07 and BZ09).  These modes generally have different
pattern speeds, and these modes also emerge and transform in morphology
as the basic state mass distribution evolves.  So for a single basic
state distribution there should in general be a compatible ``modal system set''
rather than a single mode, but such nested-modal-set is not
the same as the random and transient wave trains that Sellwood is advocating.
\item Regarding the dominant mode in that particular
simulation (which has its OLR at 42 in our units, or 4.2 in Sellwood's
units), Sellwood suggested that even this inner mode is itself composed of 
different appearing and disappearing transient waves. He argues for this
from the vertical spread of the power spectrum in the last frame 
(sum of time history)of Figure 8 in Sellwood (2011).  A hint about 
the reason of this spread, however, can be found from the two frames 
above the last frame: there one sees that the width of the power spectrum 
is a lot narrower after one divides the time history into two segments.  
This shows, that the real 
cause of the spread of power spectrum in the summed-history plot is that 
the modal characteristics are slowly changing in time as a result 
of the secular evolution of the basic state.  This phenomenon can be made 
analogous to the growth of a human being from infancy to adulthood to old age.  
At each age-range the physical characteristics will be different.  If one puts 
a stack of pictures of the individual at the different ages side by side 
(similar to Sellwood's summed spectrum), one will see a spread of appearances.  
But the underlying person is the same one nonetheless, he/she is just 
going through the different evolution stages dictated by the programming 
of the genes (similar to the fundamental dynamical laws in galaxies 
plus the initial condition of galaxy formation), as well as by the 
environment he/she is brought up in (similar to the boundary conditions 
a galaxy will be experiencing in its lifetime).  In the case of a person 
we do not say that he/she is constantly dying in the old self and 
constantly being regenerated in the new self (except as a way of speech),
but rather a single person evolves from young to old (in the case of
galactic density wave modes, this evolution includes the possibility
of one mode evolving into nested multiple modes -- but that process
happens on longer timescale than what we are concerned about in
establishing the quasi-steady nature of the spiral mode -- for which
we require the mode to be steady on local dynamical timescale).
The analogy between the growth of a person and the evolution of galaxies 
in fact goes beyond superficial: Both the galaxies and the biological 
entities can be shown to be instances of the "dissipative structures" 
of I. Prigogine.
\item Another piece of supporting evidence for the modal nature of galaxies 
is our recent work (ZB07; BZ09) 
on applying the potential-density phase shift method to determining 
the corotation radii in spiral and barred galaxies.  For the phase shift 
method to be effective, the modes in question must have reached 
approximate quasi-steady state: otherwise the Poisson equation alone 
should not be able to be used to forecast kinematics properties.  We
found from analyzing the data for more than 150 nearby galaxies that 
the majority of them show good correspondence between the CRs 
determined by this method and the resonant features in galaxy images.  
The CRs determined by the phase shift method also agree well with that determined 
by TW method when there is only one primary CR in a single galaxy (as 
the TW method is most reliable in such instances), as well as with CRs determined 
by stellar isochrone methods of Martinez-Garcia et al. (2009, 2011).  
\item Regarding the dynamical mechanisms for secular evolution, Sellwood
has made the claim that transient spirals are more effective than 
quasi-steady modes.  This claim is erroneous.  Z98 have shown that the 
secular mass flow rate due to collective effects produced by density wave modes 
depends mainly on the wave amplitude and on pattern pitch angle 
(which in turn determines the potential-density phase shift), 
and quasi-steady density-wave modes are extremely effective not only 
in terms of providing the correct sense of mass flow for bulge-building,
but also in terms of soliciting 
the help of advective torques, which were shown in our current paper for 
observed galaxies, as well as in Z98 for simulated
spiral modes, to be in the same sense of angular momentum transport
as that of gravitational torques, and the total angular momentum transport is
in fact dominated by the advective torques due to collisionless shocks.  
Similarly, quasi-steady modes had been shown to induce significant 
secular heating of the disk, and can explain the age-velocity dispersion 
relation of the solar-neighborhood stars (Z98, Z99) and the size-linewidth
relation of the Galactic molecular cloud complexes (Zhang et al. 2001).
The collective instabilities at the spiral arms are also expected to
underlie the observed non-Schwarzschild
distribution of Milky Way stars (Dehnen 1998), since the collisionless
shocks completely decorrelate the phases of the otherwise regular
orbits under the influence of smooth applied spiral potential.
The mixing in the collisionless shocks could also explain
the spread in metallicity for stars of a given age (since the
spiral shock is on the order of 1 kpc in width as shown in Z96,
thus leads to the mixing of stars and ISM from neighboring regions.
The continuous inflow and outflow of matter due to successive nested
resonances also facilitate the mixing process, and it does not
require the presence of repeated transient wave trains as Sellwood had argued).
\item Through our work using observed galaxy images, we found that
in many instances the real physical galaxies in fact show a higher degree 
of coordinated behavior than the modes we were able to produce in 
N-body simulations.  This shows up for example in the plots of the bell-shaped
C$_g$ versus R curves, which turn out to be in general much 
cleaner in observed grand-design galaxies than those in simulations. This
we think is partially the result that observed galaxies have gone through a 
long history of natural evolution, so the degree of self consistency 
between the basic state and the modal characteristics is extremely good 
(at least for the grand-design ones).  In simulations, we start 
the initial simulation somewhat abruptly (i.e. from a given axisymmetric 
basic state that does not contain a density wave mode that such a 
basic state should possess).  And the subsequent boundary conditions
enforced (that of a rigid bulge and halo in 2D simulations) are also
artificial.   So once again using the analogy of 
the growth of a human being, this is akin to a person born with a 
birth defect, and subsequently having to struggle in an environment that is not
totally compatible with his natural inclinations.
Such a person thus will never truly feel at ease throughout his life.  Therefore, 
we should not take the results of simulations too literally.
The appearance of spurious modes in these simulations could simply be a result of 
the unrealistic initial-boundary conditions assumed due to the limitations
of computational facilities. The correct approach is to compare the results 
of these simulations and the results of observations with the aim to 
extract the essences of the underlying dynamical processes, going beyond the
superficial spurious details.
\item The conclusion of quasi-steady co-evolution of the basic state
of the disk and the density wave modes is also supported by another
aspect of our current work, which shows that for galaxies that are 
still at the initial stage of the secular
evolution sequence, such as the case of the late-type galaxy NGC 0628,
there is a poor coherence in both the phase shift plot as well as in the
galaxy image, and poor correspondence between the phase shift
zero crossings with any galaxy resonance features.  As the galaxy settles
down while the secular evolution proceeds, once galaxies reach
intermediate Hubble types, such as galaxy NGC 4321, there is
a very nice correspondence between the phase-shift-plot organization
and the galaxy-image resonance organization, signaling that an internal
dynamical equilibrium state has been reached. The delicacy of this
balance (i.e., if we use an M/L-corrected {\em stellar image} alone
without adding the gas maps in calculating the phase shift distribution,
we would not get as well-organized phase shift curves as when
we use the {\em total mass image}), shows that the equilibrium state is
well negotiated between the mass distribution and the kinematic
distribution of the density wave pattern, which can only come as
a result of the long-term evolution of spontaneously-formed
density wave modes, and not possible for random transient wave trains.
The observed correlation between basic state characteristics (i.e.,
the bulge-to-disk ratio) and pattern characteristics (i.e. the pattern
pitch angle), already known at Hubble's time and is part of the
basis of his classification scheme, is also most naturally
explained within the modal framework (e.g., Bertin et al. 1989),
at least qualitatively -- even though the details of the modal
theory may be challenged as Sellwood had shown in the first part
of his 2011 paper.
\end{itemize}

To summarize, we note that to be able settle the question of
transient versus modal nature of the density wave patterns in
disk galaxies, we have to first be clear what we regard as
a satisfactory definition of such a mode.  We assert that
a density wave is a mode if
\begin{enumerate}
\item It is a genuine instability in the underlying basic state
of the disk.
\item As a result it spontaneously emerges out of an originally
featureless disk, for which the bell-shaped distribution of
total angular momentum flux (or total torque coupling) of the mode
is chiefly responsible (Z98).
\item The properties of the emerged density wave patterns are
determined solely by the basic state characteristics and not
by the accidentals of noise.
\item The emergence of these patterns greatly accelerates the
entropy evolution of the parent basic state of the disk, as
is the requirement for all dissipative structures formed in
far-from-equilibrium systems.
\end{enumerate}

Upon examining the evidence in both the simulated and observed
disk galaxies possessing grand-design density wave patterns,
we conclude that these patterns do conform to the definition
of density wave modes as stated above.  Furthermore, we emphasize
that the modes in galaxies are indeed only quasi-steady,
and not eternal and unchanging, since the secular mass flow
they induce changes the basic state distribution, which in
turn affects what kinds of modes are compatible with it.
So these mode may change morphology continuously on time scales
a few times of the local dynamical time scale (the exact rate
of modal evolution differs depending on the density wave amplitude
and pattern opening angle, etc., which will set the mass flow rates).
In this context we once again bring out the connection with
nonequilibrium phase transitions, and the fact that the coherent
patterns formed in these transitions are in a state of dynamical
rather than statical equilibrium.

Many of the more chaotic looking density wave patterns in 
late type disks, on the other hand, are likely to be
in the process of evolving into grand-design
ones due to the secular evolution connection among galaxies 
along the Hubble sequence as we are trying to establish in
this paper.

\subsection{Implications on the Secular Morphological Transformation
of Galaxies}

In the results presented so far, a picture emerges of the secular
morphological transformation of galaxies driven mostly by
collective effects mediated by large-scale density wave modes.
This evolution is a well-coordinated quasi-steady evolution, with the 
formation and transformation of large-scale density wave modes (through
the formation of successive nested resonance patterns) in most 
circumstances compatible with the evolving basic state configurations.  

The secular evolution mediated by the density wave pattern
and its associated collective dissipation effect will slowly
change the basic state mass distribution, but this change is
accompanied by a corresponding change in the morphology, kinematics, and
other physical properties of the density wave modes (including
the formation of successive nested resonances), so at every state
of the transformation from the late Hubble type to early Hubble type, there
should be a good correspondence between the basic state properties and the
density wave modes, except during the brief periods in a certain
galaxy's life when they encounter gravitational perturbation from
a companion galaxy, such is the case for M51 or NGC 3627, where
it is seen that the coherence of the phase shift curves and the
kinematic and dynamical equilibrium are temporarily disturbed.
New equilibrium states are expected to be restored once the perturbation
ceases and the galaxy adjusts its mass distribution and modal
pattern to be once again mutually compatible -- 
the reason these new equilibrium states are always
possible is because for every basic state configuration there
is almost always a set of unstable modes corresponding to it.

As a result of this coordinated co-evolution of basic state of the
disk and the density-wave modes it supports, over the major
span of a galaxy's lifetime the effective 
evolution rate depends on the stage of life a galaxy is in.
Late- or intermediate-type galaxies, having larger-amplitude and open
density wave modes in its outer disk, would lead to 
larger mass flow rates and thus 
secular evolution rates in the outer disk.  Whereas for early type
galaxies the secular evolution activity is shifted to the central region
of a galaxy. In general the secular mass flow rate in the outer disks
of early type galaxies are low, and this is reflected in the modal
characteristics as well: i.e. early type galaxies generally have
either tightly wrapped spiral arms, or straight bars, both
correspond to small phase shift and thus small secular mass flow rate. 

The secular evolution rate also depends on the interaction state of
a galaxy, with the galaxy in group or cluster environments
generally having larger evolution rates due to the large-amplitude,
open spiral and bar patterns excited.  Zhang (2008) showed that
the interaction-enhanced, density-wave mediated evolution appear
to underlie the so-called morphological Butcher-Oemler effect 
(Butcher \& Oemler 1978) in rich clusters.  Furthermore, there is the well-known
morphology-density relation originally discovered by Dressler (1980),
and subsequent shown to hold over more than 4 orders
of magnitude in mean density spanning environments from field
and groups, onward to poor clusters and rich cluster outskirts, all the
way to dense cluster central region.  The universality of such
a correlation is in fact a powerful illustration of the unifying role
played by ``nurture-assisted-nature'' type of processes during
galaxy evolution.  The dependence of morphology on environmental
density shows that nurture plays an unquestionable role in
determining the average evolution rate, yet the well-known inverse
correlation of environmental density with merger rate shows that
the effect of the environment is not chiefly in the form  of violent
cannibalism, but rather through mild tidal perturbations.  Even
the moderately-violent and episodic ``harassment'' type of direct
damage to galaxy morphology is known to be ineffective at
transforming the morphology of large disk galaxies: they are shown
to mostly lead to the transformation of small disks to dwarf
spheroidals (Zhang 2008 and the references therein).  
These leave the most obvious driver for the observed
morphological transformation of galaxies in the different environments
and the origin of the morphology-density relation as the tidally-enhanced density
wave patterns operating over long times scales (though the time
scale can be significantly shortened during strong tidal evolution
which induce large-amplitude waves, and can thus enhance
the evolution rate by a factor of several hundred times compared
to galaxies in the quiescent environment -- since the effective
evolution rate due to the density waves depends on the wave
amplitude squared).

However, as we see in the current work, even for interacting galaxies
like M51 and NGC 3627, the environment exerts its effect through
the innate mechanism already present in individual galaxies, i.e.,
the density wave modes that are excited during the interaction
are still the intrinsic modes, the effect of interaction only
enhanced the amplitudes of these modes.

The flip side of this correlation of evolution rate with environment
is that some disk galaxies in isolated environments could have very
small secular evolution rate throughout the span of a Hubble
time, which explains the observational fact that many
disk galaxies in the fields were found to have evolved little
during the past few Gyr. Still, isolated environment cannot be
automatically equated to a slow evolution rate. 
One of the most impressive examples of mass redistribution
we have encountered is galaxy NGC 1530 (ZB07), which
lies in a surprisingly pristine environment, i.e., nearly
totally isolated.  Yet NGC 1530 has by far the largest mass flow
rate of all the galaxies we've calculated so far, on the order of
more than 100 solar mass per year, due to its large surface density
and the presence of a strong set of bar-spiral modal structure.
This shows that the initial conditions that galaxies inherited at
birth are likely to have played an important rate as well in determining
the subsequent morphological evolution rate. 

The secular morphological transformation of galaxies along
the Hubble sequence also implies that the required external gas
accretion rate to sustain the current-level of star-formation rate
in disk galaxies can be much reduced from the amount previously sought:
As a galaxy's Hubble type evolves from late to early, more and more of
its store of primordial gas will be exhausted, and its star-formation
rate will naturally decline.  But this is precisely the observed
trend of star-formation in galaxies along the Hubble sequence: i.e., the
early-type disk galaxies do not have nearly as much star
formation activity as late-type galaxies.  Galaxies thus do not
have to sustain their ``current'' level of star formation over cosmic time,
since what's current today will be history by the next phase of
their morphological evolution.

A further inference is that as galaxies evolve from late 
to early Hubble types, the central potential well will gradually 
get deeper, and more and more intricate nested resonances form 
in the nuclear region as a result 
(Zhang et al. 1993).  These successive resonances
form a continued chain of mass fueling to the central region
of a galaxy while the bulge itself grows, and this process could naturally
account for the observed correlations between the galaxy
bulge mass and central black-hole mass.

Finally, we note that the continuous evolution across the S0
boundary into disky Es may erase the distinction between pesudo bulges
and classical bulges, i.e., galaxies such as the Milky Way were 
previously thought to have classical bulges because of its
$r^{1/4}$ central density profile, yet the building up of the Galactic Bulge
is most likely through the secular mass accretion process over
the pass 10 billion years, only that the disky bulge further
relaxed into the $r^{1/4}$ shaped bulge with time.  In the end, as has
already been pointed out by Franx (1993), there might not be a clear distinction
between early type disks and disky Es, and only a gradual variation
of the bulge-to-disk ratio.  The recent result from the ATLAS3D
project of a significant disk component in all low-mass ellipticals
(Cappellari et al. 2011) also supports this continuous evolution 
trend from late type disk galaxies all the way to disky Es.

\section{CONCLUSIONS}

In this paper we presented the study of a small sample of
disk galaxies of a broad range of Hubble types in order to obtain an initial
estimate of the radial mass accretion/excretion rates of these galaxies,
to gauge the relevance of these processes to the secular morphological
transformation of galaxies along the Hubble sequence.  This study shows
that the mass flow rates obtained in typical disk galaxies are able to
produce significant evolution of their Hubble types over the
cosmic time, especially if such disk galaxies have undergone external
tidal perturbation, such as those encountered in group or cluster
environment.   We have found that the reasons past studies have concluded 
that secular evolution is only important for building up late-type pseudo-bulges
are, first of all, the neglect of the important role of stellar mass
accretion, and secondly, the neglect of the dominant role played by collective
effects enabled by self-organized density wave modes.  
The recognition that a significant fraction of disk galaxies 
occupying the present-day Hubble sequence were built from a slow 
morphological transformation process implies that the composition
of the dark halo of galaxies has to be mostly baryonic, a conclusion,
though controversial in the context of the currently-popular LCDM
paradigm, in fact naturally resolves many known paradoxes in this
paradigm, such as the core-cusp controversy of low-surface-brightness 
galaxies, as well as the difficulty of cosmological simulations to
form realistic disk galaxies when the material for disk formation
is not made predominantly of baryons.

\section*{ACKNOWLEDGMENTS}

We thank J. Sellwood, S. White and D. Pfenniger for helpful
exchanges.

\section*{REFERENCES}

Alcock, C. et al. 1997, ApJ, 486, 697 

\noindent
Bell, E. F. \& de Jong, R. S. 2001, ApJ, 550, 212

\noindent
Bell, E.F., McIntosh, D.H., Katz, N., \& Weinberg, M.D. 2003, ApJS, 149, 289

\noindent
Bershadskii, A, \& Screenivasan, K.R. 2003,
Phys. Lett. A, 319, 21

\noindent
Bertin, G., Lin, C.C., Lowe, S.A., \& Thurstans, R.P. 1989,
ApJ, 338, 78

\noindent
Binney, J., \& Tremaine, S. 2008, Galactic Dynamics, second ed. (Princeton:
Princeton Univ. Press) (BT08)

\noindent
Burgers, J.M. 1948, Adv. Appl. Mech. 1, 171

\noindent
Buta, R. 1988, ApJS, 66, 233

\noindent
Buta, R. \& Combes, F. 1996, Fundamentals of Cosmic Physics, 17, 95

\noindent
Buta, R., Corwin, H. G., de Vaucouleurs, G., de Vaucouleurs, A., \& Longo, G. 
1995, AJ, 109, 543

\noindent
Buta, R.J., Vasylyev, S., Salo, H., \& Laurikainen, E., 2005, AJ, 130, 506

\noindent
Buta, R. \& Williams, K. L. 1995, AJ, 109, 517

\noindent
Buta, R.J., \& Zhang, X. 2009, ApJS, 182, 559 (BZ09)

\noindent
Buta, R. et al. 2010, ApJS, 190, 147

\noindent
Butcher, H., \& Oemler, A., Jr. 1978, ApJ, 219,18; 226,559; 

\noindent
Cappellari, M. et al. 2011, MNRAS, 416, 1680

\noindent
Ceverino, D., Dekel, A., Mandelker, N.,
Bournaud, F., Burkert A., Genzel, R., \& Primack, J. 2011,
arXiv:1106.5587
 
\noindent
Chemin, L., Cayatte, V., Balkowski, C., Marcelin, M., Amram, P., van Driel, W.,
\& Flores, H. 2003, A\&A, 405, 89

\noindent
Chung, A., van Gorkom, J. H.; Kenney, J.D.P., Crowl, H., Vollmer, B. 2009
AJ, 138, 1741

\noindent
Cimatti, A., Daddi, E., \& Renzini, A. 2006, A\&A, 453, L29

\noindent
Clowe, D. et al. 2006, ApJ, 648, L109

\noindent
Cohen, J.G.  2002, ApJ, 567, 672

\noindent
Contopoulos, G. 1980, A\&A, 81, 198

\noindent
de Blok, W.J.G., 2010, AdAst 2010E, 5

\noindent
Devereux, N.A., Kenney, J.D., \& Young, J.S. 1992, ApJ., 103, 784

\noindent
Dehnen, W. 1998, AJ, 115, 2384

\noindent
de Grijs, R. 1998, MNRAS, 100, 595

\noindent
Donner, K.J., \& Thomasson, K., A\&A., 290, 785 (DT94)

\noindent
Dressler, A. 1980, ApJ, 236, 351

\noindent
Eyink, G.L., \& Sreenivasan, K.R. 2006, Rev. Mod. Phys. 78, 87

\noindent
Firsch, U. 1995, Turbulence: The Legacy of A.N. Kolmogorov (Cambridge: CUP)

\noindent
Flagey, N., Boulaner, F., Verstraete, L., Miville Desch\^enes, M. A., Noriega Crespo, A., \&
Reach, W. T. 2006, A\&A, 453, 969

\noindent
Foyle, K., Rix, H.-W., \& Zibetti, S. 2010, MNRAS, 407, 163

\noindent
Franx, M. 1993, in Proc. IAUS 153, 
Galactic Bulges, eds. H. Dejonghe \& H.J. Having (Dordrecht:
Kluwer), 243

\noindent
Fukugita, M., Hogan, C.J., \& Peebles, P.J.E. 1998, ApJ, 503, 518

\noindent
Genzel, R., Tacconi, L.J., Eisenhauer, F., Forster Schreiber, N.M., 
Cimatti, A., Daddi, E., Bouche, N. et al. 2006, Nature, 442, 786

\noindent
Gnedin, O., Goodman, J., \& Frei, Z. 1995, AJ, 110, 1105

\noindent
Gunn, J.E. et al. 1998, AJ, 116, 3040

\noindent
Haan, S., Schinnerer, E., Emsellem, E., Garcia-Burillo, S.,
Combes, F., Mundell, C.G., \& Rix, H. 2009, ApJ, 692, 1623

\noindent
Helfer, T. 2003, 
Thornley, M. D., Regan, M.W., Wong, T., Sheth, K., 
Vogel, S. N., Blitz, L.; Bock, D.C.-J.
ApJS, 145, 259

\noindent
Helou G. et al., 2004, ApJS, 154, 253

\noindent
Hernandez. O., Wozniak, H., Carignan, C., Amram, P.,
Chemin, L., \& Daigle, O. 2005, ApJ, 632, 253

\noindent
Hopkins, P.F., Bundy, K., Murray, N., Quataert, E., Lauer, T.R., Ma, C.P.,
2009, MNRAS, 398, 898

\noindent
Jablonka, P., Gorgas, J., \& Goudfrooij, P. 2002,
Ap\&SS, 281, 367

\noindent
Jalocha, J.,  Bratek, L., Kutschera, M. 2008, ApJ, 679, 373

\noindent
Jeans, J.H. 1928, Astronomy and Cosmology (Cambridge:CUP)

\noindent
Kendall, S., Kennicutt, R. C., Clarke, C., \& Thornley, M. 2008, MNRAS, 387, 1007

\noindent
Kennicutt, R.C. et al. 2003, PASP, 115, 928

\noindent
Koopmans, L.V.E., \& De Bruyn, A.G. 2000, A\&A, 358, 793

\noindent
Kormendy, J., \& Kennicutt, R.C. 2004, ARAA, 42, 603

\noindent
Laurikainen, E. \& Salo, H. 2002, MNRAS, 337, 1118

\noindent
Lilly, S., Abraham, R., Brinchmann, J., Colless, M., Crampton, D., Ellis, R.,
Glazebrook, K., Hammer, F., Le Fevre, O., Mallen-Ornelas, G., Shade, D.,
\& Tresse, L. 1998, in The Hubble Deep Field, eds. M. Livio, S.M. Fall,
\& P. Madau (Cambridge: CUP), 107

\noindent
Lin, D.N.C., \& Pringle, J.E., 1987, MNRAS,
225, 607

\noindent
Lopez-Sanjuan, C., Balcells, M., Perez-Gonzalez, P.G., Barro, G.,
Garcia-Dabo, C.E., Gallego, J., \& Zamorano, J. 2009,
A\&A, 501, 505

\noindent
Lynden-Bell, D., \& Kalnajs, A.J. 1972, MNRAS, 157, 1

\noindent
Mahdavi, A., Hoekstra, H., Babul, A., \& Balam, D.D. 2007, ApJ, 668, 806

\noindent
Martinez-Garcia, E.E., Gonzalez-Lopezlira, R.A., Bruzual, A. G.
2009, ApJ, 694, 512

\noindent
Martinez-Garcia, E.E., Gonzalez-Lopezlira, R.A., Bruzual, A. G.
2011, ApJ, 734, 122

\noindent
McGaugh, S.S., Schombert, J.M., de Blok, W.J.G., \& Zagursky, M.J. 2010, 
ApJ, 708, L14

\noindent
Meidt, S., et al. 2012, ApJ, 744, 17

\noindent
Nelan, J.E. et al. 2005, ApJ, 632, 137

\noindent
Nicol, M.-H. 2006, MPIA Student Workshop, http://www.
\newline mpia-hd.mpg.de/70CM/3rdworkshop/presentations/Marie-Helene$\_$Nicol$\_$dark$\_$matter$\_$SF$\_$rate.pdf

\noindent
Nicolis, G., \& Prigogine, I. 1977, Self-Organization in Nonequilibrium Systems
(NY: Wiley)

\noindent
Onsager, L. 1949, Nuovo Cimento, Suppl. 6, 279

\noindent
Ozernoy, L.M. 1974, in In: Confrontation of Cosmological Theories with Observational Data; Proc.
IAUS 63 (Dordrecht: Reidel), 227

\noindent
Peebles, P.J.E., \& Yu, J.T. 1970, ApJ, 162, 815

\noindent
Persic, M., \& Salucci, P. 1992, MNRAS, 258, 14

\noindent
Pfenniger, D. 2009, in Chaos in Astronomy, p.63 (Berlin: Springer)

\noindent
Pfenniger, D., \& Combes, F. 1994, A\&A, 285, 94

\noindent
Polyakov, A.M. 1993, Nucl. Phys. B, 396,367

\noindent
Quillen, A. C., Frogel, J. A., \& Gonzalez, R. 1994, ApJ, 437, 162

\noindent
Reach, W. T. et al. 2005, PASP, 117, 978

\noindent
Sakharov, A.D. 1965, Zh.E.T.F., 49, 345

\noindent
Schwarz, M. P. 1984, MNRAS, 209, 93

\noindent
Schwinger, J. 1951, Phys. Rev. 82, 664

\noindent
Sellwood, J.A., 2008, in Formation and Evolution of Galaxy
Disks, Eds. J.G. Funes, S.J. and E.M Corsini (SFO: ASP), 241
 
\noindent
Sellwood, J.A., 2011, MNRAS, 410, 1637

\noindent
Sellwood, J.A., \& McGaugh, S.S. 2005, ApJ, 634, 70

\noindent
Shostak, G.S., van Gorkom, J.H., Ekers, R.D., Sanders, R.H.,
Goss, W.M., \& Cornwell, F.J. 1983, A\&A, 119, L3

\noindent
Shu, F.S. 1992, The Physics of Astrophysics, vol. II.  Gas Dynamics,
(Mill Valley, Univ. Sci. Books)

\noindent
Sofue, Y. 1996, ApJ, 458, 120

\noindent
Sofue, Y., Tutui, Y., Honma, A., Tomita, A., Takamiya, T., Koda, J.,
\& Takeda, Y. 1999, ApJ, 523, 136

\noindent
Sparke, L.S., \& Sellwood, J.A. 1987, MNRAS, 225, 653

\noindent
Sunyaev,R.A., \& Zeldovich, Y.A.B. 1970, Ap\&SS, 7, 3

\noindent
Toomre, A. 1969, ApJ, 158, 899

\noindent
Tully, R.B. 1974, ApJS, 27, 415

\noindent
van Dokkum, P.G. et al. 2011, arXiv:1108.6060

\noindent
Walter, F., Brinks, E., de Blok, W.J.G., Bigiel, F.,
Kennicutt, R.C.Jr., Thornley, M.D., Leroy, A. 2008, AJ, 136, 2563

\noindent
White, S.D.M. 2009, in Galaxy Disk in Cosmological Context, Proc.
IAUS 254, eds. J. Andersen, J. Bland-Hawthorn, \& B. Nordstrom

\noindent
Wolfe, A.M., Gawiser, E., \& Prochaska, J.X. 2005, ARA\&A, 43, 861

\noindent
York, D.G. et al. 2000, AJ, 120, 1579

\noindent
Zeldovich, Y.B. 1970, A\&A, 5, 84

\noindent
Zhang, X. 1992, Ph.D. Dissertation, University of California, Berkeley

\noindent
Zhang, X. 1996, ApJ, 457, 125 (Z96)

\noindent
Zhang, X. 1998, ApJ, 499, 93 (Z98)

\noindent
Zhang, X. 1999, ApJ, 518, 613 (Z99)

\noindent 
Zhang, X. 2004, 
in "Penetrating Bars through the Masks of Cosmic Dust: 
The Hubble Tuning Fork Strikes a New Note", eds. D. Block et al.,
astro-ph/0406583

\noindent 
Zhang, X. 2008, PASP, 120, 121

\noindent
Zhang, X. \& Buta, R. 2007, AJ, 133, 2584 (ZB07)

\noindent
Zhang, X., Lee, Y., Bolatto, A., \& Stark, A.A. 2001, ApJ, 553, 274

\noindent
Zhang, X., Wright, M.C.H., \& Alexandria, P. 1993, ApJ,
418,100

\section*{APPENDIX A. DESCRIPTIONS OF THE PROCEDURES FOR OBTAINING
SURFACE MASS DENSITY MAPS}

Stellar surface mass density maps can be made from two-dimensional
images using surface colors as indicators of stellar mass-to-light
ratio (Bell \& de Jong 2001). Calibrated surface brightness maps can be
converted to units of $L_{\odot}$ pc$^{-2}$, and then multiplied by
color-inferred $M/L$ values in solar units to give the surface mass
density $\Sigma$ $(i,j)$ in units of $M_{\odot}$ pc$^{-2}$ at pixel
coordinate $(i,j)$. Thus our approach is two-dimensional and not
based on azimuthal averages of the luminosity distribution.

It is widely regarded that the best images to use for mapping stellar
mass distributions are infrared images, because these penetrate
interstellar dust more effectively than optical images and also because
such images are more sensitive to the light of the old stellar
population that defines the backbone of the stellar mass distribution.
For our study here, we used two principal types of images: (1) an
Infrared Array Camera (IRAC) image taken at 3.6$\mu$m for the {\it
Spitzer Infrared Nearby Galaxies Survey} (SINGS, Kennicutt et al.
2003); and (2) a 0.8$\mu$m $i$-band image obtained from the Sloan
Digital Sky Survey (SDSS; Gunn et al.  1998; York et al. 2000). The
pixel scales are 0\rlap{.}$^{\prime\prime}$75 for the 3.6$\mu$m images
and 0\rlap{.}$^{\prime\prime}$396 for the $i$-band images. Only four of
our six galaxies had SDSS data available.  For NGC 3351 and 3627, the
$i$-band images were rescaled to the scale of the 3.6$\mu$m images.
For NGC 5194, the 3.6$\mu$m mass map was rescaled to the scale of the
$i$-band to insure that the same area is covered on the two images.

Bell \& de Jong (2001) give linear relationships between the log of the
$M/L$ ratio in a given passband and a variety of color indices in the
Johnson-Cousins systems. Bell et al. (2003) give the same kinds of
relations for SDSS filters.  For the 3.6$\mu$m images of NGC 628, 3351,
3627, and 5194, we used $B-V$ as our $M/L$ calibration color index,
while for M100 we used $B-R$. For the SDSS $i$-band, we used $g-i$ as
our main color. Using different color indices to scale the two
base images from array units to solar masses per square parsec
means that independent photometric calibrations are used, allowing
us to examine effects that might be due to $M/L$ uncertainties or
systematics.

The base images we have used have different advantages and
disadvantages for mass map calculations. For example, IRAC 3.6$\mu$m
images have the advantages of much greater depth of exposure than most
ground-based near-IR images and also they give the most extinction-free
view of the old stellar background. Nevertheless, 3.6$\mu$m images are
affected by hot dust connected with star-forming regions and by a
prominent 3.3$\mu$m emission feature due to a polycyclic aromatic
hydrocarbon compound that also is associated with star-forming regions
(see Meidt et al. 2012).  These star-forming regions appear as
conspicuous ``knots" lining spiral arms in 3.6$\mu$m images, such that
the appearance of a galaxy at this mid-IR wavelength is astonishingly
similar to its appearance in the $B$-band, minus the effects of
extinction (e. g., Buta et al. 2010).

The advantages of the SDSS $i$-band are the shear quality of the SDSS
images in general (especially with regard to uniformity of background),
the reduced effect of star-forming regions compared to the $B$-band and
the 3.6$\mu$m band, and the pixel scale which is almost a factor of two
better than for the 3.6$\mu$m IRAC band. Nevertheless, extinction at
0.8$\mu$m is more than 40\% of that in the $V$-band, so $i$-band images
are considerably more affected by extinction than are 3.6$\mu$m images.

We use the 3.6$\mu$m and $i$-band mass maps as consistency checks on
our results since neither waveband is perfect for the purpose intended.
In practice, the star-forming region problems in the 3.6$\mu$m
band can be reduced using an 8.0$\mu$m image if available (Kendall et al.
2008). These ``contaminants" can also be eliminated using Independent
Component Analysis (Meidt et al. 2012) if no 8.0$\mu$m image is
available. 

\subsection*{A1. Procedure}

{\it SDSS images} - Images were downloaded from the SDSS archive using
the on-line DAS Coordinate Submission Form. For the large galaxies in
our sample, it was necessary to download multiple images per filter to cover
the whole object (ranging from 3 for NGC 3351 to 9 for NGC 4736). These
were mosaiced, cleaned of foreground and background objects, and then
background-subtracted using routines in the Image Reduction and
Analysis Facility (IRAF). SDSS images were available for 5 of the 6
sample galaxies, excluding NGC 628.

The zero points for the $g$ and $i$-band images were obtained using
information given on the SDSS DR6 field pages. The airmasses $x_g$ and
$x_i$, calibration zero points $aa_g$ and $aa_i$, and extinction
coefficients $kk_g$ and $kk_i$, were extracted from this page and the
zero points appropriate to the main galaxy fields were derived as

$$zp_g = -(aa_g+kk_g x_g)+2.5log(a_{pix}t)$$
$$zp_i = -(aa_i+kk_i x_i)+2.5log(a_{pix}t)$$

\noindent
where $a_{pix}$ is the pixel area equal to (0.396)$^2$ or (0.75)$^2$
if rescaled to the 3.6$\mu$m image and $t$ is the integration time of
53.907456s.  For the galaxies that were rescaled, we
matched the coordinates of the SDSS $g$- and $i$-band images to the
system of the 3.6$\mu$m image, which made it necessary to modify $zp_g$
and $zp_i$ to account for the new pixel size. This was done using IRAF
routines GEOMAP and GEOTRAN. Foreground stars were selected that were
well-defined on all three of the images. GEOMAP gave the rotation,
shifting, and scale parameters, while GEOTRAN performed the actual
transformations. Before the final transformations, the point spread
function (PSF) of the images was checked. If the PSFs of the $g$- and
$i$-band images were significantly different, the image with the best
seeing was matched to the other using IRAF routine GAUSS. For M100, the
higher resolution of the SDSS images relative to the 3.6$\mu$m image
was retained.

The next step was to deproject these images in flux-conserving mode.
For this purpose, IRAF routine IMLINTRAN was used with the adopted
orientation parameters (see Table 1). When possible, we used kinematic
orientation parameters for the deprojections. In the case of NGC 3351,
we also used isophotal ellipse fits to deduce these parameters.
No photometric decomposition was used for the deprojections; the bulges
were assumed to be as flat as the disk. 

The SDSS $i$-band mass map was derived as follows. The absolute magnitude of
the Sun was taken to be $M_i$ = 4.48 (CNA Willmer), with which the zero
point needed to convert $i$-band surface brightnesses to solar $i$-band
luminosities per square parsec is $zp_{\odot i}$ = 26.052.  To convert
from these units to solar masses per square parsec, Bell et al. (2003)
give a simple relationship:

$$log{M\over L_i} = -0.152 + 0.518(\mu_g - \mu_i)_o$$

\noindent
where $(\mu_g - \mu_i)_o$ is the reddening-corrected $g-i$ color index.
The only correction made was for Galactic reddening. 
This was judged using information from the NASA/IPAC Extragalactic Database
(NED).\footnote{This research has made use of the NASA/IPAC Extragalactic
Database (NED), which is operated by the Jet Propulsion Laboratory, California
Institute of Technology, under contract with the National Aeronautics
and Space Administration.} Although NED lists extinction in the
broad-band Johnson filters like $B$ and $V$, it does not list the
extinctions in $g$ and $i$. To get these, the York Extinction Solver
(YES; McCall 2004) was used on the NED website.  The actual use of the
above equation requires that some account be made of noise, because
SDSS images are not as deep as 3.6$\mu$m images. For our analysis, it
is important to use the two-dimensional color index distribution, not
an azimuthal average except in the outer parts of the disk where there
is little azimuthal structure. Our procedure was to derive
azimuthally-averaged surface brightness and color index profiles, and
to interpolate colors from these profiles in the outer regions where
noise made the actual pixel color too uncertain to be used in the above
equation. Smoothing was also used even in intermediate regions.
Usually, the colors in the inner regions were used without smoothing,
and then annular zones of increasing radius used an $n$$\times$$n$
median box smoothing.

The median-smoothed raw $i$-band counts, $C_i (i,j)$, at array position
$(i,j)$, were then converted to surface mass density $\Sigma (M_{\odot}
pc^{-2})$ through

$$\Sigma(i,j) = C_i(i,j) \times 10^{-0.4(zp{_i}
- A{_i} - zp_{\odot i})} $$
$$\times 10^{-0.152 + 0.518[\mu_g(i,j) - \mu_i(i,j)
- A_g + A_i)]}$$

The final SDSS mass maps usually would have left-over foreground stars
that had not been removed, or dark spots in areas of star-formation
where the color index was affected by too much blue light. These spots
were removed using IRAF routine IMEDIT.

{\it 3.6$\mu$m Images} - The procedure for these is in many respects
similar to the procedure used for the SDSS images, except that
different color index maps are used and these are not directly linked
to the $M/L$ ratio at 3.6$\mu$m. For 3.6$\mu$m maps, we used $B-V$ or
$B-R$ colors where the individual images are calibrated using
photoelectric multi-aperture photometry. The sources and error analysis
of such photometry, originally used to derive total magnitudes and
color indices for RC3, are described in Buta et al. (1995) and Buta \&
Williams (1995).\footnote{A full catalogue of $UBV$ measurements may be
found at http://kudzu.astr.ua.edu/devatlas/revUBV.ecat.txt.} The
typical uncertainty in the $B$-band zero point from this approach is
0.017 mag while for $V$ it is 0.015 mag, based on 8-27 measurements.
Fewer measurements are available for $R$-band calibrations from this
approach; the uncertainties in these are described by Buta \& Williams
(1995). The $B$ and $V$ images used were downloaded from the SINGS
database webpage. $B-R$ was used only for M100 and is based on images
due to B. Canzian from observations with the USNO 1.0m telescope.

The surface brightnesses in the 3.6$\mu$m images were derived using a
common zero point of 17.6935, based on the calibration from Reach et
al. (2005). IRAC images are in units of MegaJanskys per steradian.

For our study here, we have made a ``hot dust correction" to the
3.6$\mu$m images using a procedure similar to that outlined by Kendall
et al. (2008) since all of our galaxies have an 8.0$\mu$m image
available. The first step is to match the coordinate systems of the 3.6
and 8.0$\mu$m images and then subtract a fraction (0.232) of the
3.6$\mu$m flux from the 8.0$\mu$m image to correct the latter for
continuum emission (Helou et al. 2004). Then, a fraction $R_{3.6/8.0}$
of the net dust map is subtracted from the 3.6$\mu$m map to give an
image corrected for the hot dust emission. In general this method did
improve our mass maps. The factor we used, $R_{3.6/8.0}$ = 0.059, is at
the low end of the flux ratios found by Flagey et al. (2006) for the
Galactic diffuse interstellar medium.

For converting 3.6$\mu$m surface brightnesses to solar luminosities per
square parsec, the absolute magnitude of the Sun was taken to be 
$M_{3.6}$ = $M_L$ = 3.27 for the $L$-band, which is close to the same
wavelength (Worthey 1994). This corresponds to $zp_{\odot,3.6}$ =
24.842.  To convert from these units to solar masses per square parsec,
two steps are used. The first is to derive the $K$-band $M/L$ ratio
from the corrected broadband colors. For $B-R$, the
relation used is (Bell et al. 2003):

$$log{M\over L_K} = -0.264 + 0.138(\mu_B - \mu_R)_o$$

\noindent
where $(\mu_B - \mu_R)_o$ is the Galactic reddening-corrected color
index based on extinctions listed in NED. This relation differs substantially
from that for the same color and near-IR band listed in Table 1
of Bell \& de Jong (2001), which Bell et al. (2003) suggest is due
to a larger metallicity scatter than accounted for in the earlier paper.

The second step is to convert ${M\over L_K}$ into ${M\over L_{3.6}}$. We
used a simple relation due to Oh et al. (2008), based on stellar population
synthesis models with a range of metallicities and star formation histories;

$${M\over L_{3.6}} = 0.92{M\over L_K} - 0.05$$

\noindent
Because of the higher signal-to-noise in the 3.6$\mu$m image, it was
not necessary to use the staggered median smoothing approach used for
the SDSS images. The surface mass densities were then derived from the
array values $C_{3.6}$ using

$$\Sigma(i,j) = C_{3.6}(i,j) \times
10^{-0.4(zp_{3.6} - A_{3.6} - zp_{\odot,3.6})} \times
$$
$$
0.92(10^{-0.264 + 0.138[\mu_B(i,j) - \mu_R(i,j) - A_B + A_R)]}) - 0.05$$

\noindent 
For those cases where $B-V$ was used instead (NGC 628, 3351, 3627, and
5194), the $M/L$ relation applied was (Bell et al. 2003)

$$log{M\over L_K} = -0.206 + 0.138(\mu_B - \mu_V)_o$$

\noindent
where against the Galactic extinction corrections were taken from NED.

{\it Addition of gas} - Our analysis requires total mass maps, and thus
it is essential to add in the contributions of atomic and molecular
gas. Five of our sample galaxies were observed in The HI Nearby Galaxy
Survey (THINGS; Walter et al.  2008), while M100 was observed as part
of the VLA Imaging of Virgo Spirals in Atomic Gas (VIVA) program (Chung
et al. 2009). All 6 are included in the BIMA Survey of Nearby Galaxies
(BIMA SONG, Helfer et al. 2003). VIVA provides an HI image with
resolution 31\arc\ $\times$ 28\arc\ and pixels 10\arc\ $\times$ 10\arc.
THINGS provides HI maps with a resolution of
$\approx$6$^{\prime\prime}$ and pixels 1\rlap{.}$^{\prime\prime}$5 in
size. The maps in all cases are publicly available, and the procedure
for adding both HI and CO into the mass maps was the same. For the HI
map, the total flux in the image was integrated to a radius consistent
with the HI size of the object using IRAF routine PHOT. The map was
then scaled to the measured total flux $S_{HI}$ given in Table 5 of
Walter et al. (2008) and in Table 3 of Chung et al. (2009). With this
scaling, each pixel in the image then has the same units, Jykm
s$^{-1}$, and can be converted to mass using $ M_{HI}(i,j) = 2.36\times
10^5 D^2 S_{HI}(i,j)$, where $D$ is the distance in Mpc. Dividing each
value by the number of square parsecs in a pixel, this gives the
distance-independent surface mass density of HI gas in units of
$M_{\odot}$ pc$^{-2}$.

For the CO map, Table 4 of Helfer et al. (2003) gives the global CO
flux, $S_{CO}$, for each galaxy. The same procedure as for the HI map
gives the scaling of each pixel, such that the mass in each pixel is
$M_{H_2}(i,j) = 7845.0 D^2 S_{CO}(i,j)$, where a conversion factor of
$X$ = 2$\times$10$^{20}$ has been used (Helfer et al.).  The scale of
the publicly available images is 1\arcs 0 per pixel.

The pixel sizes of the two gas maps were different from the pixel sizes
of the 3.6$\mu$m and $i$-band images. IRAF routine IMLINTRAN was used
to create scaled maps with the same pixel sizes, outputted to an
appropriate center of the galaxy. Each paper gave the right ascension
and declination of the pointing center, which was compared with the
coordinates in the RC3 to judge where we should set the centers in our
mass maps. Each scaled map had its own flux-scale
factor to keep the total masses the same as published by Helfer et al.
(2003), Walter et al. (2008), and Chung et al. (2009).

\noindent
{\it Gravitational Potentials} - The potentials were calculated using the
2D Cartesian approach outlined by Binney \& Tremaine (2008) as described
in ZB07 (and similar to the approach used by Quillen et al. 1994).
An important parameter needed in this calculation is the vertical
scale height, assuming an exponential vertical density distribution.
We used the approximate radial scalelengths listed in Table 1 and
information from de Grijs (1998) to judge scale heights. Being bright
galaxies, there are other sources of radial scale-length determinations
for our sample. A literature search showed good agreement between
our estimated values and other sources except for NGC 4736, whose
complex structure causes a large spread in values, and for NGC 5194,
which is complicated by its companion.

\noindent
{\it Uncertainties in mass map determinations and derived results} -
The uncertainties in our mass maps come from a variety of sources. In
general, photometric calibration uncertainties are small, and less than
0.05 mag. The principal uncertainties come from the $M/L$ calibrations,
effects of dust, and from deprojection and orientation parameter
uncertainties. According to Bell et al. (2003), typical uncertainties
in a color-dependent $M/L$ involving a near-IR band (such as $M/L_K$)
is $\pm$0.1 dex for redder $B-V$ and $B-R$ colors, and $\pm$0.2 dex for
bluer colors. We have shown that the Bell et al. (2003) $B-V$/$B-R$
calibrations with 3.6$\mu$m as the base stellar mass image give
azimuthally-averaged radial surface mass density profiles very similar
to those given by the the Bell et al. (2003) $g-i$ calibration with
the $i$-band as the base stellar mass image. From comparisons between
the two base images and using also the Bell \& de Jong (2001)
$B-V$/$B-R$ calibrations, we found that phase shift distributions are
more robust to $M/L$ uncertainties than are mass flow rates, although
even the latter agree fairly well between the two base images.

On the issue of orientation parameters, we did not experiment with different
values but simply point to section 5.2 of BZ09 where tests were made of
the impact of such uncertainties, including problems of bulge deprojection.
The CR radii we list in Table 2 are for the assumed orientation parameters
in Table 1.

Uncertainties in the assumed vertical scale-heights were examined for M100.
Reducing $h_z$ from 12\arcs 6 in Table 1 to 3\arcs 8 had almost no effect
on the phase shift distribution. In the outer disk near 7 kpc, the flow
rate is increased by about 10\% but at 1 kpc the difference is about 30\%. 
Thus even a drastic difference in $h_z$ has only a relatively small effect
on our results.

The issue of dust enters in the uncertainties in two ways: through the
significant impact of dust on the optical $M/L$ calibrating colors, and
through the ``hot dust correction" to the 3.6$\mu$m image. The former
is less of a problem than might be thought. As noted by Bell et al.
(2003), the effects of dust approximately cancel out to 0.1-0.2 dex
when estimating color-derived $M/L$ values because, in most passbands,
stellar populations and dust predict about the same amounts of
reddening per unit fading. That is, while dust reddens the starlight,
redder colors imply higher $M/L$, which effectively can reduce the
impact of dust lanes. Our results here basically verify this point. 
The uncertainty in the hot dust correction lies mainly in the
factor R$_{3.6/8.0}$. We chose a low end value from Flagey et al.
(2006) for this correction, but higher values may be appropriate
for some galaxies (e.g., Kendall et al. 2008).

\section*{APPENDIX B. CATEGORIES OF BAR-SPIRAL MODAL MORPHOLOGY}

In this appendix, we give the schematics of the phase shift
distributions of the various types of bar-spiral morphology, followed
by examples of real galaxies that we have analyzed in BZ09.  The
sequence we present with these examples agrees roughly with the order
of Hubble sequence from late to early, in order to show possible
evolutionary connections between the various morphological patterns.

\begin{figure}
\vspace{120pt}
\centerline{
\includegraphics{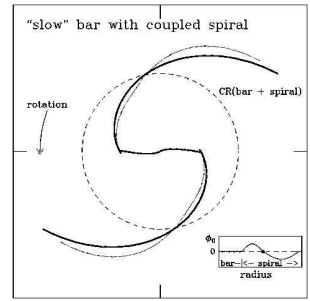}
}
\caption{Schematic of the phase shift distribution for the slow bar.}
\label{fg:Fig22}
\end{figure}

\begin{figure}
\vspace{120pt}
\centerline{
\includegraphics{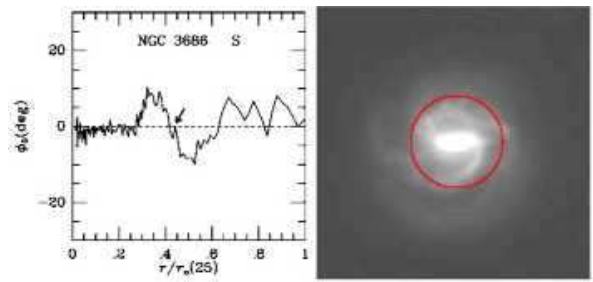}
}
\caption{Examples of the phase shift distribution and morphology
for the slow bar}
\label{fg:Fig23}
\end{figure}

In Figures \ref{fg:Fig22} and \ref{fg:Fig23},
we present the schematic and an example of a slow bar.  These bars
are characterized by the fact that the bar ends within CR, with possible
spiral structure emanating from the bar ends.  The hosts are
predominantly late-type galaxies, with a
significant flocculent pattern in the outer regions.  Closer
inspection of the phase shift plot for NGC 3686 shows that around the
location of the bar end, a new phase shift P/N transition
is in the process of forming. Thus the slow bar appears to be
a short-lived phase in the process of evolving into bar-driven
spirals, which we will analyze next.

\begin{figure}
\vspace{120pt}
\centerline{
\includegraphics{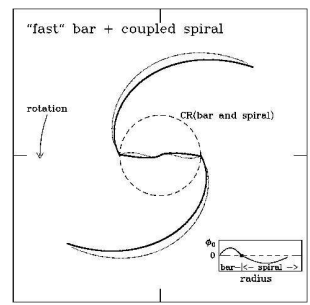}
}
\caption{Schematic of the phase shift distribution for the
fast bar with coupled spiral pattern.}
\label{fg:Fig24}
\end{figure}

\begin{figure}
\vspace{120pt}
\centerline{
\includegraphics{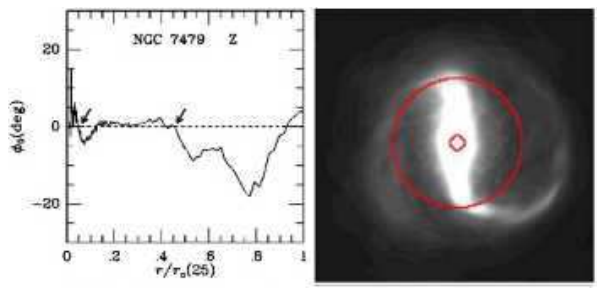}
}
\caption{Examples of the phase shift distribution and morphology for the
fast bar with coupled spiral pattern.}
\label{fg:Fig25}
\end{figure}

In Figures \ref{fg:Fig24} and \ref{fg:Fig25},
we present the schematic and an example of fast bars with coupled
spirals, commonly referred to as bar-driven sprials. 
These patterns most often appear in intermediate-type galaxies,
and the bar-driven spiral appears to have evolved through stages
of either a slow bar or else a skewed long bar.

\begin{figure}
\vspace{120pt}
\centerline{
\includegraphics{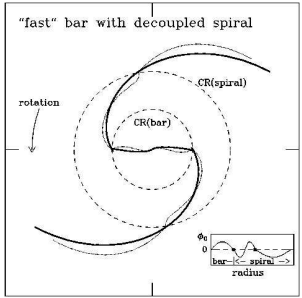}
}
\caption{Schematic of the phase shift distribution for the
fast bars with decoupled spiral pattern.}
\label{fg:Fig26}
\end{figure}

\begin{figure}
\vspace{220pt}
\centerline{
\includegraphics{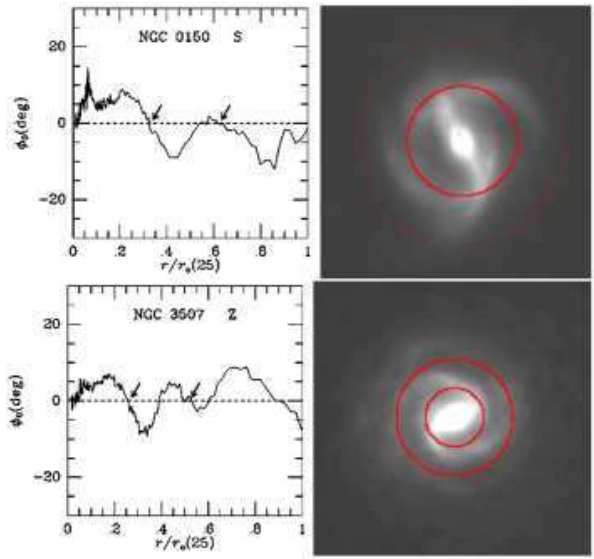}
}
\caption{Examples of the phase shift distribution and morphology for the
fast bars with decoupled spiral pattern.}
\label{fg:Fig27}
\end{figure}

In Figures \ref{fg:Fig26} and \ref{fg:Fig27},
we present the schematic and two examples of fast bars with decoupled
spirals. On the one hand, these appear to be the further evolutionary stage
of a bar-driven spiral, with the inner bar decoupled from the
outer spirals and with the bar end coinciding roughly with the
CR radii.  The signature of the decoupling (at the N/P
phase shift crossing) shows up as branching of the spiral arms
disconnected from the bar-end.  Close inspection of the images
and the phase shift plots show that
the two examples we give will evolve into somewhat different configurations
later on:  NGC 3507 appears to be evolving towards an inner bar-driven
spiral followed by an outer spiral, whereas,
NGC 0150 appears to be evolving
towards superfast bar with decoupled spiral that we will discuss next.

\begin{figure}
\vspace{120pt}
\centerline{
\includegraphics{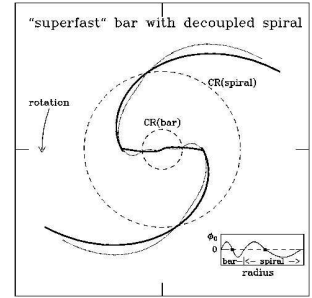}
}
\caption{Schematic of the phase shift distribution for the
super-fast bars with decoupled spiral pattern.}
\label{fg:Fig28}
\end{figure}

\begin{figure}
\vspace{120pt}
\centerline{
\includegraphics{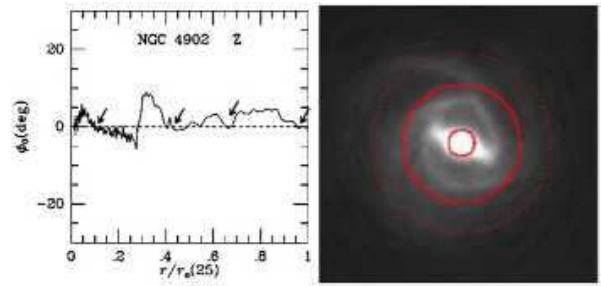}
}
\caption{Examples of the phase shift distribution and morphology for the
super-fast bars with decoupled spiral pattern.}
\label{fg:Fig29}
\end{figure}

In Figures \ref{fg:Fig28} and \ref{fg:Fig29}, we present the schematic 
and an example of the so-called super-fast bars with decoupled spirals.
The phrase super-fast bars are used to denote the fact that the bars
extend significantly beyond their CR radii. In ZB07, we have shown a
case of a single isolated bar in the galaxy NGC 4665, where the bar ends
extend about 10-20\% beyond the CR radius, and argued that in this
case it is reasonable to expect the bar to be longer than the CR radius
because the SWING amplified over-reflected waves from the inner disk
must penetrate CR into the outer disk as transmitted wave in order to
have the overall angular momentum budget balance.  In our current plot,
the super-fast section of the bars are straight segments emanating from the location
of an inner oval.  The end of the bar coincides not with CR but with the
next N/P phase shift crossing after the CR.  This is reasonable
because modal growth requires that for a complete self-sustained
mode there must be a positive phase shift packet followed by the
negative phase shift packet, with the two packets joining at CR.  This
is because the density wave/mode has negative energy and angular
momentum density inside CR (Shu 1992), and for its spontaneous growth the potential
must lag the density -- which leads to the positive 
potential-density phase shift --
in order for the wave to torque the disk matter in the correct sense to
lead to its own spontaneous growth by losing angular momentum to the disk matter.
Vice versa for the modal content outside CR.  Therefore we see that for
the kind of twin-bars joining the central oval the mode has little choice
but to have the N/P phase shift crossing be at the end of the bar.

Evidence supporting the claim that an N/P crossing is the location 
where the modal pattern speed changes discontinuously is that 
there is a pronounced ring-like structure at the radius of the N/P crossing, obviously
caused by the snow-plough effect of the interaction of the inner and outer modes.
Similar configuration of a central oval joined by straight super-fast bars
are also observed in NGC 3351, NGC 1073, NGC 5643, and in the central region of
NGC 4321.  As a matter of fact, since this configuration requires
the central oval connecting to the straight bars, it is always found
in the central configuration of the nested modes.

Comparing the last two types of morphologies, we can clearly see
that the fast-bars-with-decoupled-spirals appear to be evolving
into super-fast-bars-with-decoupled spirals (i.e., in the phase shift plot
for NGC 0150 we see that a new P/N crossing at the end of inner
oval in in the process of forming, or dropping down to zero.  When
it is fully formed this will become a super-fast bar.

It is no coincidence that the super-fast bars, which often appear
in early type galaxies, have a more rounded nuclear pattern and
very straight two segment of bars connecting to it: both of these
patterns are not much skewed and they correspond to very small
phase shift, and slow secular evolution rate as befits the
early type galaxies.  The fact that bar-driven spirals mostly 
have skewed nuclear bar patterns followed by trailing spiral segments
that taper into narrow tails, whereas super-fast bars most often
have rounded nuclear patterns followed by very straight bar segments
which broaden into dumb-bell shaped pile-up of material at the mode-decoupling
radii shows that super-fast bars are real, and are of completely different
modal category than bar-driven spirals.

Note that these four categories are the main types of bar-spiral
associations, but they do not exhaust all the morphological
types encountered in real galaxies.  For example, the above
categories did not include the cases of either pure spiral
(i.e. NGC 5247 analyzed in ZB07), or pure bar
(i.e. NGC 4665 also analyzed in ZB07).

From the above analysis we see that distinctive phase shift patterns
seem to delineate distinctive galaxy morphology proto-types, with the
morphology of galaxies within a type category repeatable to a high
degree.  These morphological features also seem to correlate with 
the Hubble types of the basic state of the galactic disks (i.e., its
being early, intermediate, or late).  These correlations are naturally 
explained under modal theory of density wave patterns since the
density wave modal morphology are determined by the basic state
characteristics (i.e., the radial distribution of surface density.
rotation curve, and velocity dispersion in galaxy).  The fact that
the PDPS method can consistently pick out the given set of modal
morphology and classify its resonance structure shows that its
success is not an accident, but rather supported by the underlying
modal structure of the density wave patterns.

\end{document}